\documentclass[11pt,preprint]{aastex}

\pdfoutput=1
\usepackage{graphicx, graphics}

\usepackage[normalem]{ulem}
\usepackage{color}
\usepackage{natbib}

\received{}
\revised{}
\accepted{}

\shorttitle{Local Absorption Line}
\shortauthors{Fang et al.}

\begin{document}

\title{{\sl XMM}-Newton Survey of Local {\rm {O{\scriptsize VII}}} Absorption Lines in the Spectra of Active Galactic Nuclei}

\author{Taotao~Fang\altaffilmark{1}, David Buote\altaffilmark{2}, James Bullock\altaffilmark{2}, Renyi Ma\altaffilmark{1}}

\altaffiltext{1}{Department of Astronomy and Institute for Theoretical Physics and Astrophysics, Xiamen University, Xiamen, Fujian~361005, China; fangt@xmu.edu.cn}
\altaffiltext{2}{Department of Physics \& Astronomy, 4129 Frederick Reines Hall, University of California, Irvine, CA~92697, U.S.A.}

\begin{abstract}

Highly ionized, $z=0$ metal absorption lines detected in the X-ray spectra of background active galactic nuclei (AGNs) provide an effective method to probe the hot ($T\sim10^6$ K) gas and its metal content in and around the Milky Way. We present an all-sky survey of the $K_{\alpha}$ transition of the local \ion{O}{7} absorption lines obtained by Voigt-profile fitting archival {\sl XMM}-Newton observations. A total of 43 AGNs were selected, among which 12 are BL Lac-type AGNs, and the rest are Seyfert 1 galaxies. At above the $3\sigma$ level the local \ion{O}{7} absorption lines were detected in 21 AGNs, among which 7 were newly discovered in this work. The sky covering fraction, defined as the ratio between the number of detections and the sample size, increases from at about 40\% for all targets to 100\% for the brightest targets, suggesting a uniform distribution of the \ion{O}{7} absorbers. We correlate the line equivalent width with the Galactic coordinates and do not find any strong correlations between these quantities. Some AGNs have warm absorbers that may complicate the analysis of the local X-ray absorber since the recession velocity can be compensated by the outflow velocity, especially for the nearby targets. We discuss the potential impact of the warm absorbers on our analysis. A comprehensive theoretical modeling of the X-ray absorbers will be presented in a later paper.\\

\end{abstract}


\section{Introduction}

\begin{figure*}[t]
\center
\includegraphics[height=0.5\textheight,width=1.\textwidth]{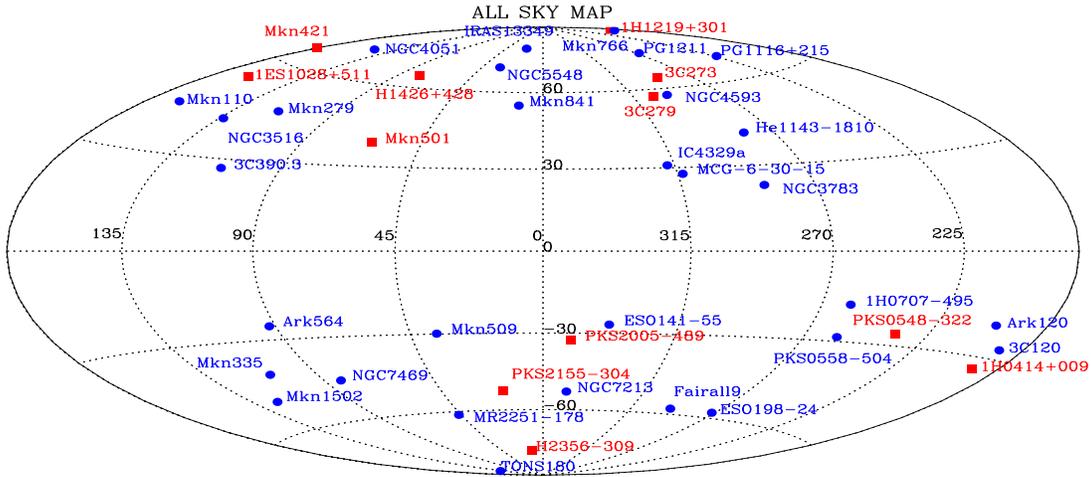}
\vskip-1.5cm
\caption{All-sky Hammer-Aitoff projection of our targets in the Galactic coordinates. Red squares are BL Lac-type targets, and blue circles are Seyfert 1 galaxies.}
\label{fig:map}
\end{figure*}

Since the launch of the {\sl Chandra} and {\sl XMM}-Newton X-ray Observatories, a number of $z\sim0$ absorption lines, produced by highly ionized metal species, were detected in the spectra of background active galactic nuclei (see, e.g., \citealp{nicastro2002, fang2002b,rasmussen2003}). These highly ionized metal absorption lines, mostly in the form of \ion{O}{7} and with a column density on the order of $\sim 10^{16}\rm cm^{-2}$, imply a hot, metal enriched gas distribution in and around our Galaxy. It stirs a great amount of interest for the potential impact on our understanding of galaxy formation and evolution.

However, ever since the first detection, a crucial debate is the origin of the X-ray absorbers. For the local absorption lines seen in the quasar spectra, even with the high resolution spectrometers on-board {\sl Chandra} and {\sl XMM}-Newton, we can only limit the location of the absorbers to within a few Mpc, therefore it is still unclear whether this gas is associated with our disk, an extended galactic halo (Wang et al. 2006; Fang et al. 2006), or even permeates the Local Group in the form of the intragroup medium (e.g., Nicastro et al. 2002; Williams et al. 2005). Identifying the origin of this gas has significant implications on many fronts. A disk origin would benefit the study of the three-phase model of the interstellar medium (ISM, Ostriker \& McKee~1978). If the absorption is produced by the hot, circumgalactic medium (CGM) in the distant halo, it would imply a significant reservoir of baryons may account for the ``missing Galactic baryons" (see, e.g., \citealp{fang2006, gupta2012}). Furthermore, the hot gas could reside in the hot intergroup medium in the Local Group, which would potentially explain the so-called warm-hot intergalactic medium (WHIM, see, e.g, \citealp{nicastro2002}).    

The existence of the hot gas in and around the Milky Way has long been established through the observations of the soft X-ray background emission (see, e.g., McCommon \& Sanders~1990; Wang \& Yu~1995; Snowden et al.~1998; Kuntz \& Snowden~1998; Galeazzi et al.~2007; Henley \& Shelton~2008, 2013; Yoshino et al.~2009). However, current data cannot distinguish between a disk and a halo origin of the X-emitting gas. While most studies have argued for a disk morphology of the hot gas (see, e.g., \citealp{yao2007, hagihara2010}), other studies have suggested that a halo origin is also possible \citep{fang2013, henley2013}.

Substantial progress has been made since these absorbers were first identified. \citet{mckernan2004}  studied the {\sl Chandra} spectra of 15 type I AGNs and argued that the observed metal absorption lines should be local to our Galaxy instead of high velocity outflows intrinsic to the AGNs. Fang et al.~(2006) further investigated 20 AGNs from both {\sl Chandra} and {\sl XMM}-Newton observations and concluded the absorption line systems must be in/around the Milky Way instead of in the intragroup medium of the distant Local Group. \citet{bregman2007} reached a similar conclusion by studying the correlation between the X-ray absorption line strength and the {\sl ROSAT} background emission measure. Furthermore, based on a survey of 26 AGNs, \citet{bregman2007} and \citet{miller2013} analyzed the structure of the CGM and argued that the hot gas accounts for at most 10-50\% of the missing Galactic baryons.

\begin{figure}[t]
\center
\includegraphics[height=0.35\textheight,width=0.5\textwidth]{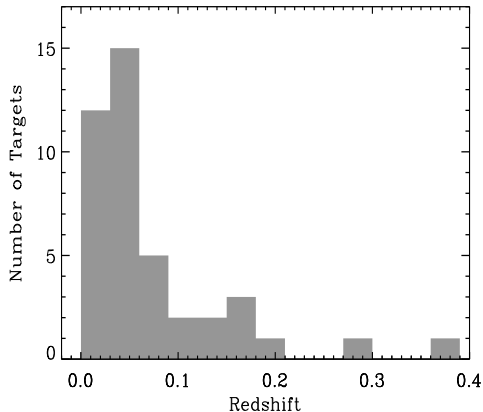}
\vskip-1cm
\caption{Redshift distribution of the sample AGNs.}
\label{fig:z}
\end{figure}

Despite all the progress, a critical, unresolved issue is still the relative contribution of the disk and the distant halo to the observed absorption seen in X-ray. Inspired by the disk morphology of the X-ray emission of nearby galaxies, several authors (see, e.g., Yao et al. 2009; Hagihara et al. 2010) argued that most of the X-ray absorbing gas is confined within a few kpc of the Milky Way disk. X-ray absorption lines were also detected in the nearby Galactic X-ray binaries (see, e.g., \citealp{miller2004, yao2005}). Such absorption lines have been compared with those detected in the AGNs to determine the relative contribution of the disk and halo (see, e.g., \citealp{wang2005a, yao2007}). However, A complication arises in these studies because the X-ray absorpriton lines in the XRBs may be contaminated by the circumstellar medium intrinsic to the XRBs \citep{miller2004,cackett2008}. While a disk origin for the hot gas is attractive, a halo-origin cannot be definitely excluded (see, e.g., \citealp{fang2006, gupta2012, miller2013}, but also see \citealp{wang2012}).

Analysis of the complete global distribution of the $z\sim0$ absorbers becomes necessary to resolve these issues. We have performed a comprehensive analysis of all the available data in the {\sl XMM}-Newton archive. Our targets include a total of 43 AGNs. In this first paper, we present the results of our data analysis. We also report the first detection of $z\sim0$ \ion{O}{7} absorption lines in 7 background AGNs at more than the $3\sigma$ level. Prior to this work, the most comprehensive study was performed by \citet{miller2013}. Their sample includes 26 AGNs, and 10 targets showed $z\sim0$ \ion{O}{7} absorption lines at more than the $3\sigma$ level. Therefore our work is a significant increase in both the sample size and the newly detections of $z\sim0$ \ion{O}{7} absorption lines.

This paper is organized as follows. In section~\S2 we describe the observations and data analysis procedures. We present our main results in this section. We discuss the consistency of our data analysis by comparing our work with previous studies in section~\S3, as well as the impact of warm absorbers in some Seyfert 1 galaxies. We also comment on several targets with newly detected, $z\sim0$ \ion{O}{7} lines in this section. The last section provides a summary.

\section{Observations and Data Analysis}

\subsection{Target Selection}

Three high resolution X-ray spectrometers are suitable for the study of the narrow X-ray absorption lines: the low and high energy transmission gratings (LETG and HETG)\footnote{See http://asc.harvard.edu/} on-board the {\sl Chandra} X-ray Observatory, and the The Reflection Grating Spectrometer (RGS)\footnote{See http://xmm.esa.int/} on-board the {\sl XMM}-Newton X-ray Telescope. We focus on the RGS data mainly due to its high collecting area at the wavelength region we are interested in. There are two RGS units: RGS 1 and RGS 2. However, due to operation failures the CCD 7 of RGS 1 and CCD 4 of RGS 2 are not working. Particularly, the CCD 4 of RGS 2 covers the He-like oxygen region which is our primary focus in this paper. Therefore in this work we will focus on the RGS 1 unit only.

The $z=0$ X-ray absorption lines from various ion species have been reported previously. These lines were produced by metal species in a variety of ionization stages, ranging from neutral to the H-like. Most reported ion species are \ion{Ne}{9} at 13.44 \AA, \ion{O}{8} at 18.98 \AA, and \ion{O}{7} at 21.6 \AA\ (see, e.g, \citealp{nicastro2002, fang2002b,rasmussen2003, williams2005}). In this work, we will mainly focus on highly ionized metals, in particular the He-like oxygen, \ion{O}{7}, for several reasons. First, \ion{Ne}{9} $K_{\alpha}$ is located in a region where the RGS 1 has no effective area, and also there are several bad pixels near the \ion{O}{8} $K_{\alpha}$ region. Second, Under collisional ionization, the temperature for the peak \ion{O}{7} fraction is between $10^{5.5}$--$10^{6.5}$ degrees, providing an effective way to probe the hot gas in and around the Milky Way. 

We searched the entire {\sl XMM}-Newton archive of the AGNs with enough RGS photon counts to warrant a spectral analysis. The RGS resolution has a full width of half maximum (FWHM) of $\sim 50$ m\AA. We define one resolution element as half of the FWHM, or 25 m\AA.\ Our past experience indicates that the minimum requirement for a useful spectral fit is that the counts per resolution element, or CPRE, must be at least 20 photons. This selection criterion results in a total of 43 AGNs in our sample. Among them 12 are BL Lac-type target. The remaining 31 are various Seyfert galaxies with a subclass of between 1 and 1.5. For convenience we broadly categorize them as Seyfert 1 AGNs\footnote{See the NASA Extrgalactic Database at http://ned.ipac.caltech.edu/.}. In Figure~\ref{fig:map} we plot the all-sky Hammer-Aitoff projection of our targets in the Galactic coordinate. In Table~\ref{tab:log} we list basic properties of these targets. Column (1) is the target name. Column (2) lists the AGN types. We broadly divide the entire sample into BL Lac targets (Type 1) and Seyfert 1 Galaxies (Type 2). Columns (3), (4), and (5) are the Galactic latitude, longitude and redshift, respectively. Column (6) is the Galactic neutral hydrogen column density \citep{dickey1990}. Columns (7) and (8) are the total exposure time (in units of $ksec$) and the CPRE. Columns (9) -- (15) are line parameters which will be described in detail later. Columns (9), (10), (11) are the  \ion{O}{7} column density,  the Doppler-$b$ parameter, and the velocity shift of the line center, respectively. Columns (12) (13), and columns (15), (16) are the line line equivalent width (EW) and the signal-to-noise ratio (S/N) of the detection, for two different evaluation methods, respectively. We do not list the line parameters (denoted as ``...") if the line is not detected (defined as the $S/N$ falls below 1), except the 3$\sigma$ upper limits of the EW. We also list the $C$-statistic and degree of freedom in column (14) (see next section). In the last column we list the references that first reported the detection of the target. All the errors are $1\sigma$ unless otherwise mentioned. We also plot the redshift distribution of our sample targets in Figure~\ref{fig:z}. The majority of the targets are located at $z<0.1$.

\begin{figure*}[t]
\center
\includegraphics[height=0.23\textheight,width=0.47\textwidth]{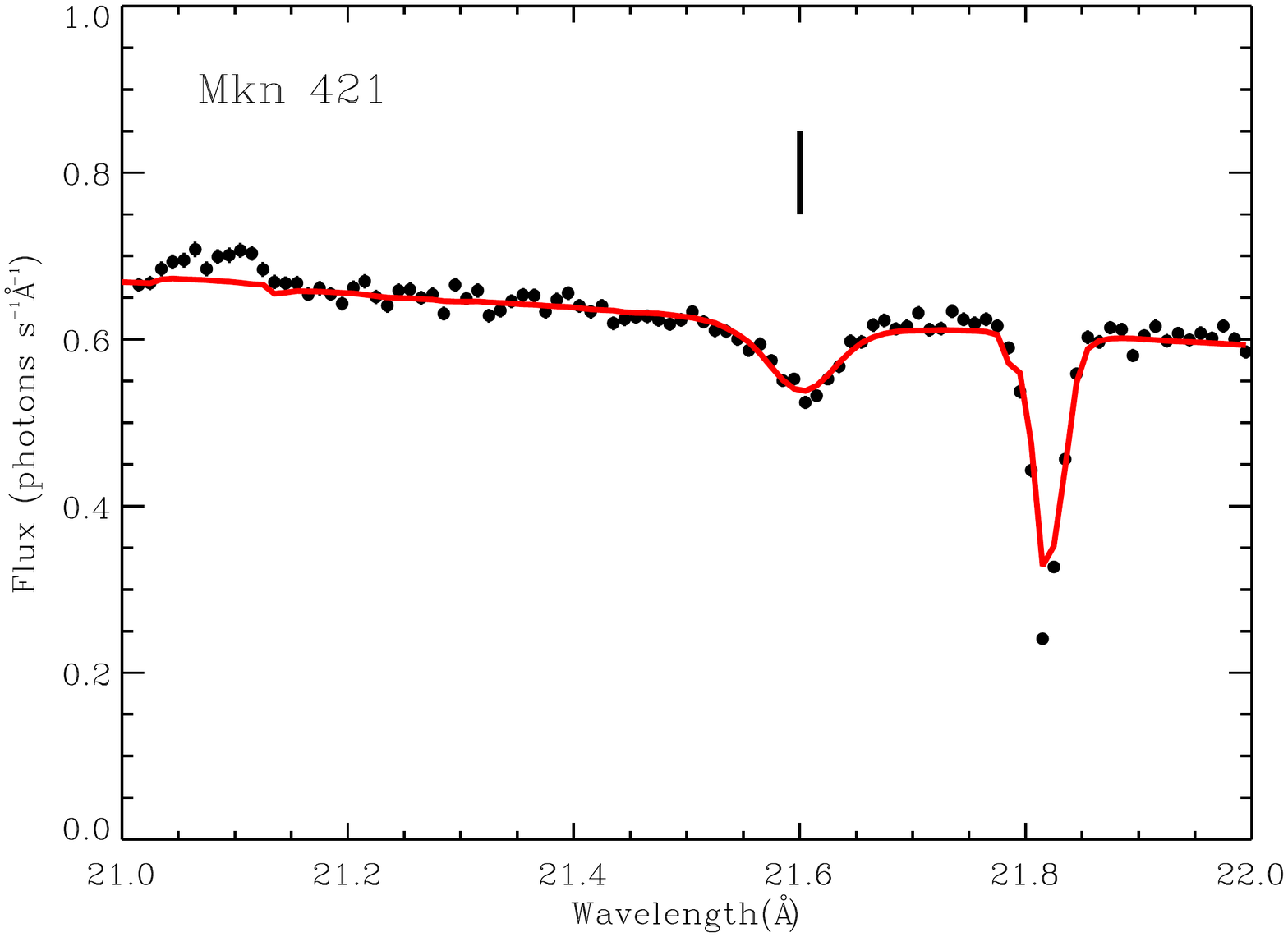}
\includegraphics[height=0.23\textheight,width=0.47\textwidth]{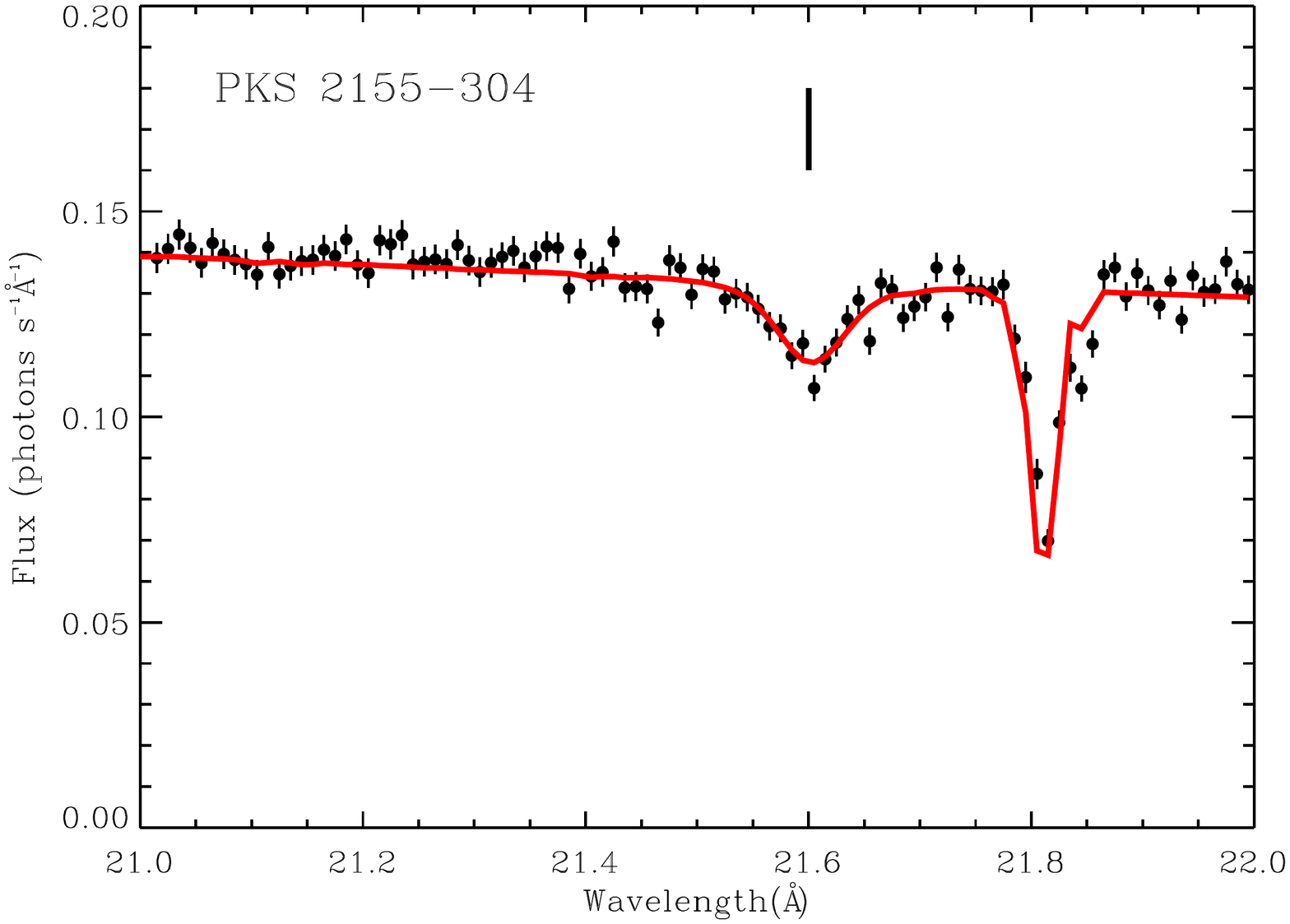}
\vskip-0.8cm
\includegraphics[height=0.23\textheight,width=0.47\textwidth]{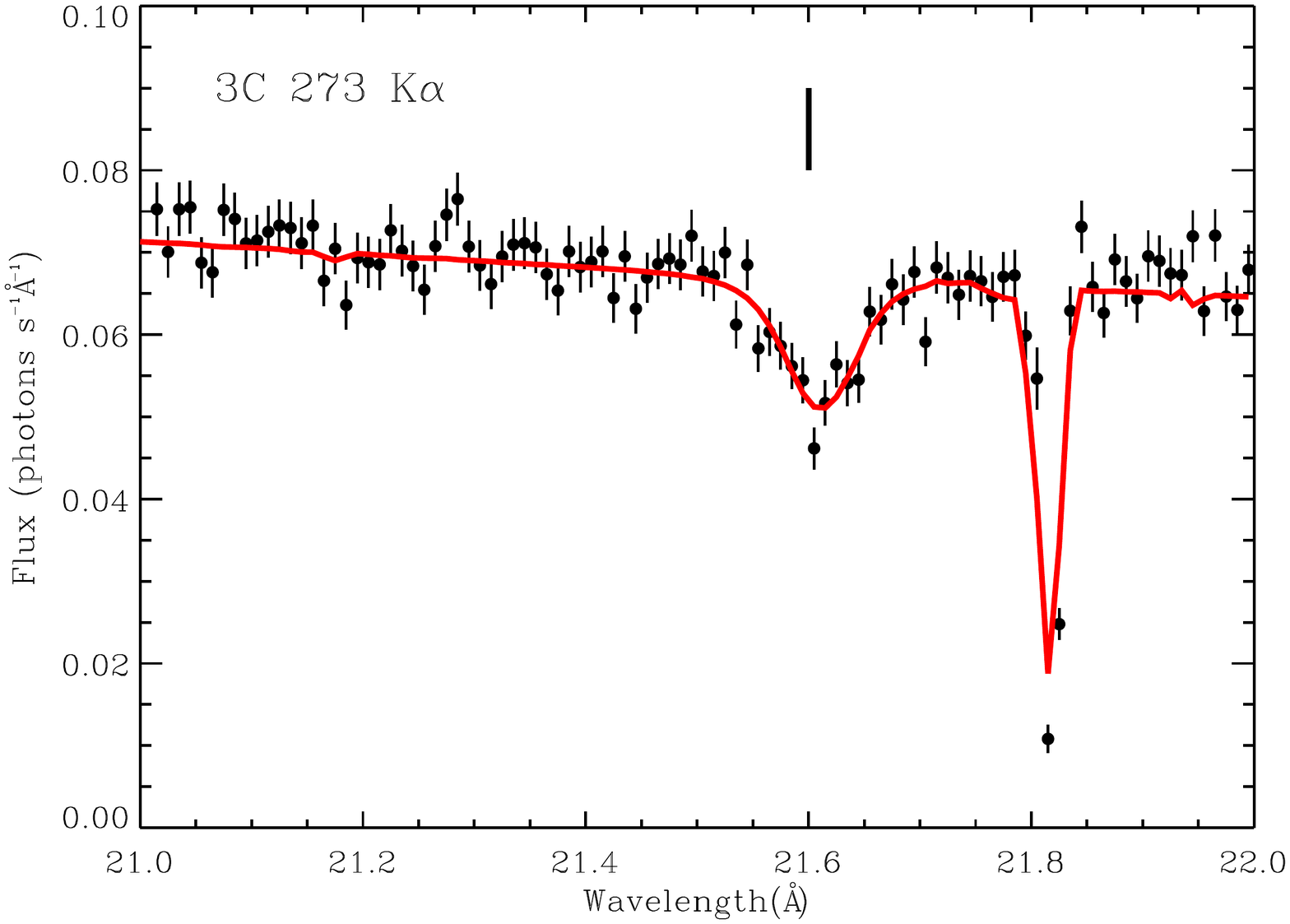}
\includegraphics[height=0.23\textheight,width=0.47\textwidth]{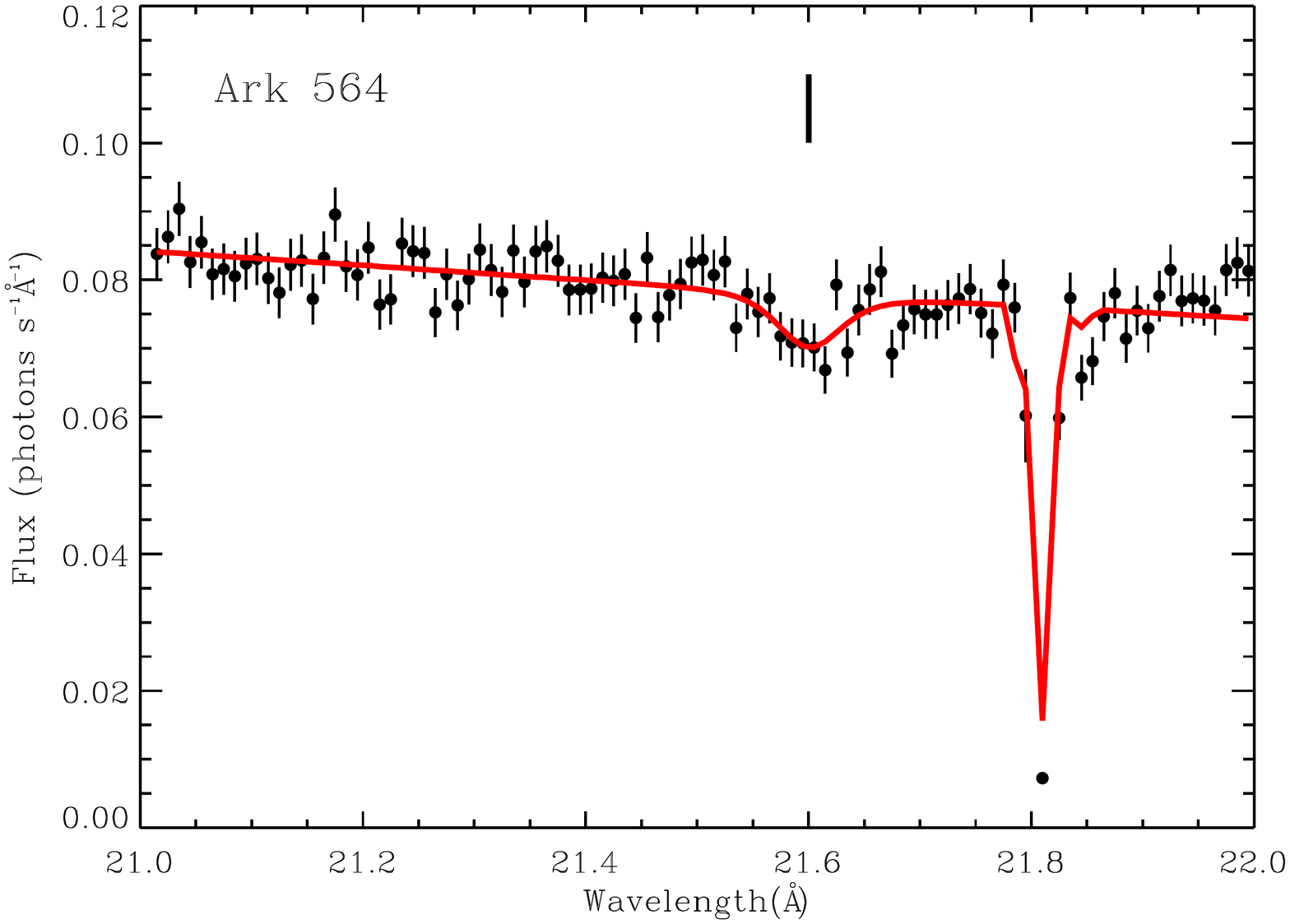}
\vskip-0.8cm
\includegraphics[height=0.23\textheight,width=0.47\textwidth]{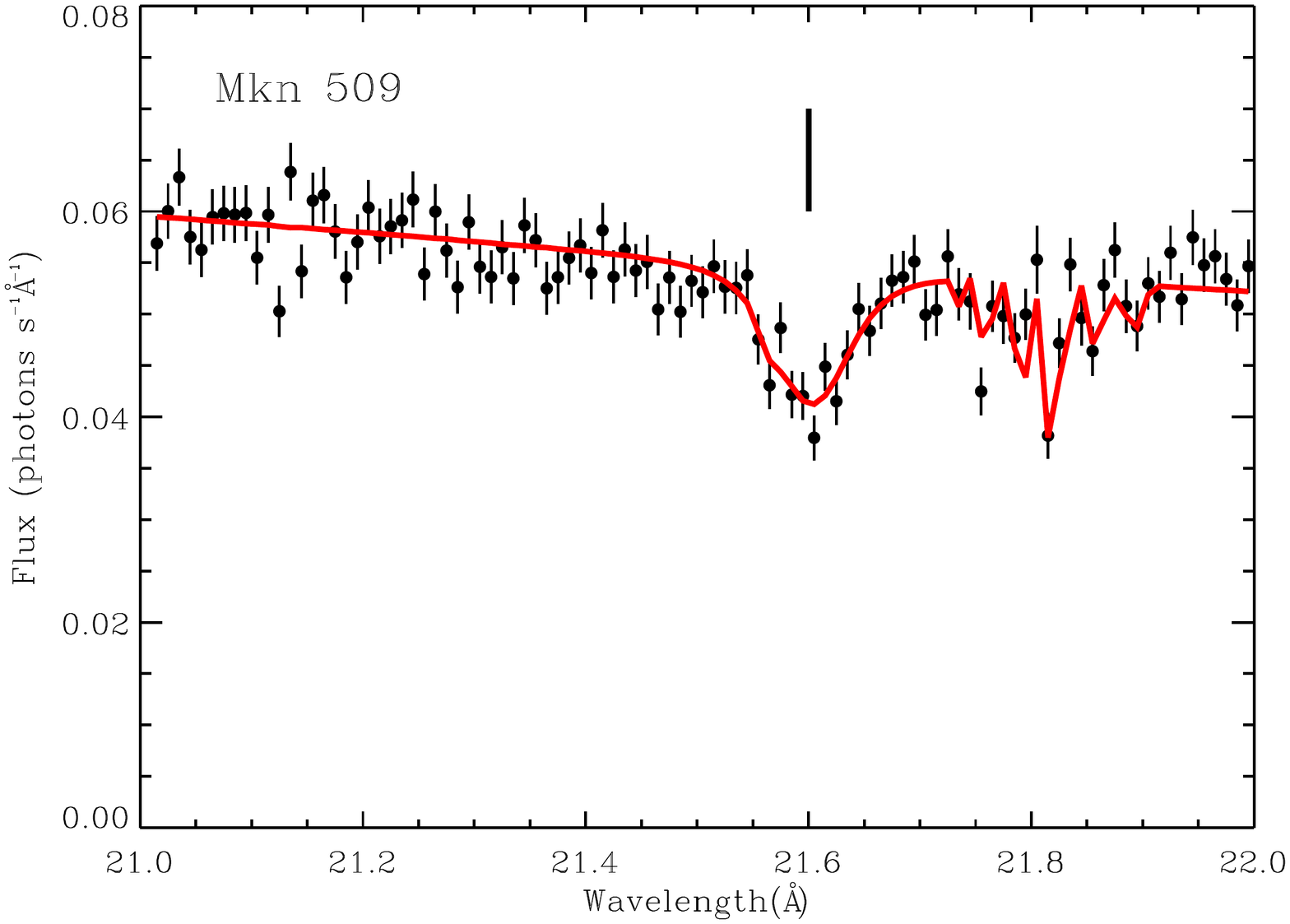}
\includegraphics[height=0.23\textheight,width=0.47\textwidth]{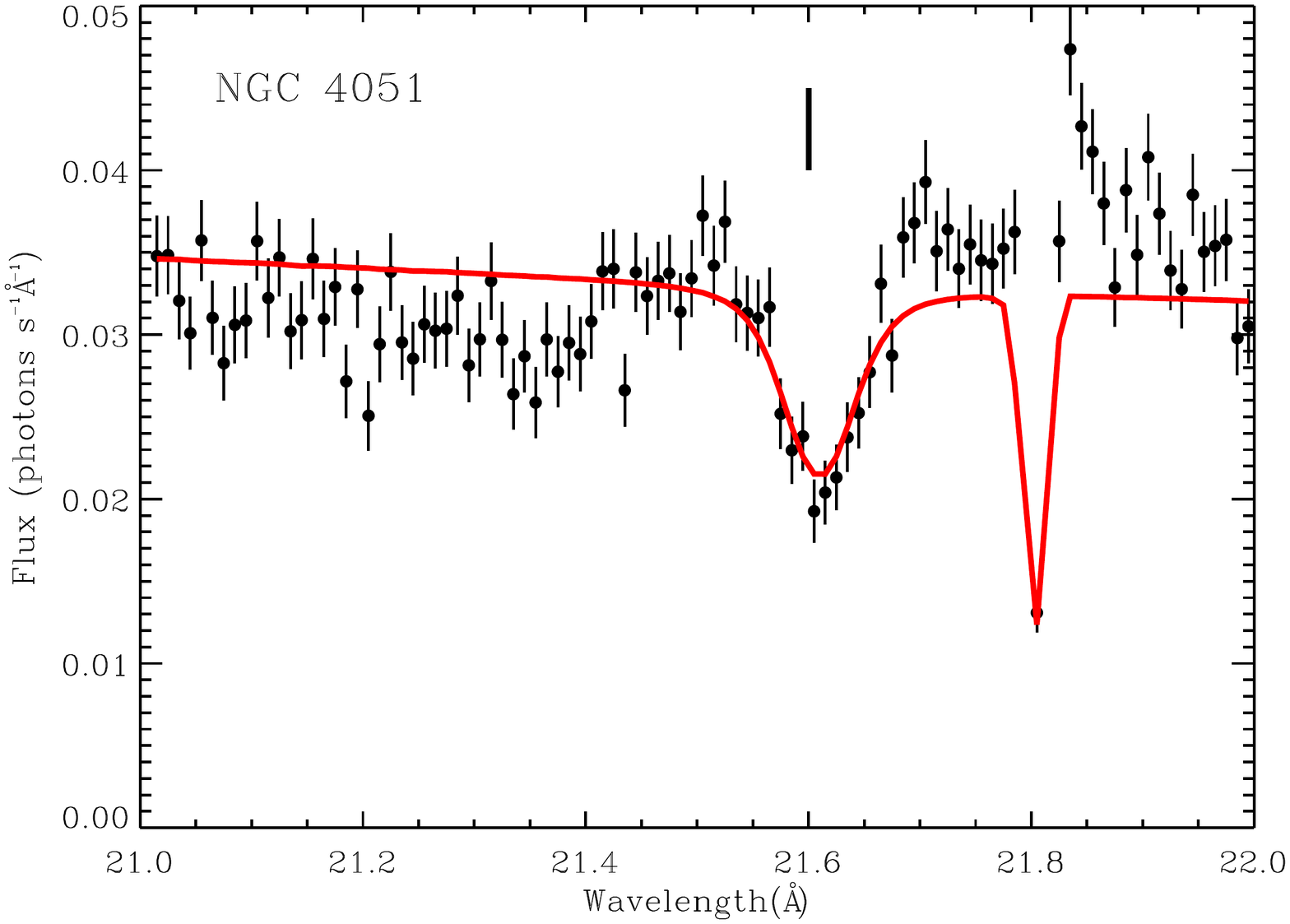}
\vskip-0.8cm
\caption{RGS spectra of Mkn~421, PKS~2155-304, 3C~273, Ark~564, Mkn~509, and NGC~4051. The location of the  \ion{O}{7} $K_{\alpha}$ line is labeled with a vertical line. The red line in each panel is the model. The structures at 21.82 \AA\ are instrumental features.}
\label{fig:spec1}
\end{figure*}

\begin{figure*}[t]
\center
\includegraphics[height=0.23\textheight,width=0.47\textwidth]{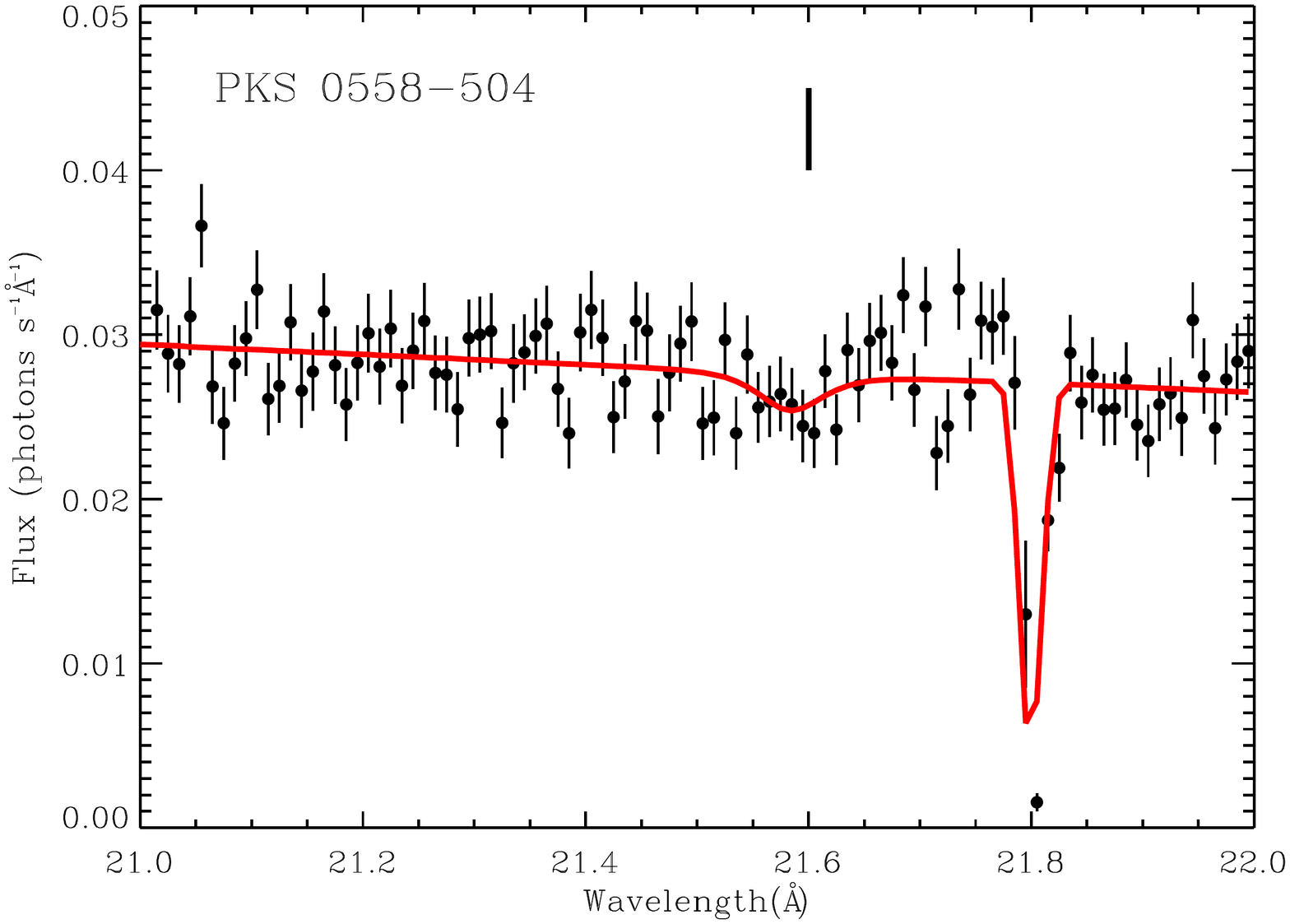}
\includegraphics[height=0.23\textheight,width=0.47\textwidth]{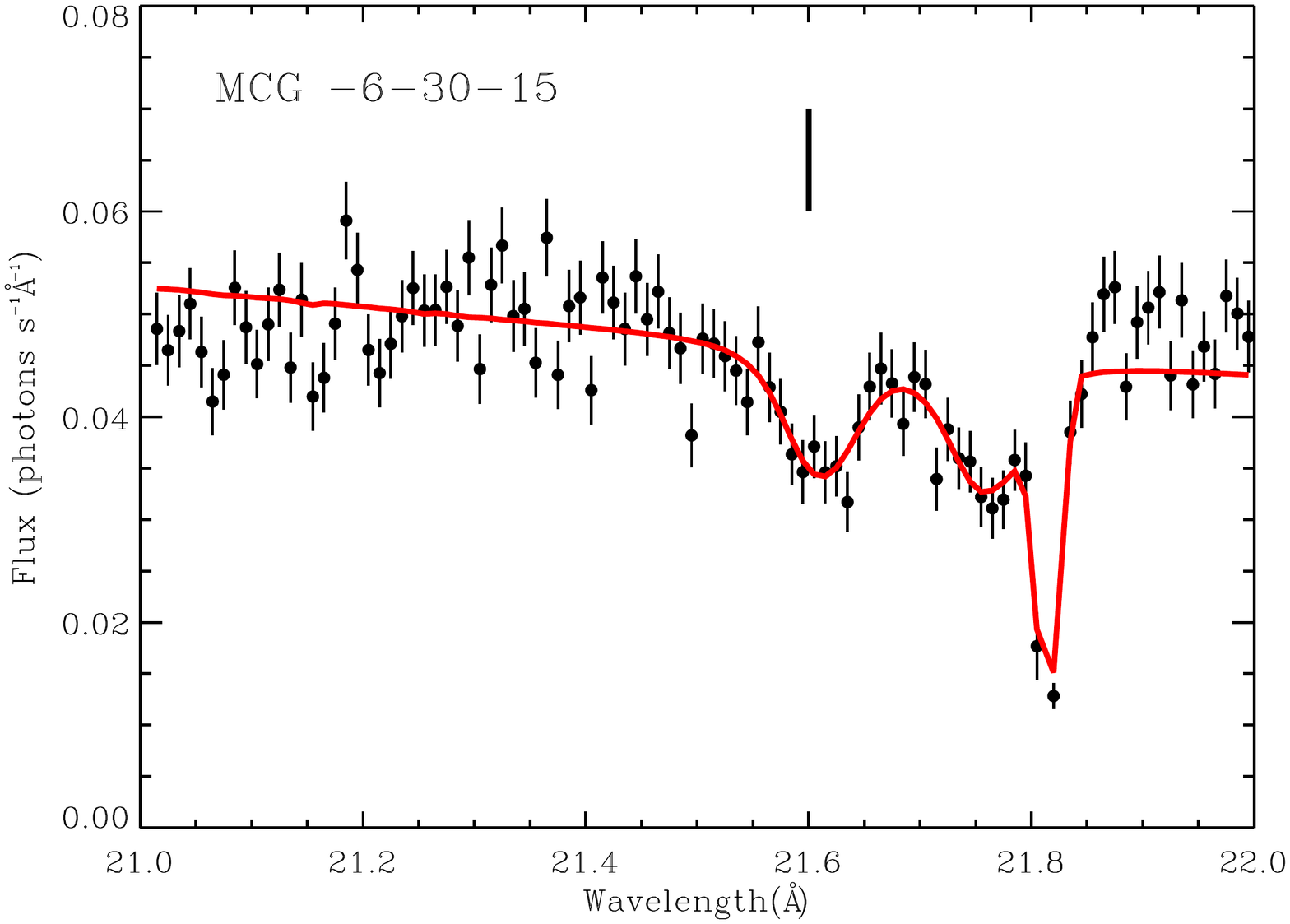}
\vskip-0.8cm
\includegraphics[height=0.23\textheight,width=0.47\textwidth]{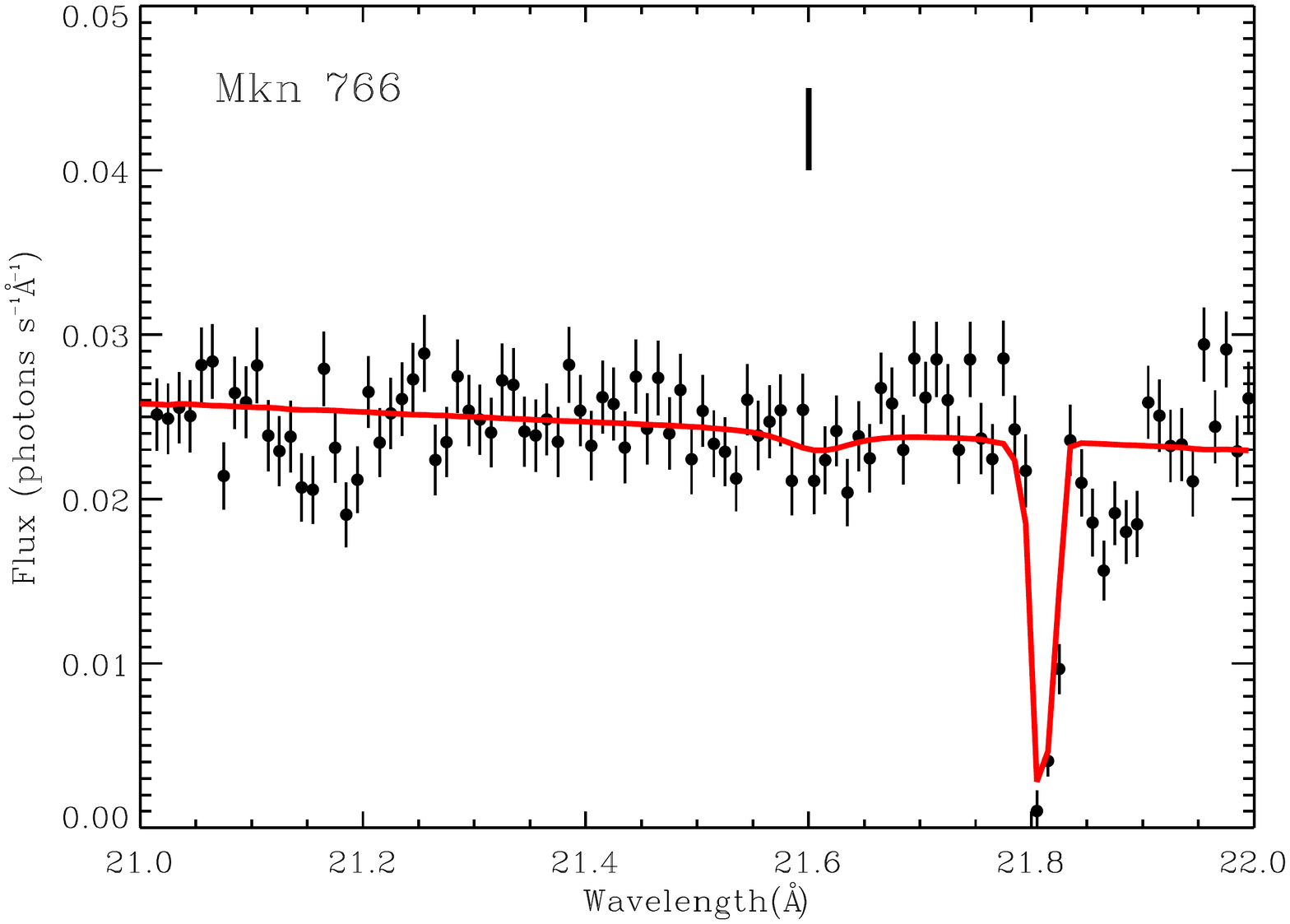}
\includegraphics[height=0.23\textheight,width=0.47\textwidth]{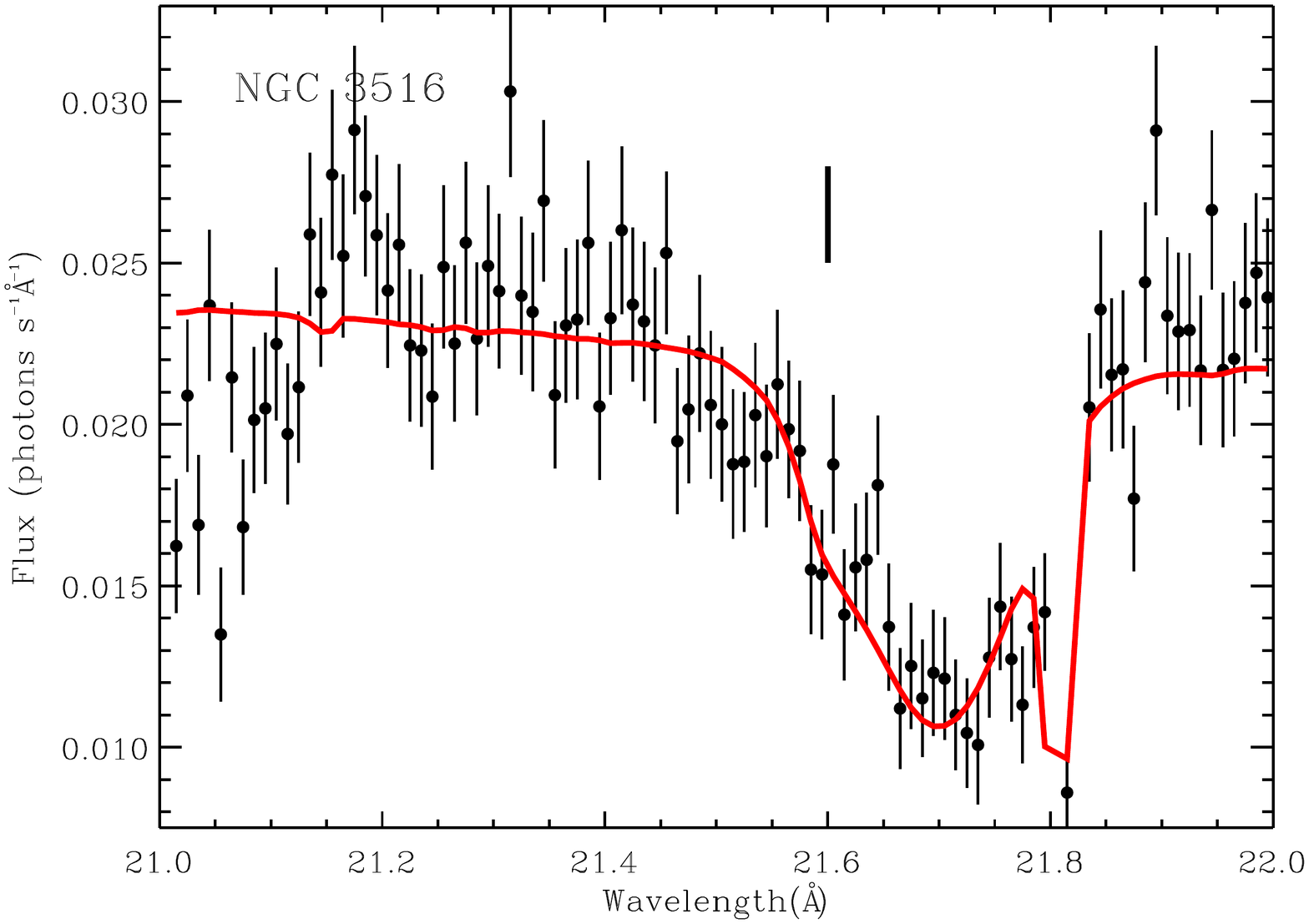}
\vskip-0.8cm
\includegraphics[height=0.23\textheight,width=0.47\textwidth]{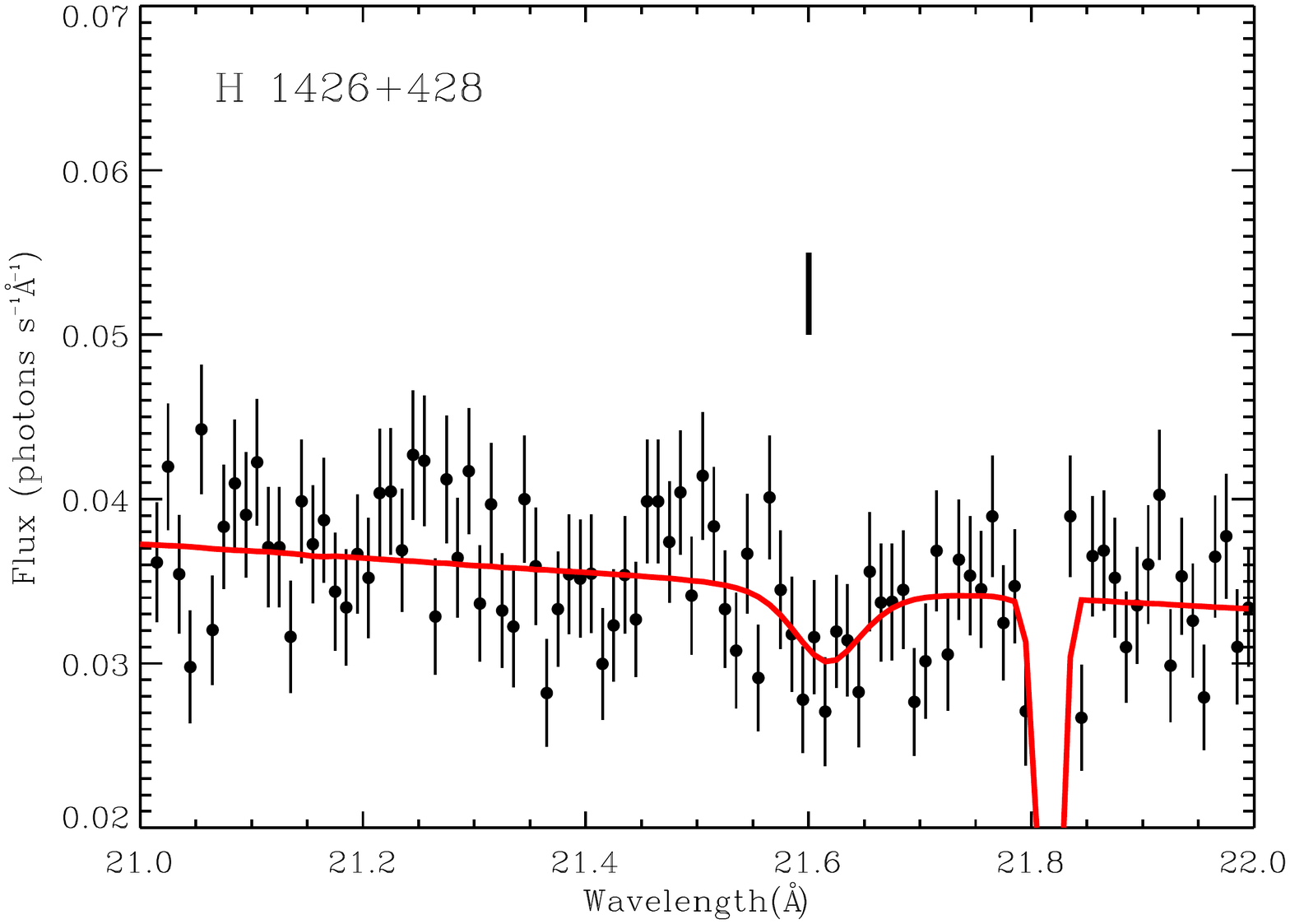}
\includegraphics[height=0.23\textheight,width=0.47\textwidth]{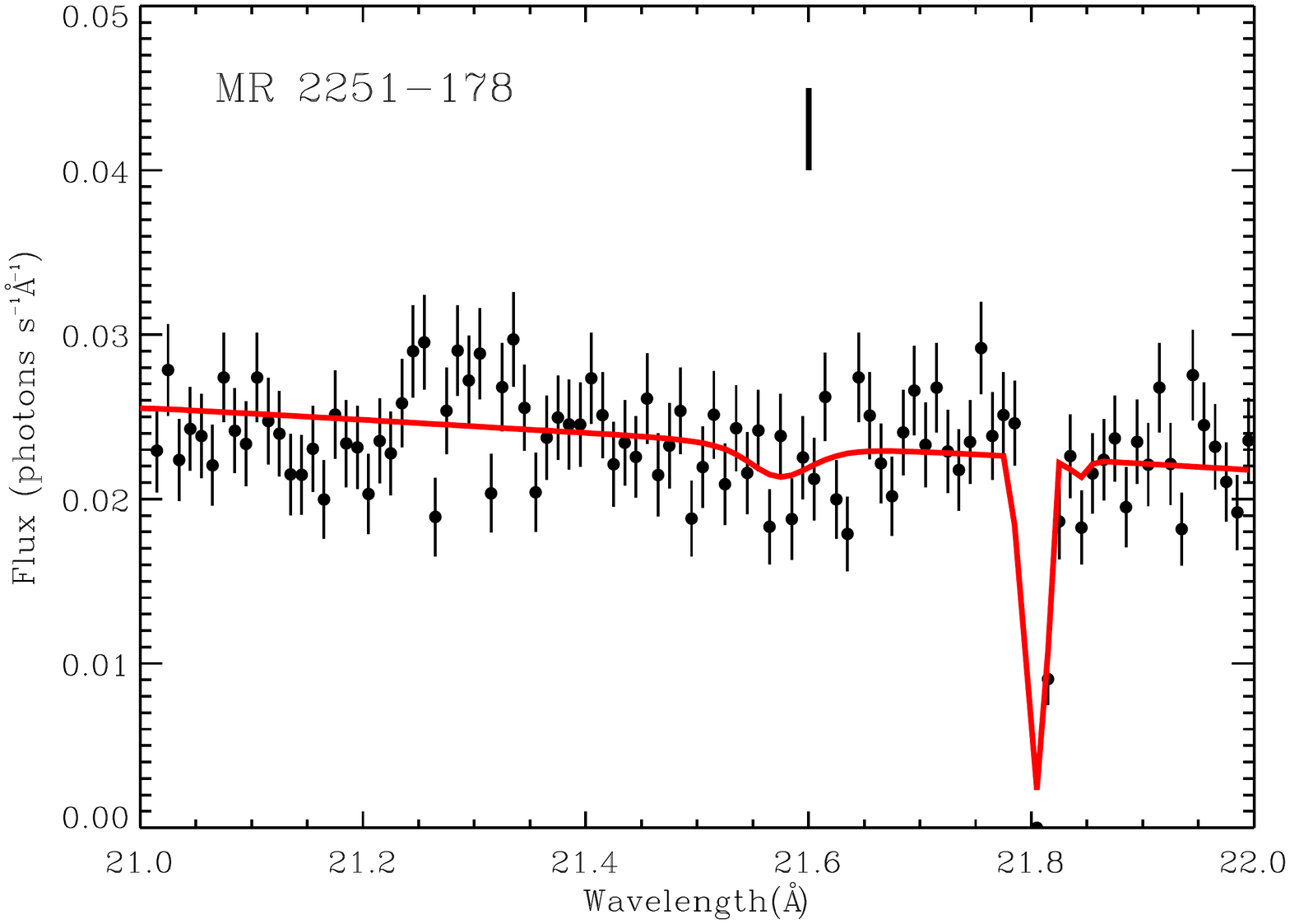}
\vskip-0.8cm
\caption{Same as the Figure~\ref{fig:spec1}, but for RGS spectra for PKS~0558-504, MCG-6-30-15, Mkn~766, NGC~3516, H~1426+428, and MR~2251-178.}
\label{fig:spec2}
\end{figure*}

\begin{figure*}[t]
\center
\includegraphics[height=0.23\textheight,width=0.47\textwidth]{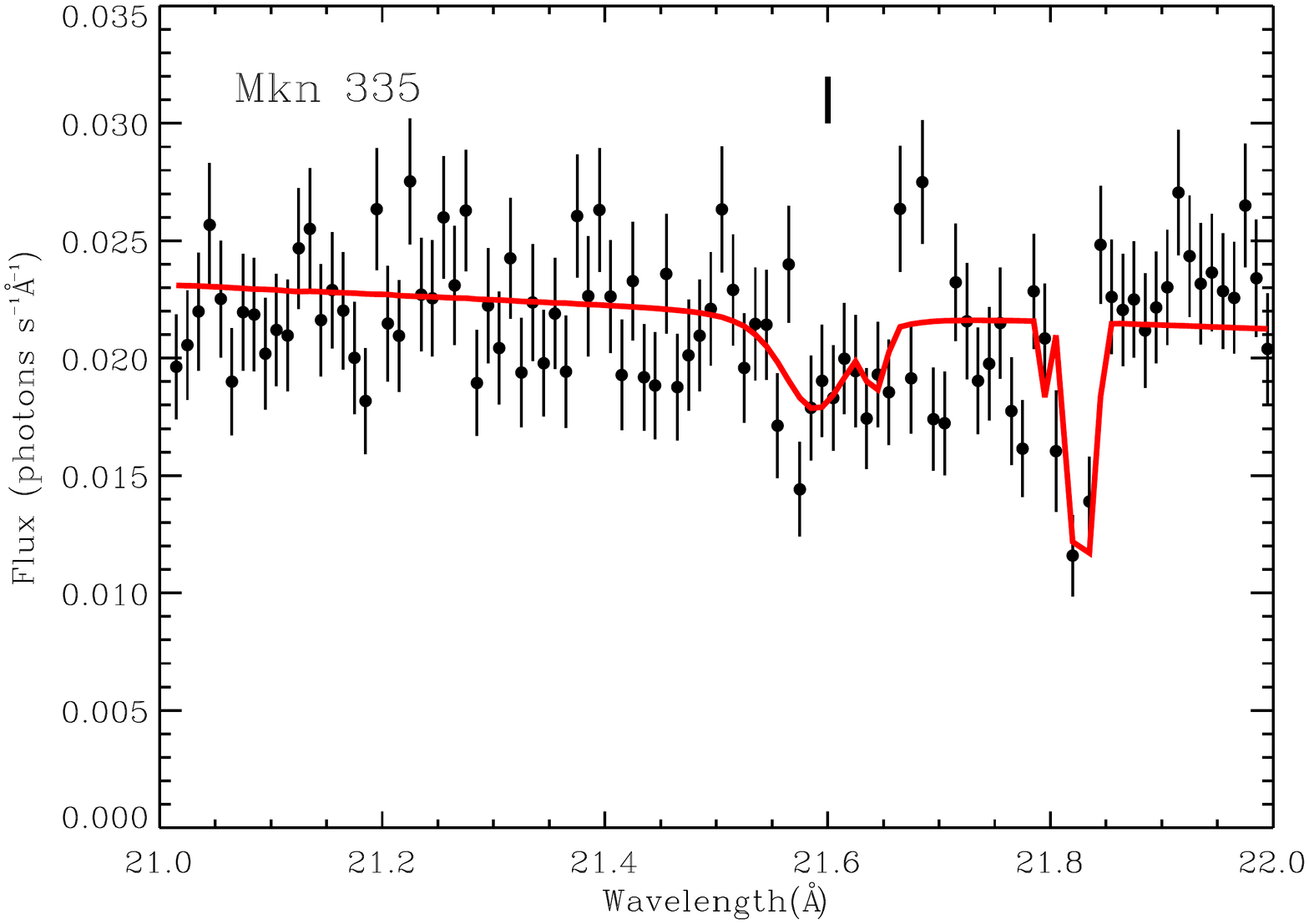}
\includegraphics[height=0.23\textheight,width=0.47\textwidth]{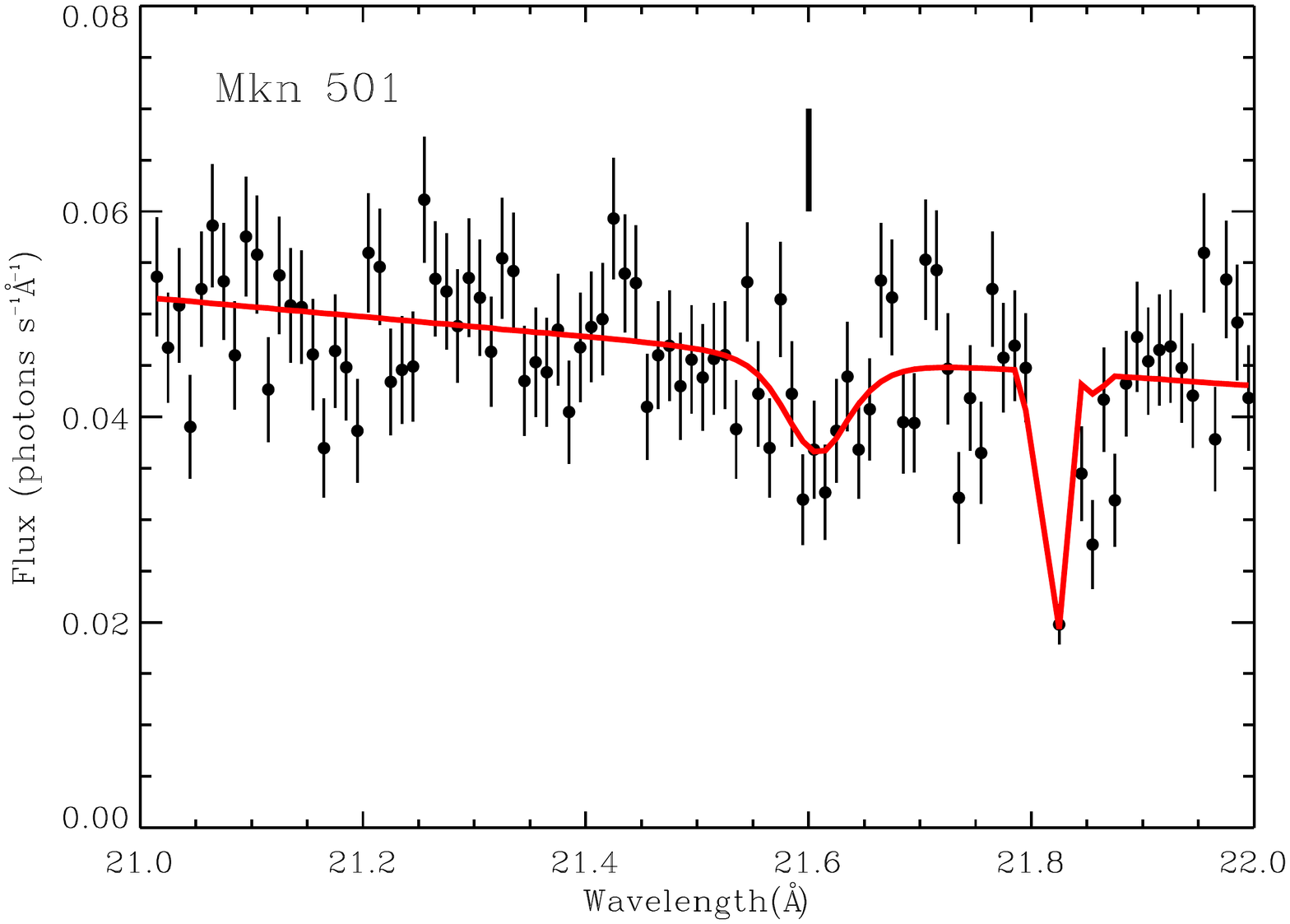}
\vskip-0.8cm
\includegraphics[height=0.23\textheight,width=0.47\textwidth]{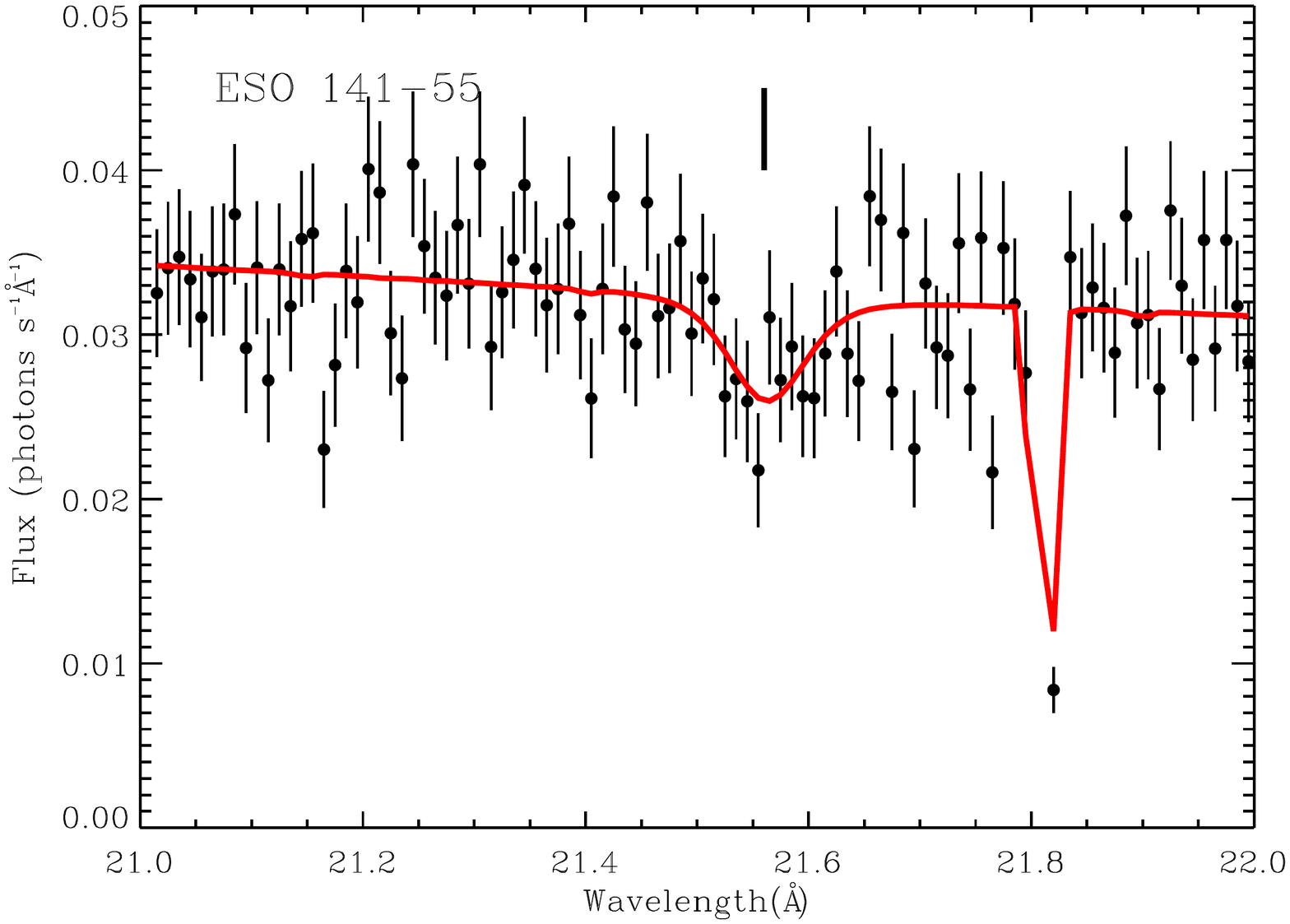}
\includegraphics[height=0.23\textheight,width=0.47\textwidth]{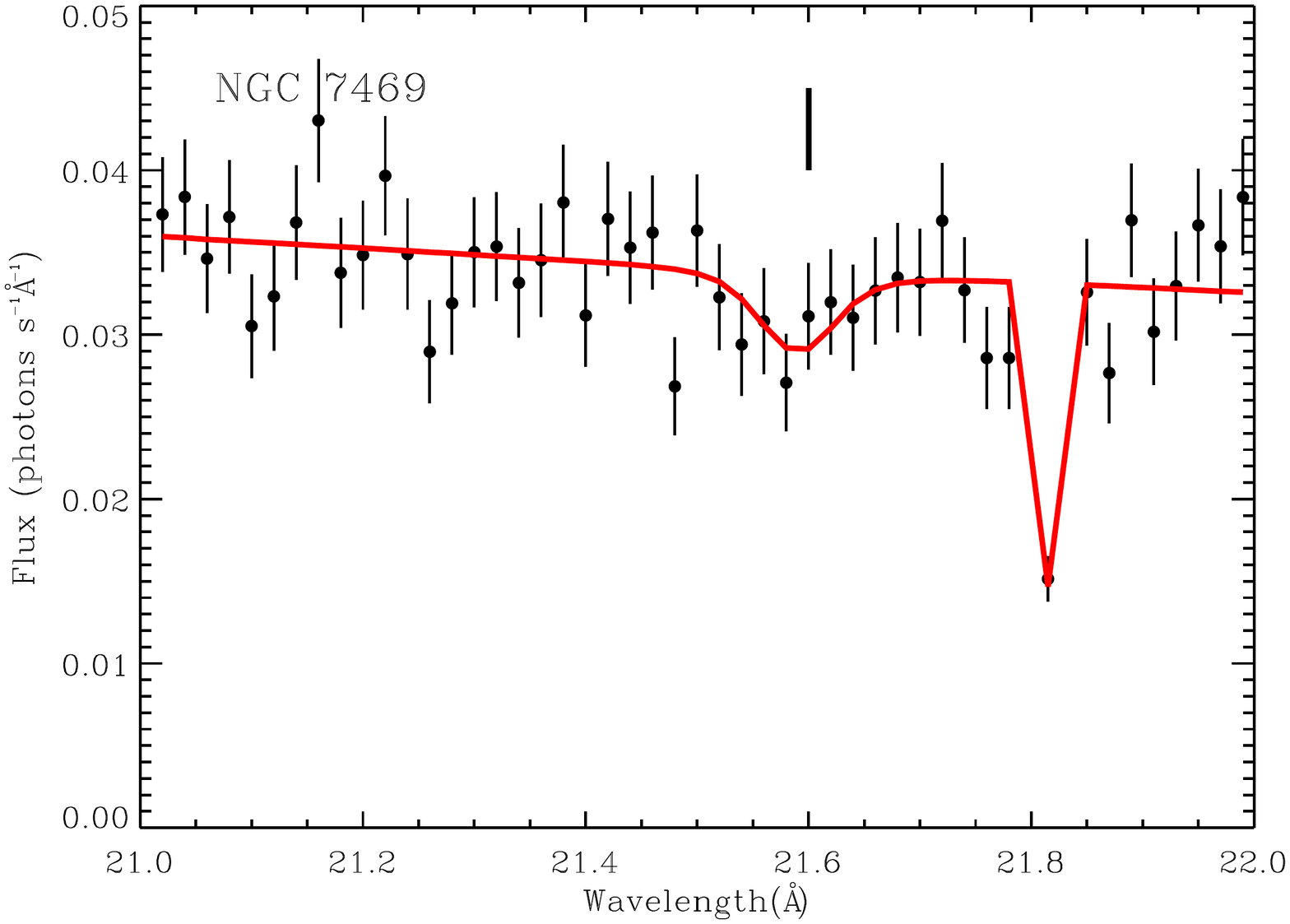}
\vskip-0.8cm
\includegraphics[height=0.23\textheight,width=0.47\textwidth]{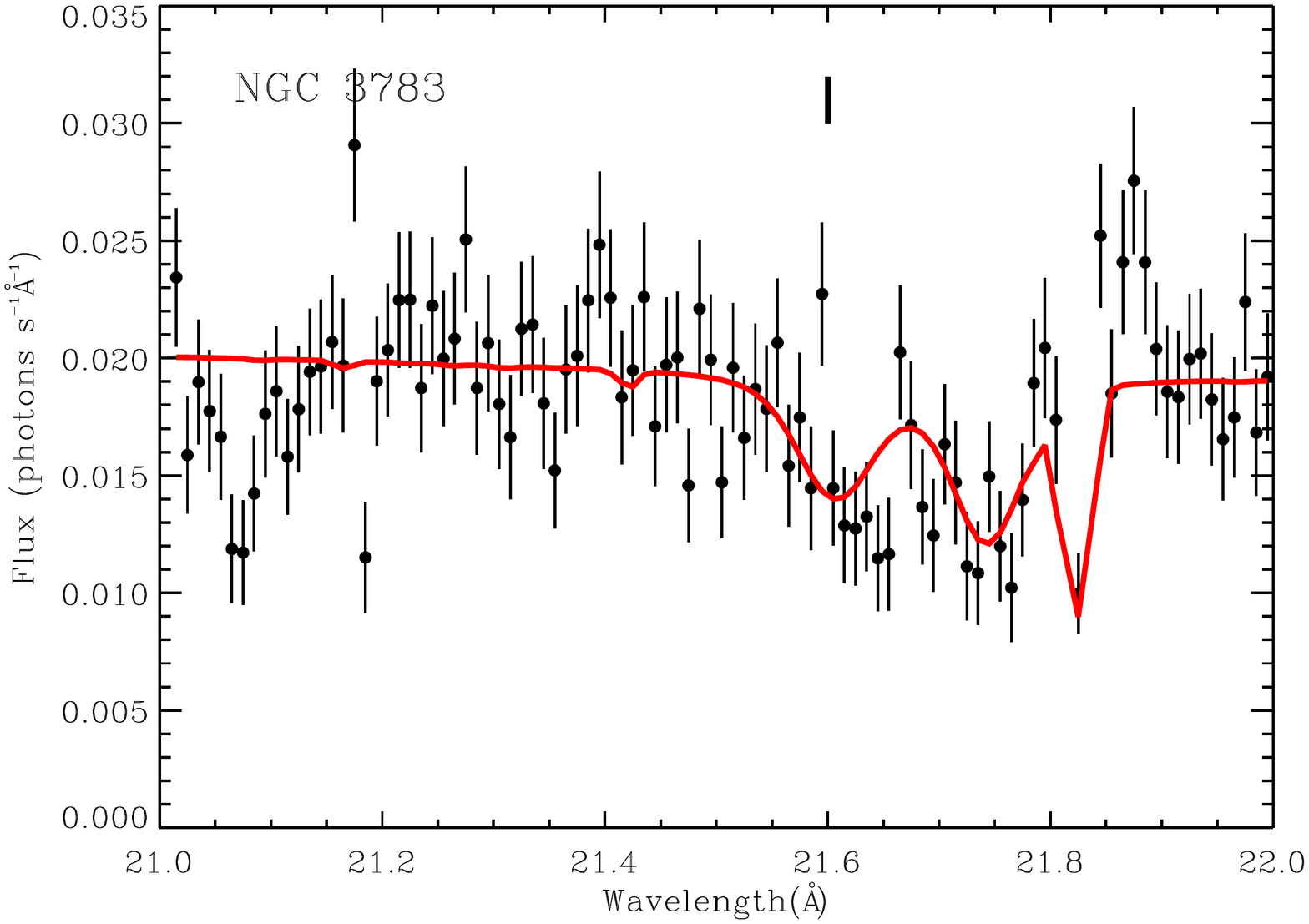}
\includegraphics[height=0.23\textheight,width=0.47\textwidth]{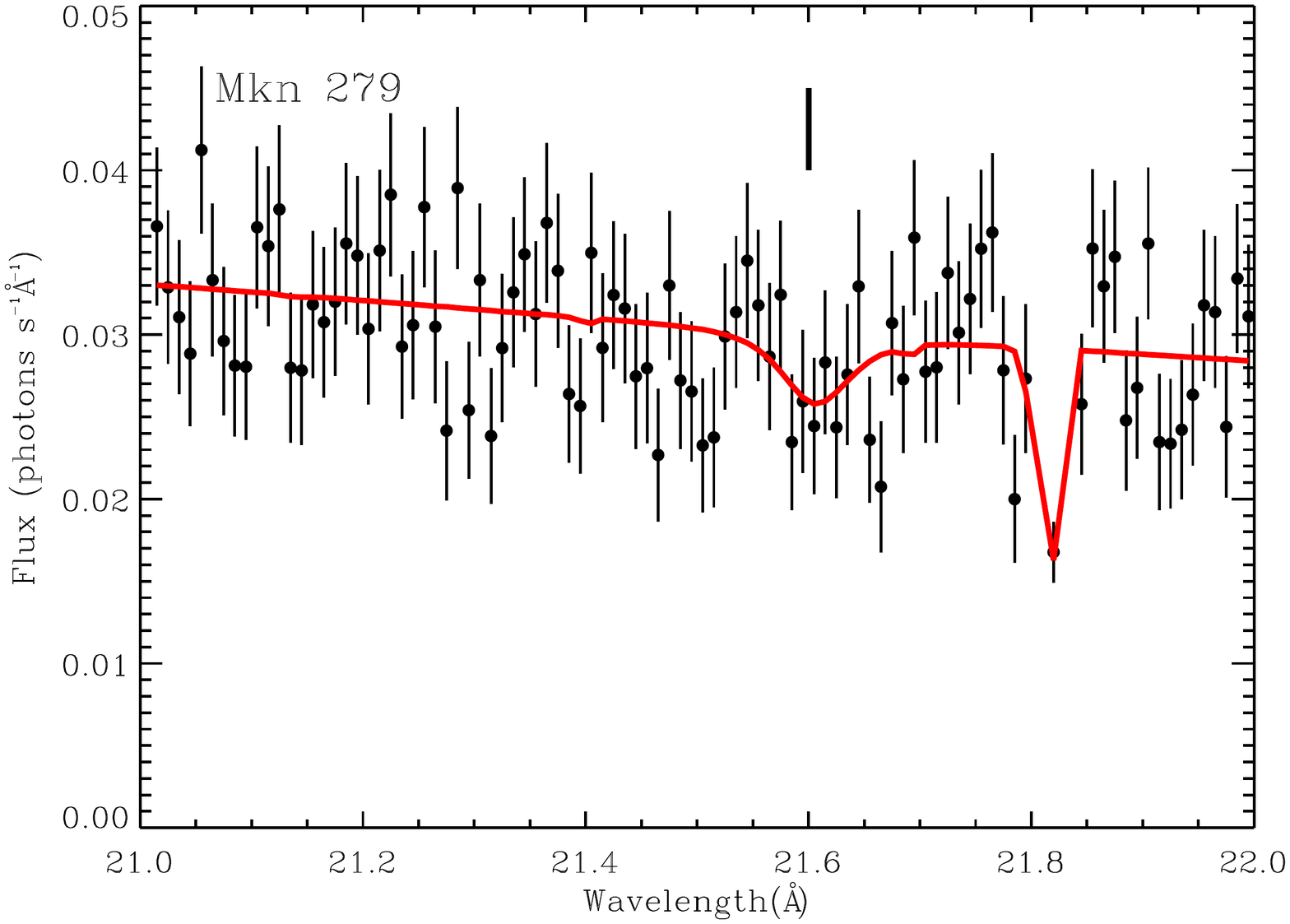}
\vskip-0.8cm
\caption{Same as the Figure~\ref{fig:spec1}, but for Mkn~335, Mkn~501, ESO~141-55, NGC~7469, NGC~3783, and  Mkn~279.}
\label{fig:spec3}
\end{figure*}

\begin{figure*}[t]
\center
\includegraphics[height=0.23\textheight,width=0.47\textwidth]{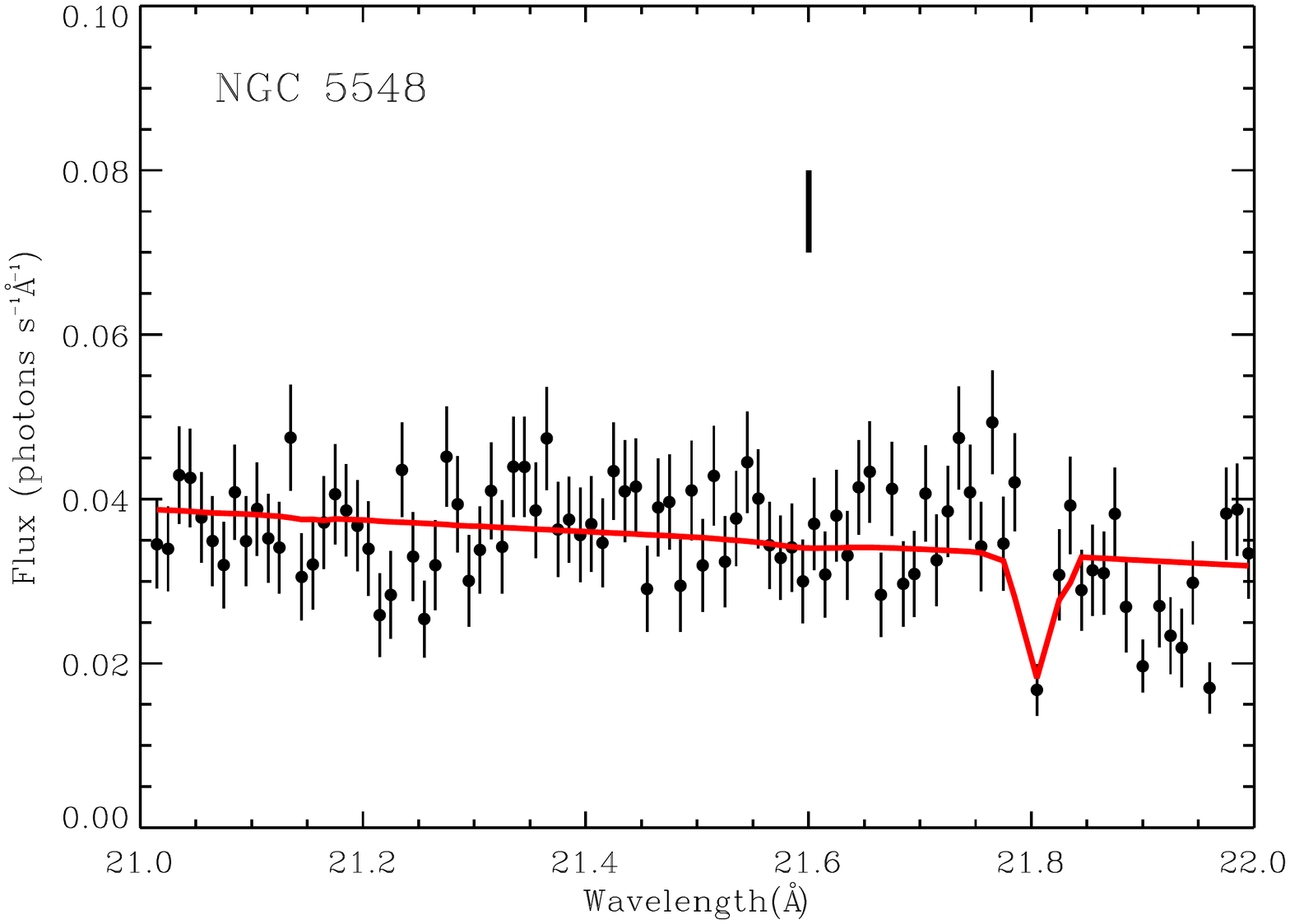}
\includegraphics[height=0.23\textheight,width=0.47\textwidth]{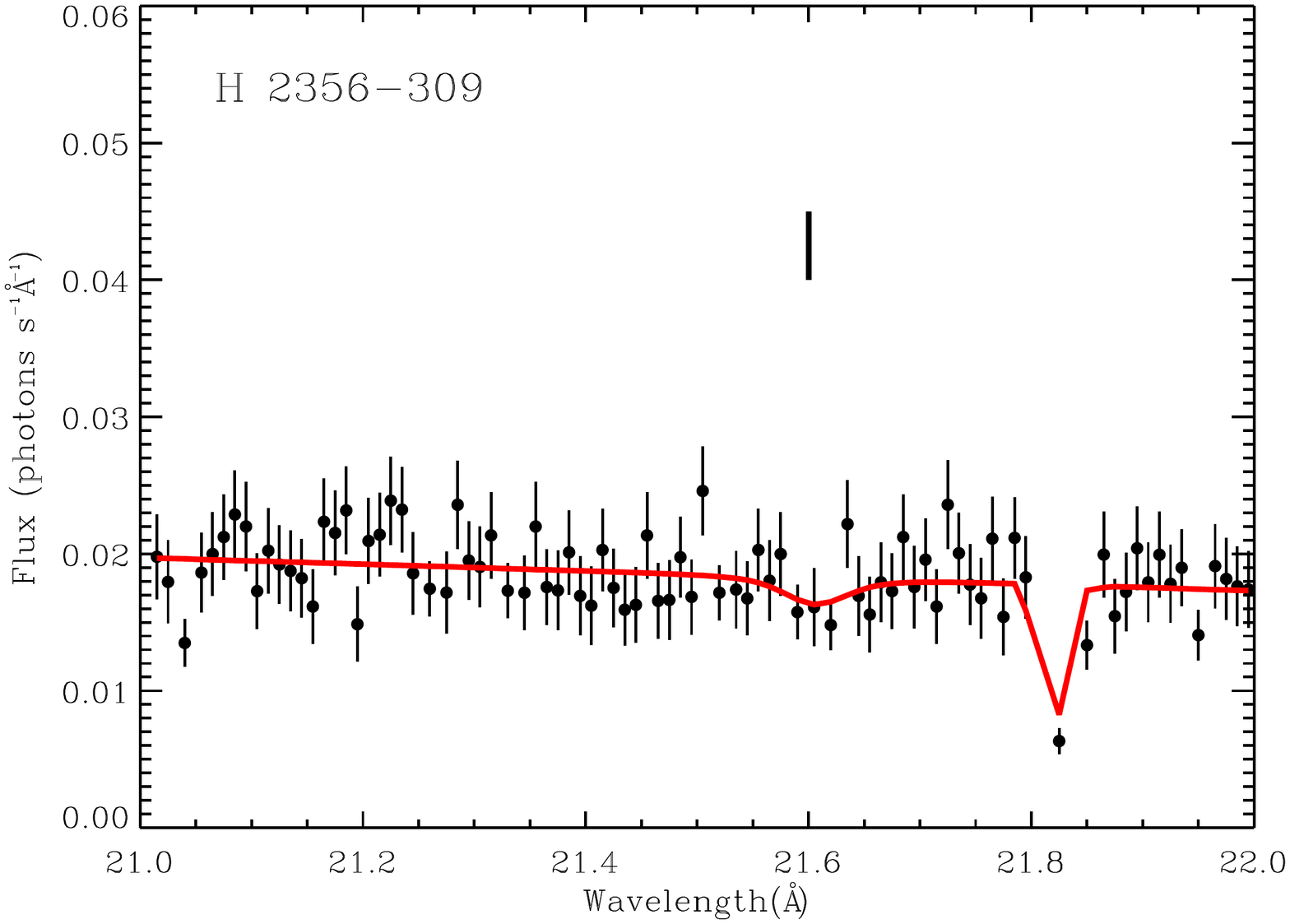}
\vskip-0.8cm
\includegraphics[height=0.23\textheight,width=0.47\textwidth]{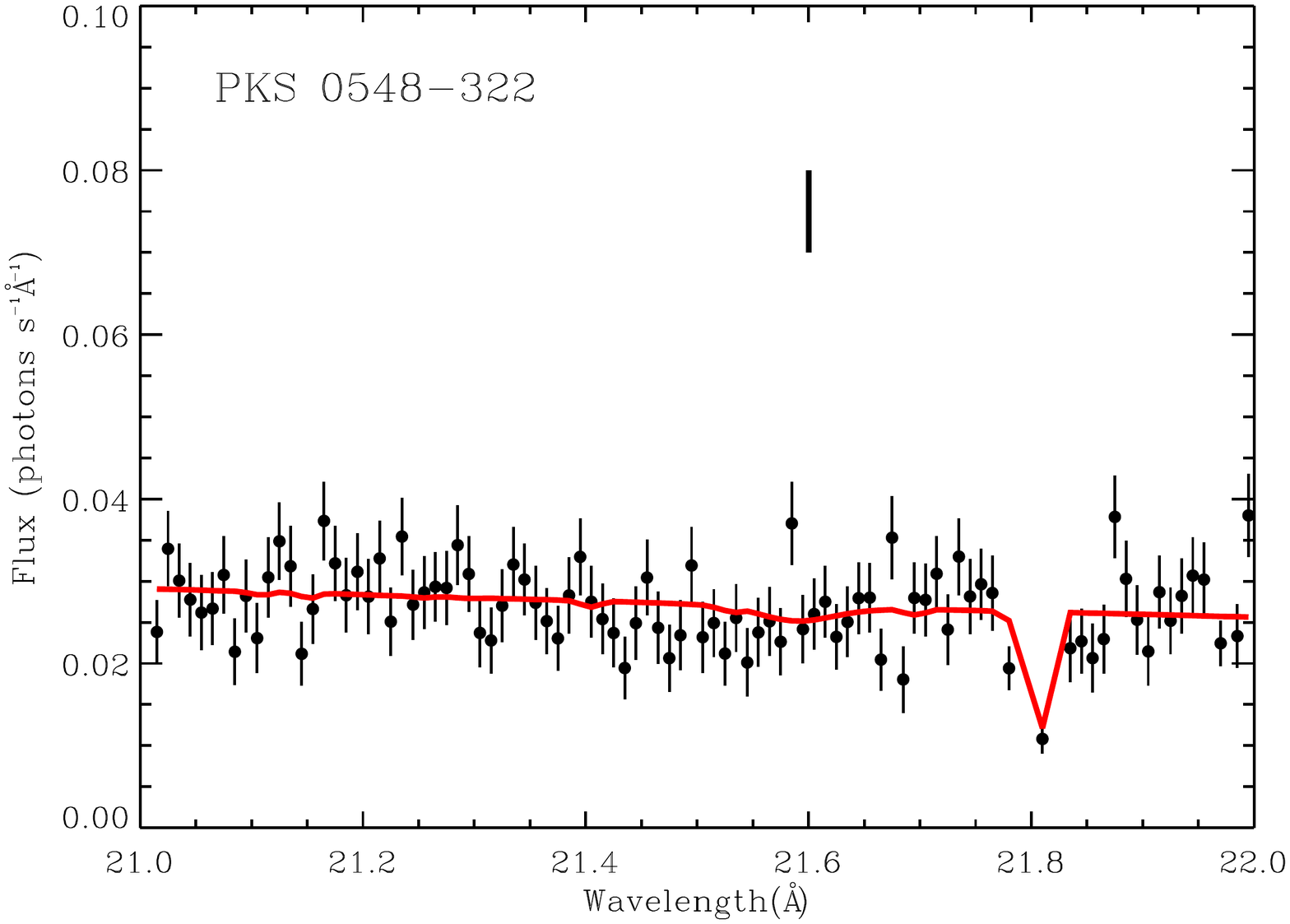}
\includegraphics[height=0.23\textheight,width=0.47\textwidth]{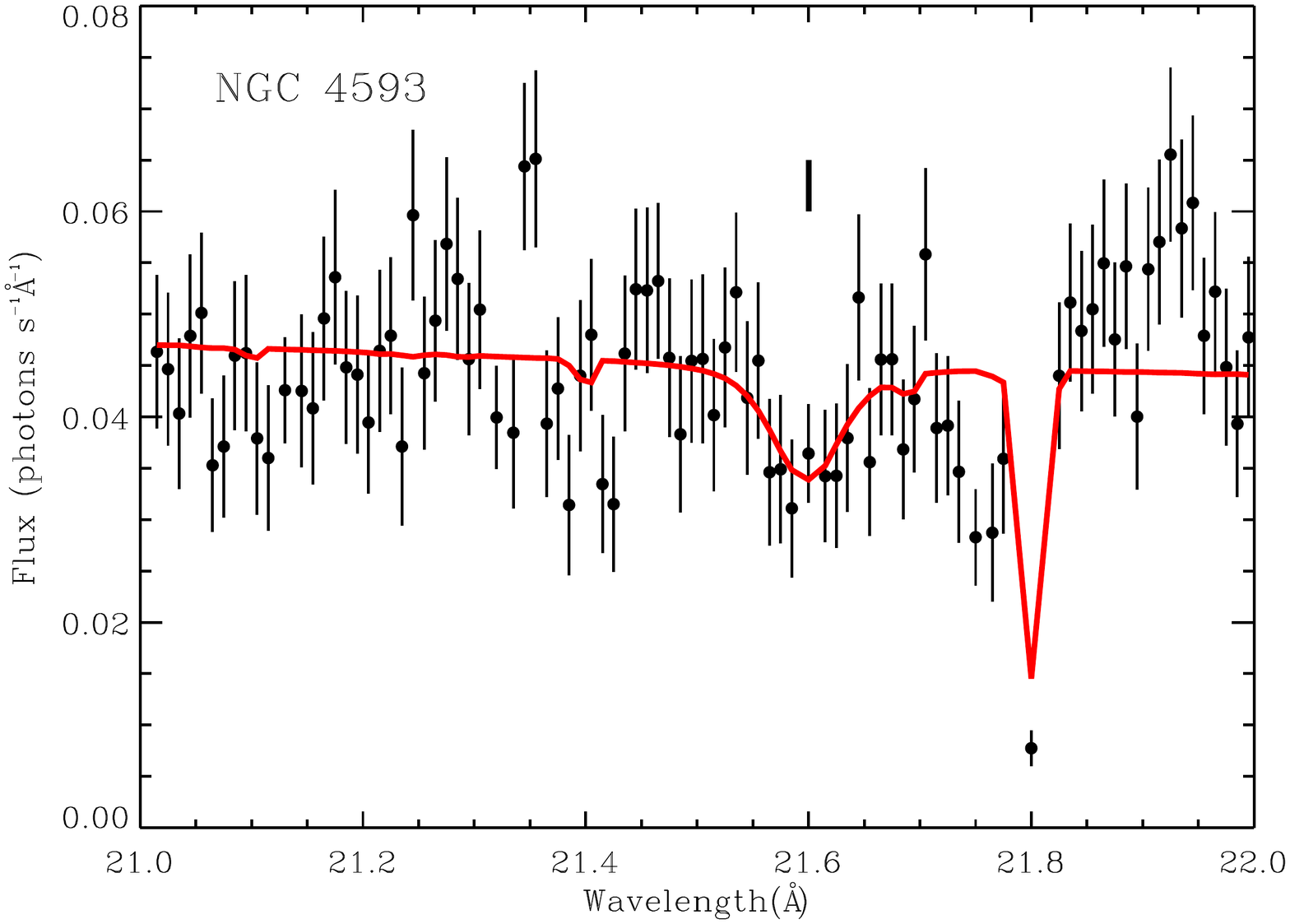}
\vskip-0.8cm
\includegraphics[height=0.23\textheight,width=0.47\textwidth]{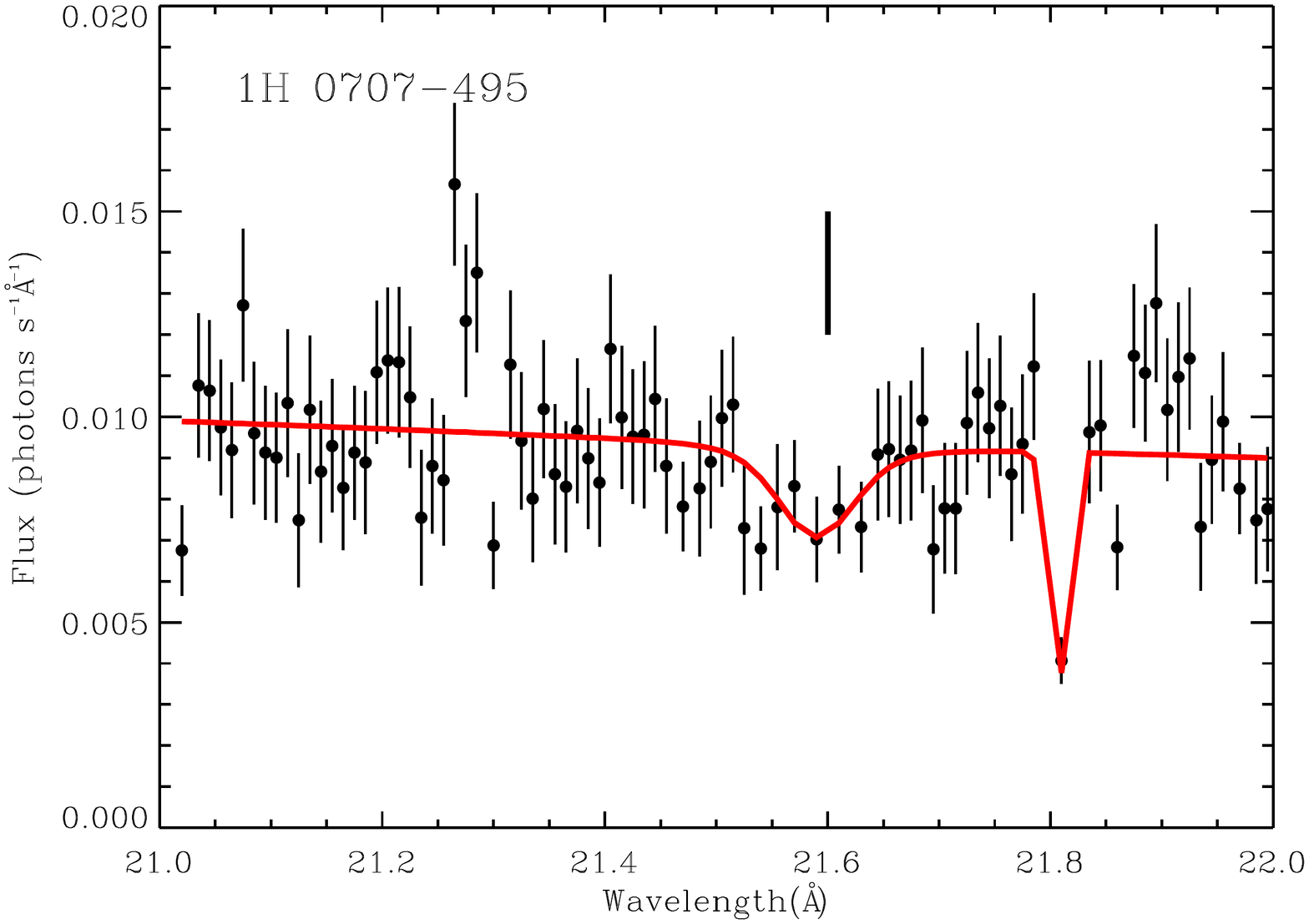}
\includegraphics[height=0.23\textheight,width=0.47\textwidth]{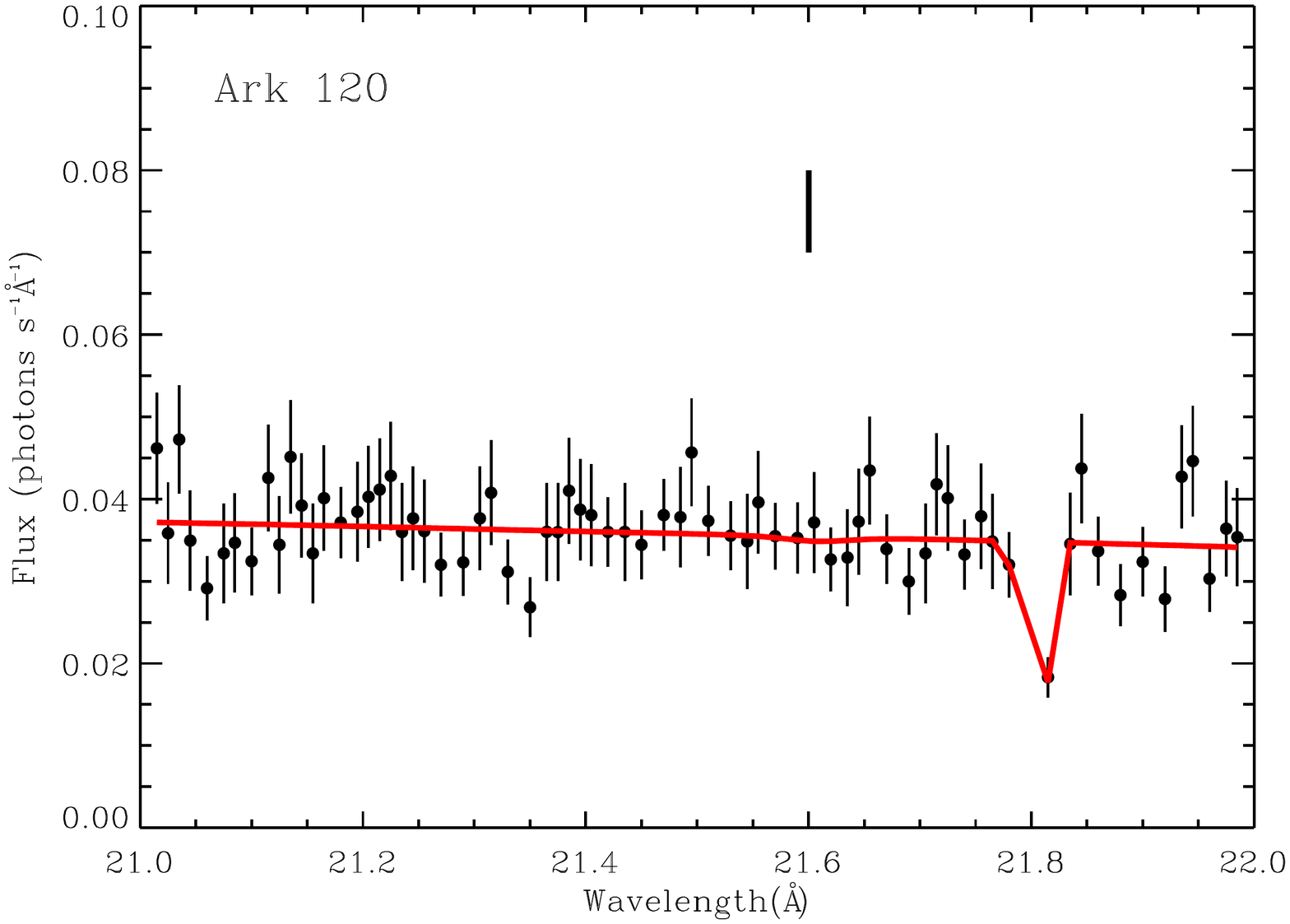}
\vskip-0.8cm
\caption{Same as the Figure~\ref{fig:spec1}, but for NGC~5548, H~2356-309, PKS~0548-322, NGC~4593, 1H~0717-495, and Ark~120.}
\label{fig:spec4}
\end{figure*}

\begin{figure*}[t]
\center
\includegraphics[height=0.23\textheight,width=0.47\textwidth]{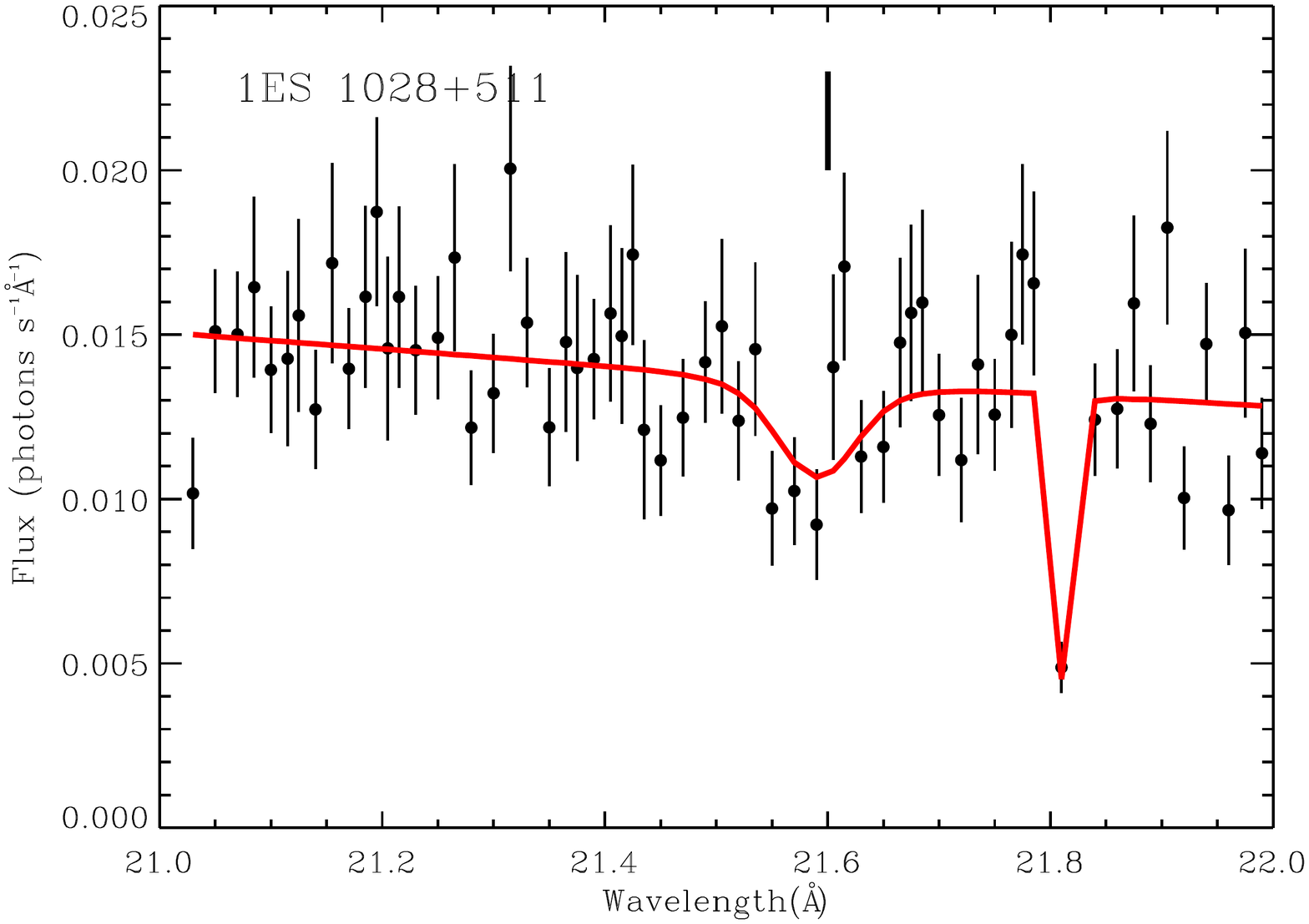}
\includegraphics[height=0.23\textheight,width=0.47\textwidth]{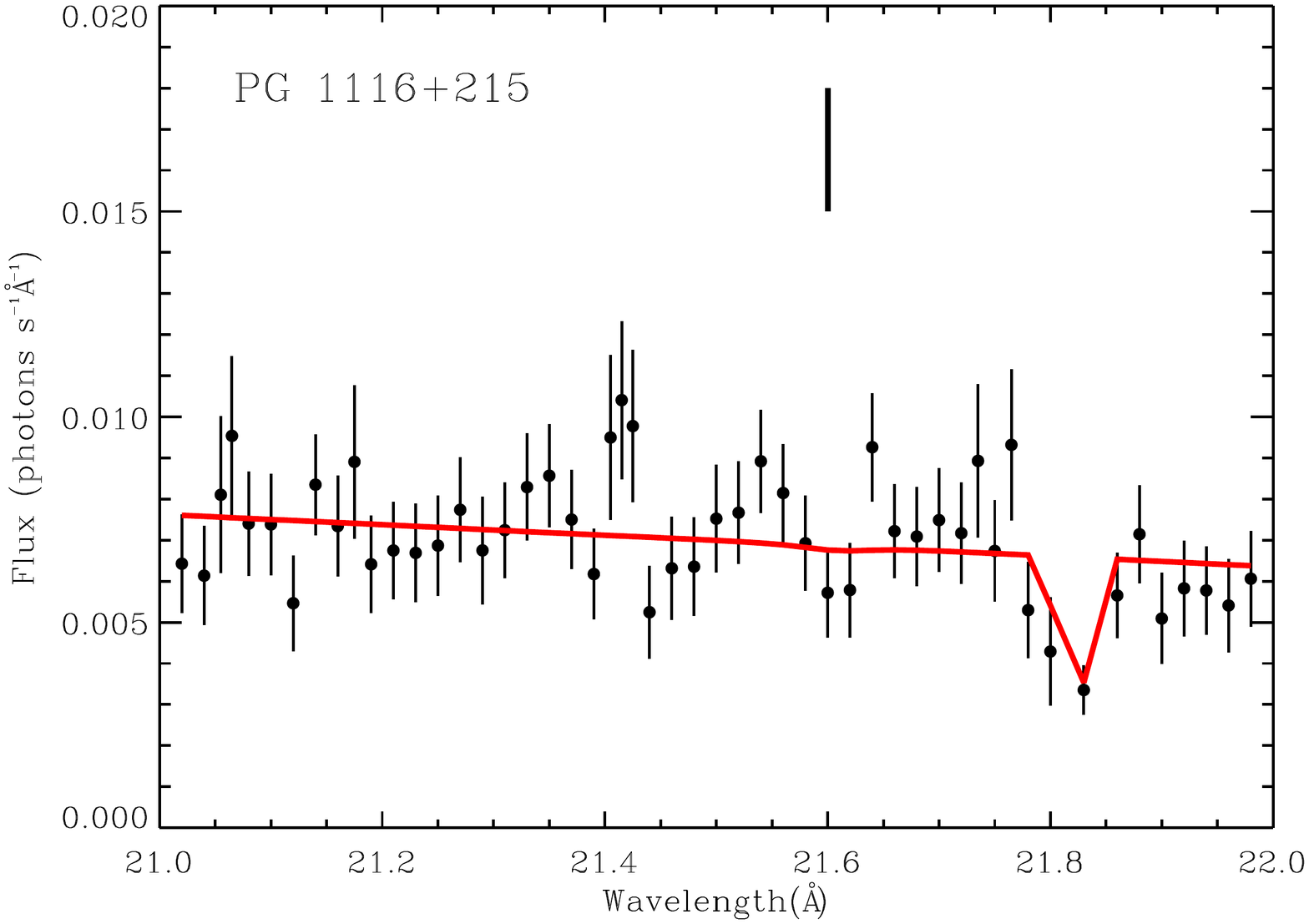}
\vskip-0.8cm
\includegraphics[height=0.23\textheight,width=0.47\textwidth]{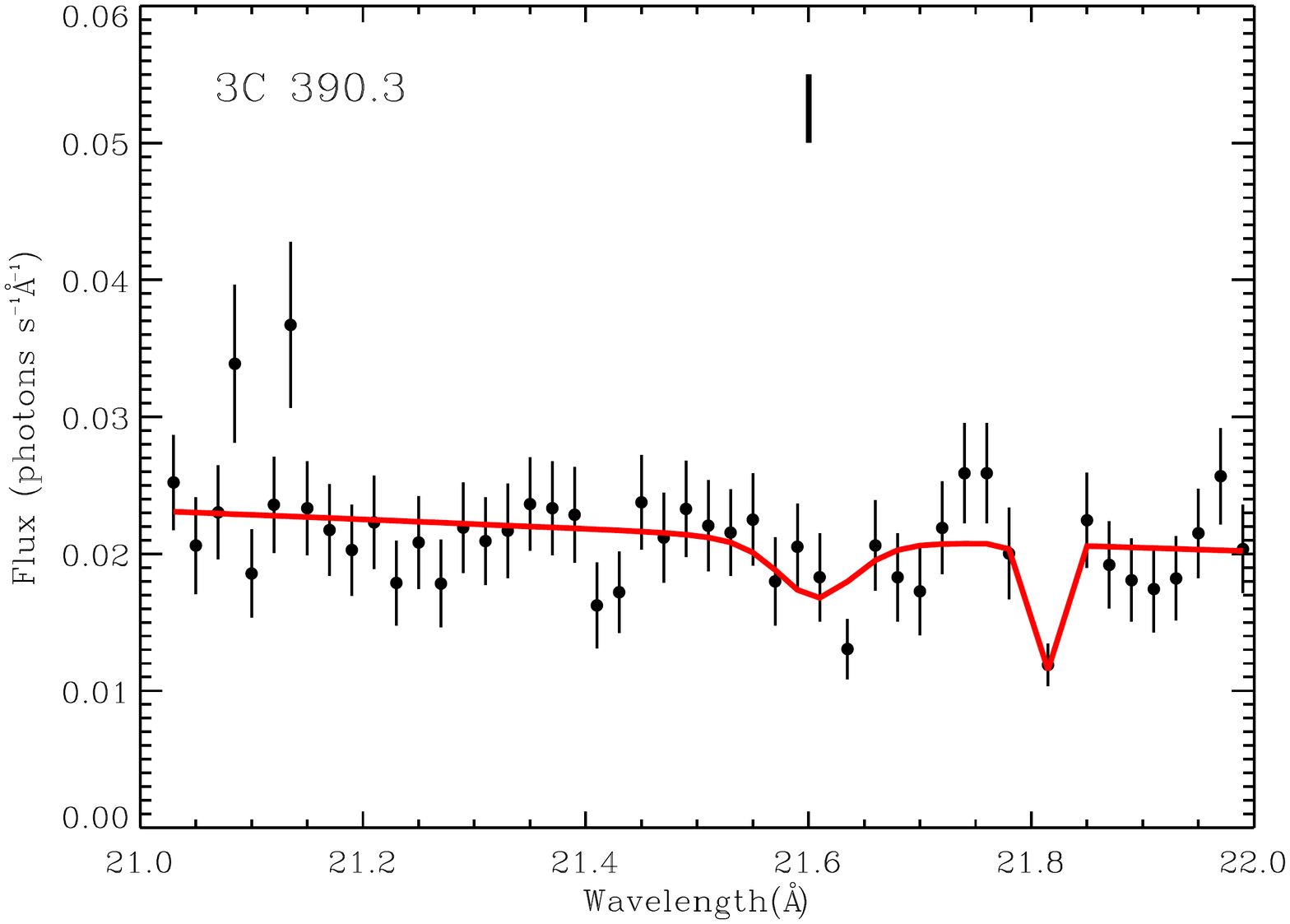}
\includegraphics[height=0.23\textheight,width=0.47\textwidth]{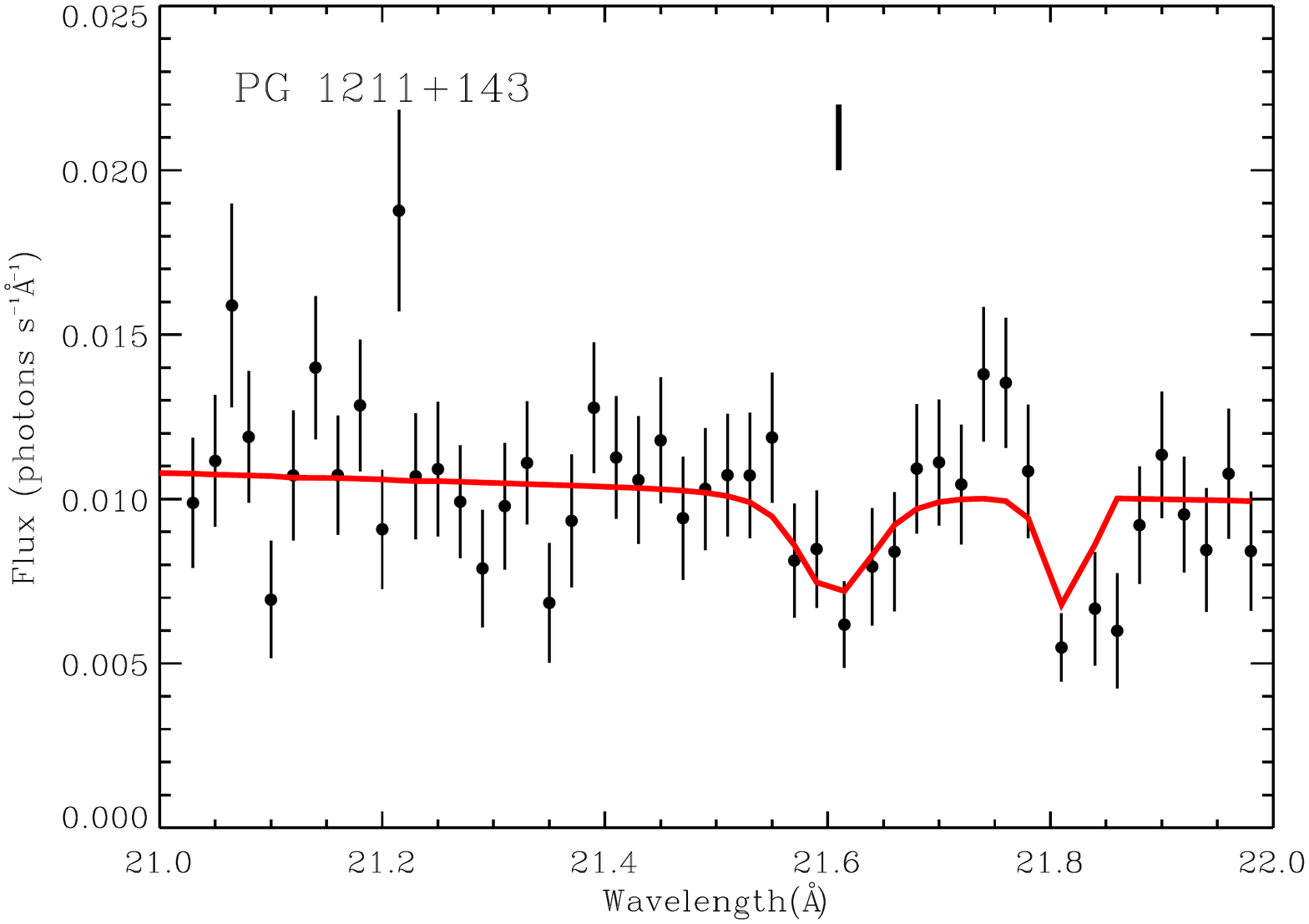}
\vskip-0.8cm
\includegraphics[height=0.23\textheight,width=0.47\textwidth]{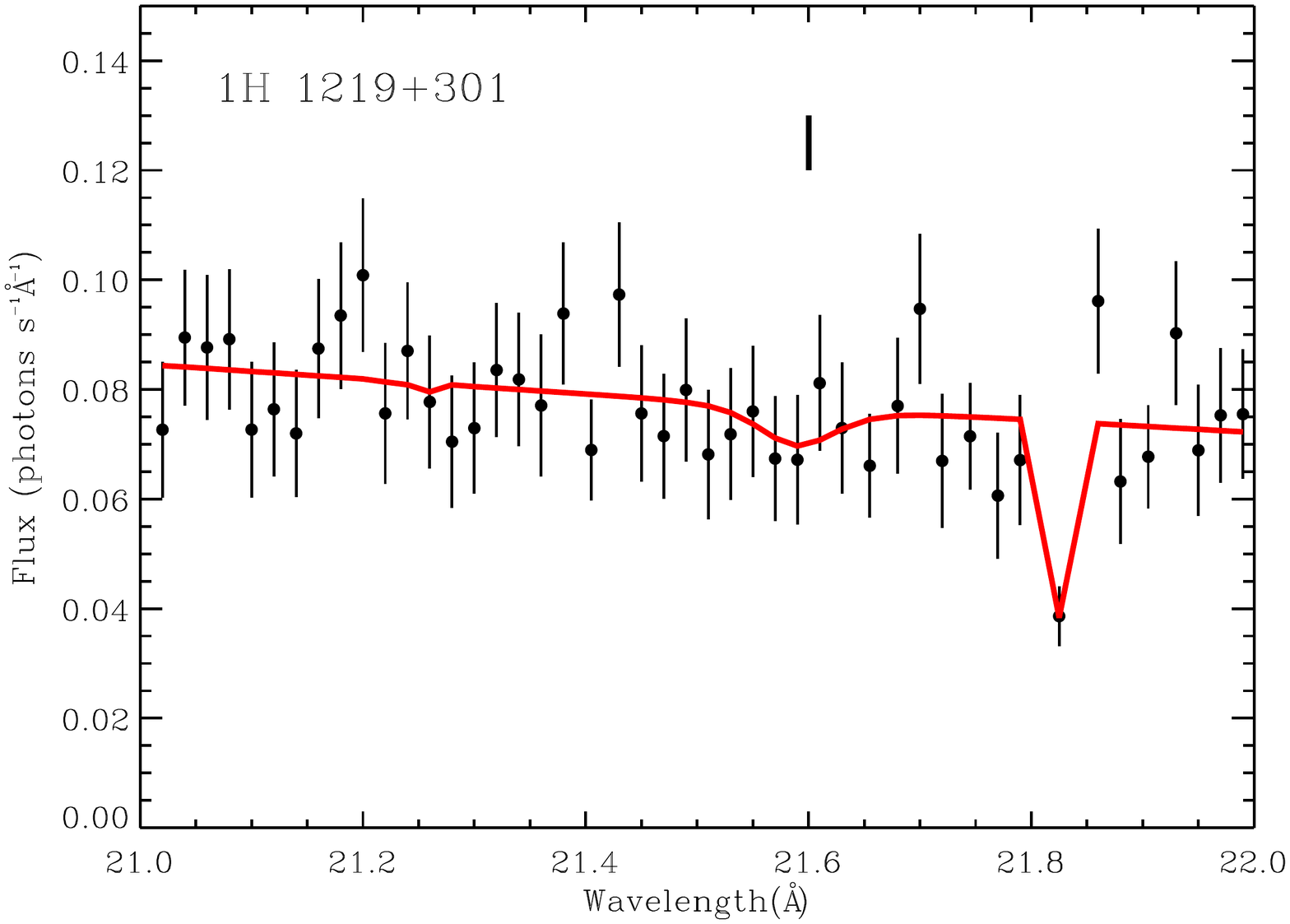}
\includegraphics[height=0.23\textheight,width=0.47\textwidth]{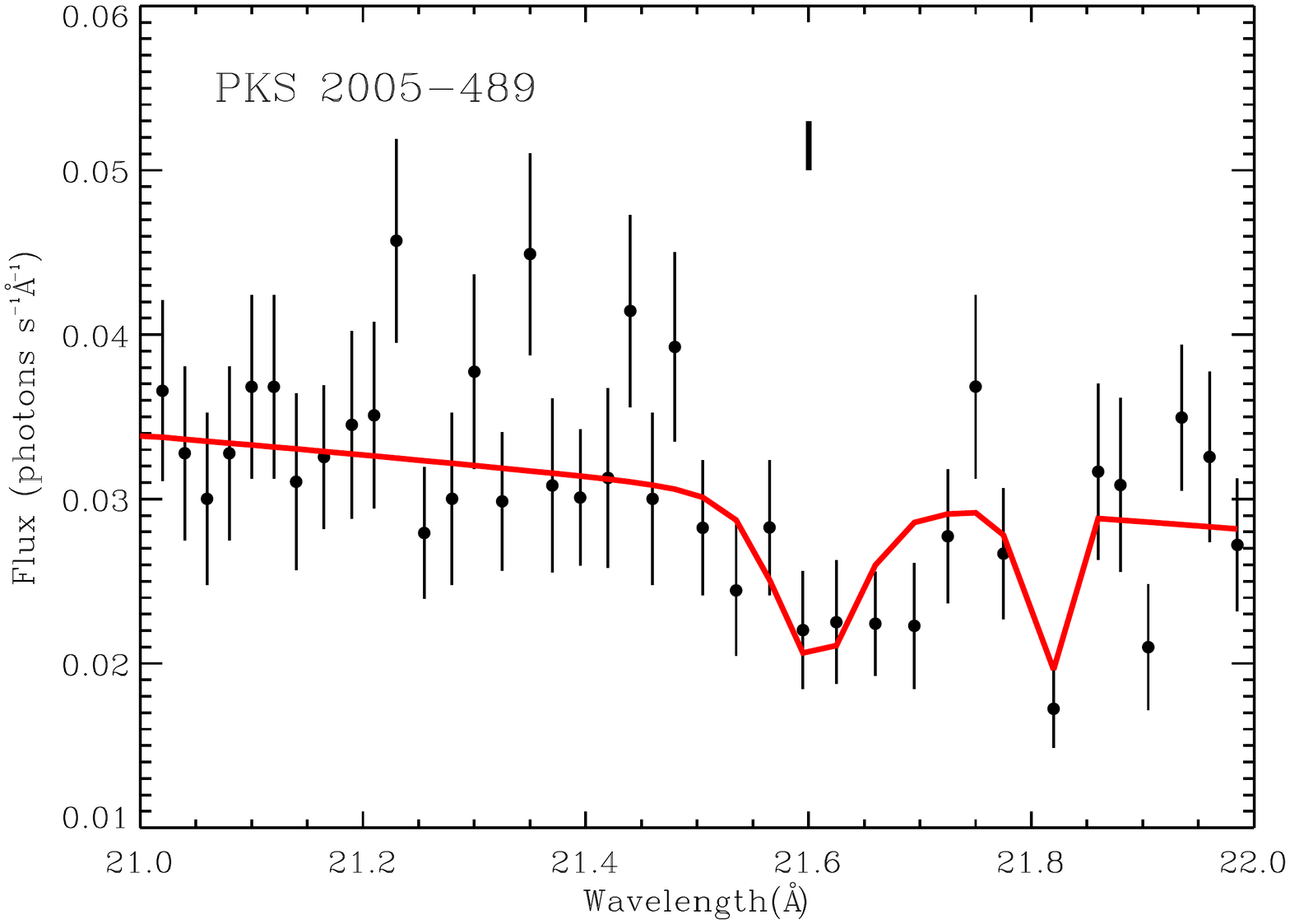}
\vskip-0.8cm
\caption{Same as the Figure~\ref{fig:spec1}, but for 1ES~1028+511, PG1116+215, 3C390.3, PG~1211+143, 1H~1219+301, and PKS~2005-489.}
\label{fig:spec5}
\end{figure*}

\begin{figure*}[t]
\center
\includegraphics[height=0.23\textheight,width=0.47\textwidth]{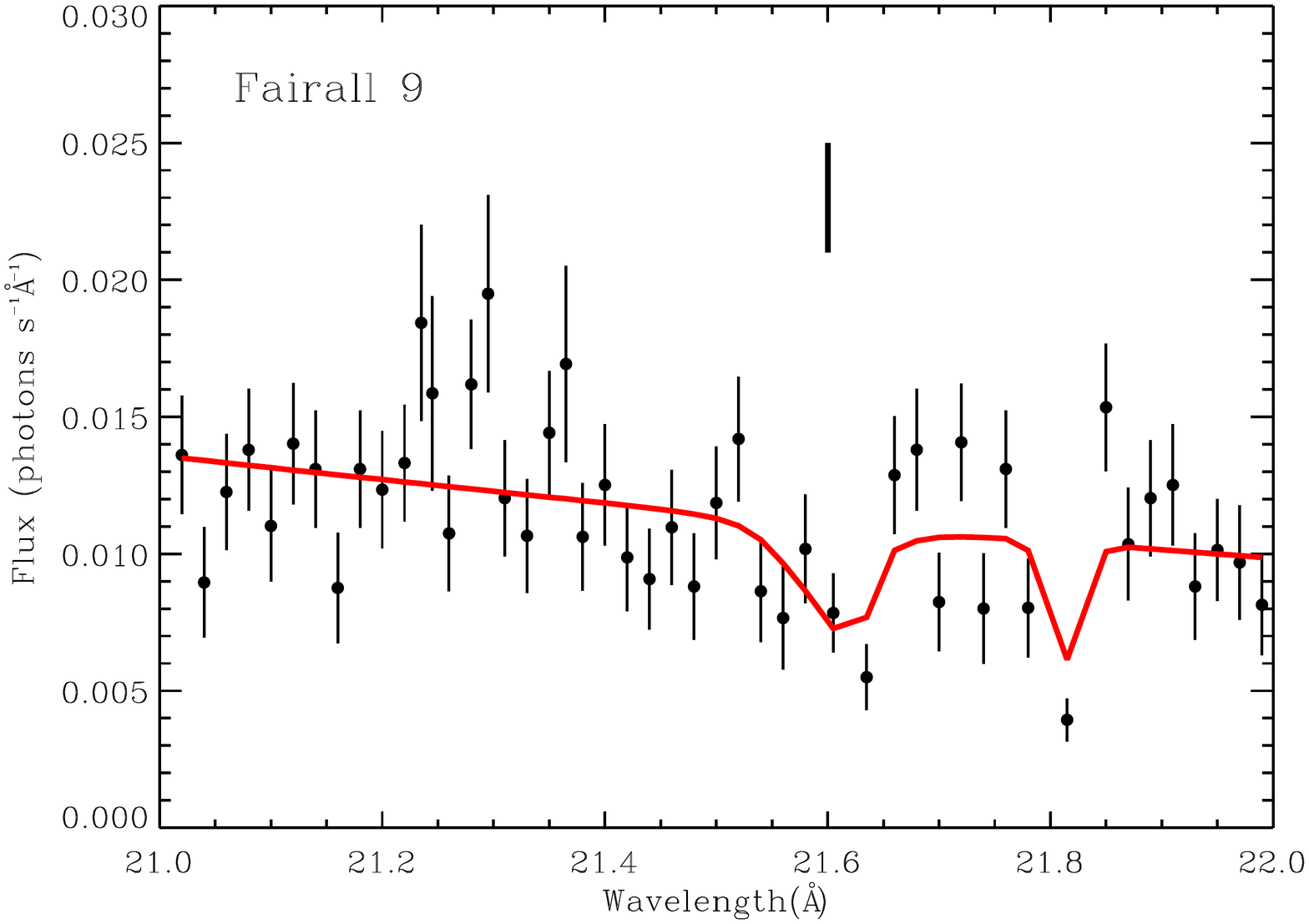}
\includegraphics[height=0.23\textheight,width=0.47\textwidth]{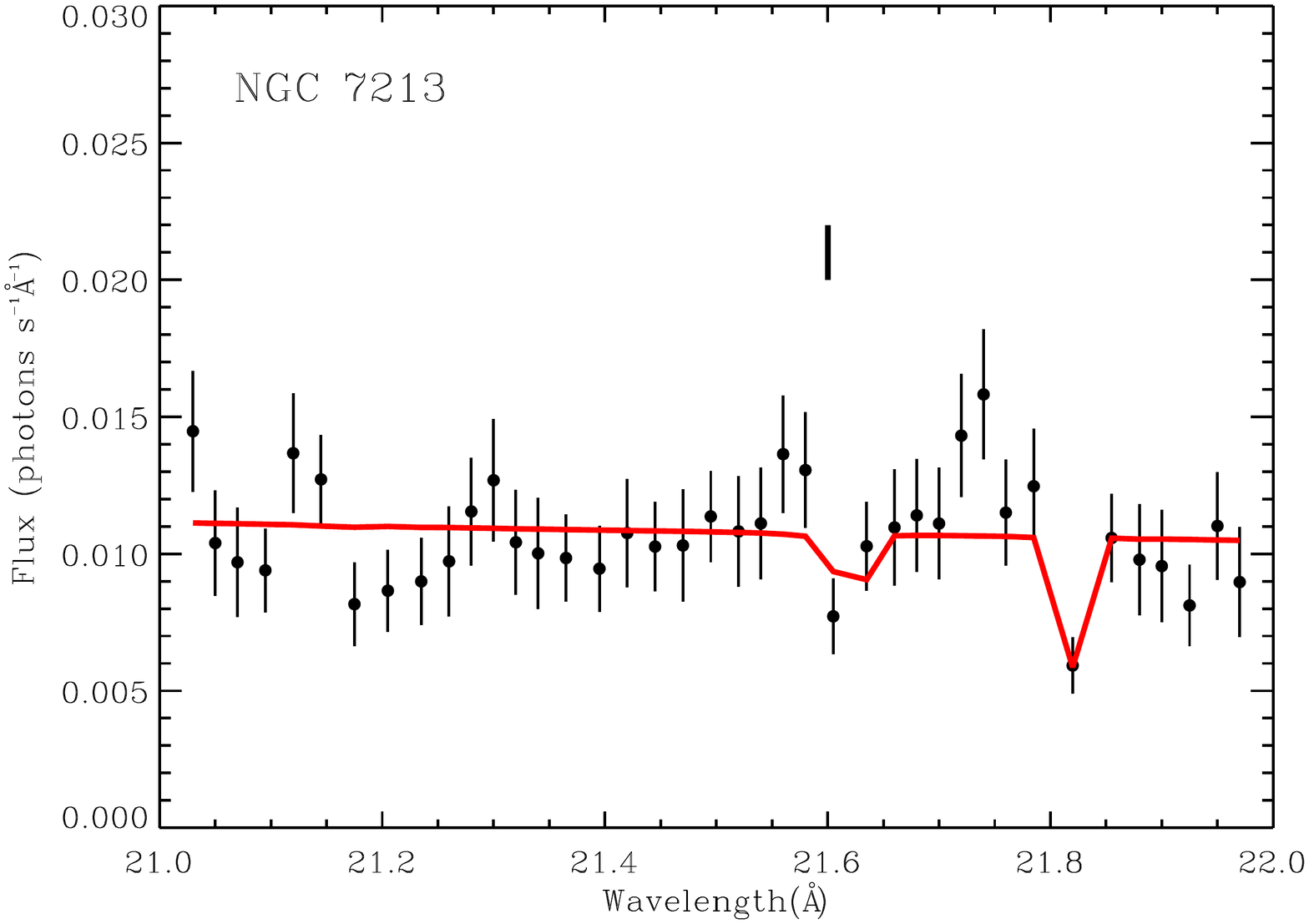}
\vskip-0.8cm
\includegraphics[height=0.23\textheight,width=0.47\textwidth]{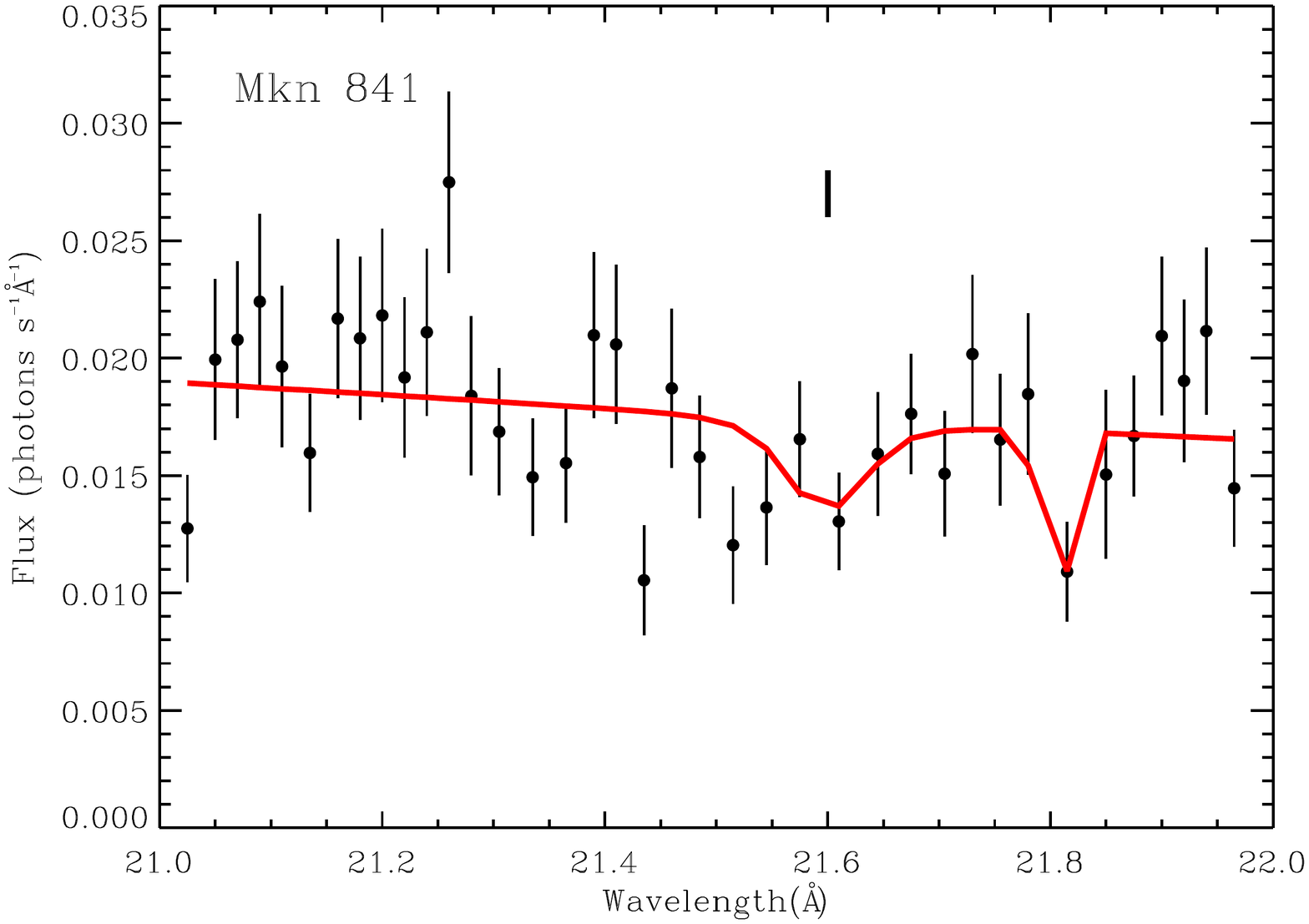}
\includegraphics[height=0.23\textheight,width=0.47\textwidth]{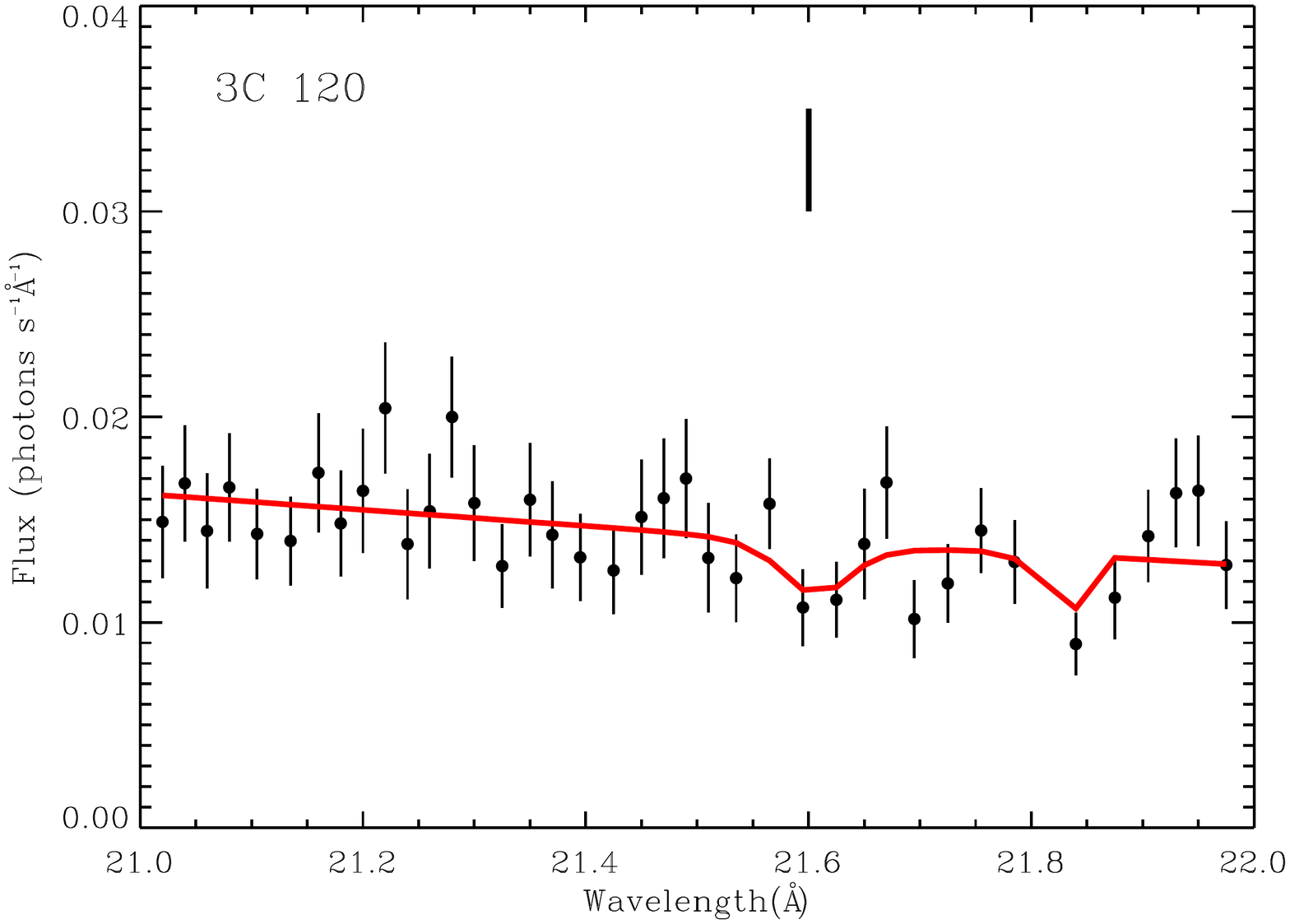}
\vskip-0.8cm
\includegraphics[height=0.23\textheight,width=0.47\textwidth]{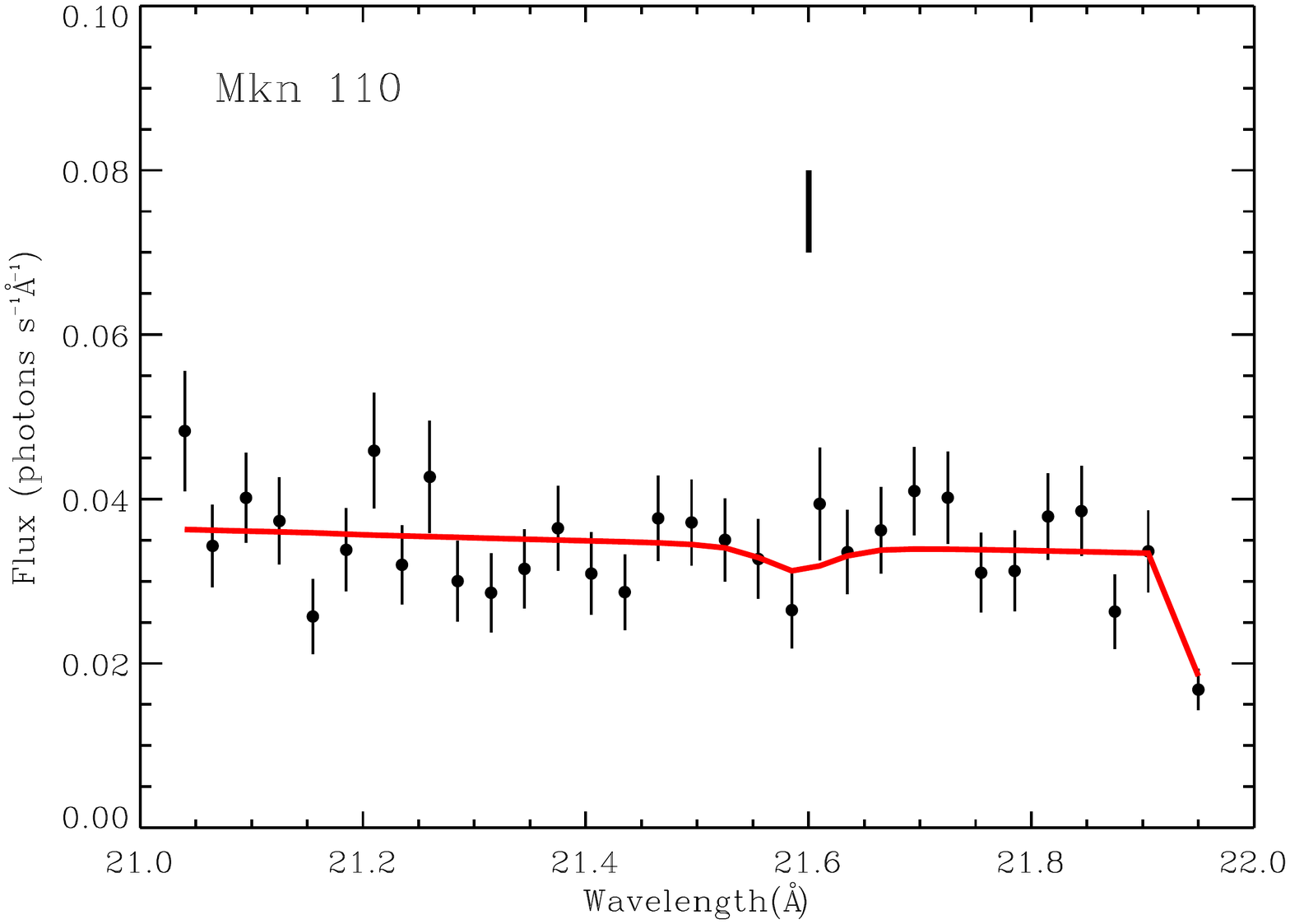}
\includegraphics[height=0.23\textheight,width=0.47\textwidth]{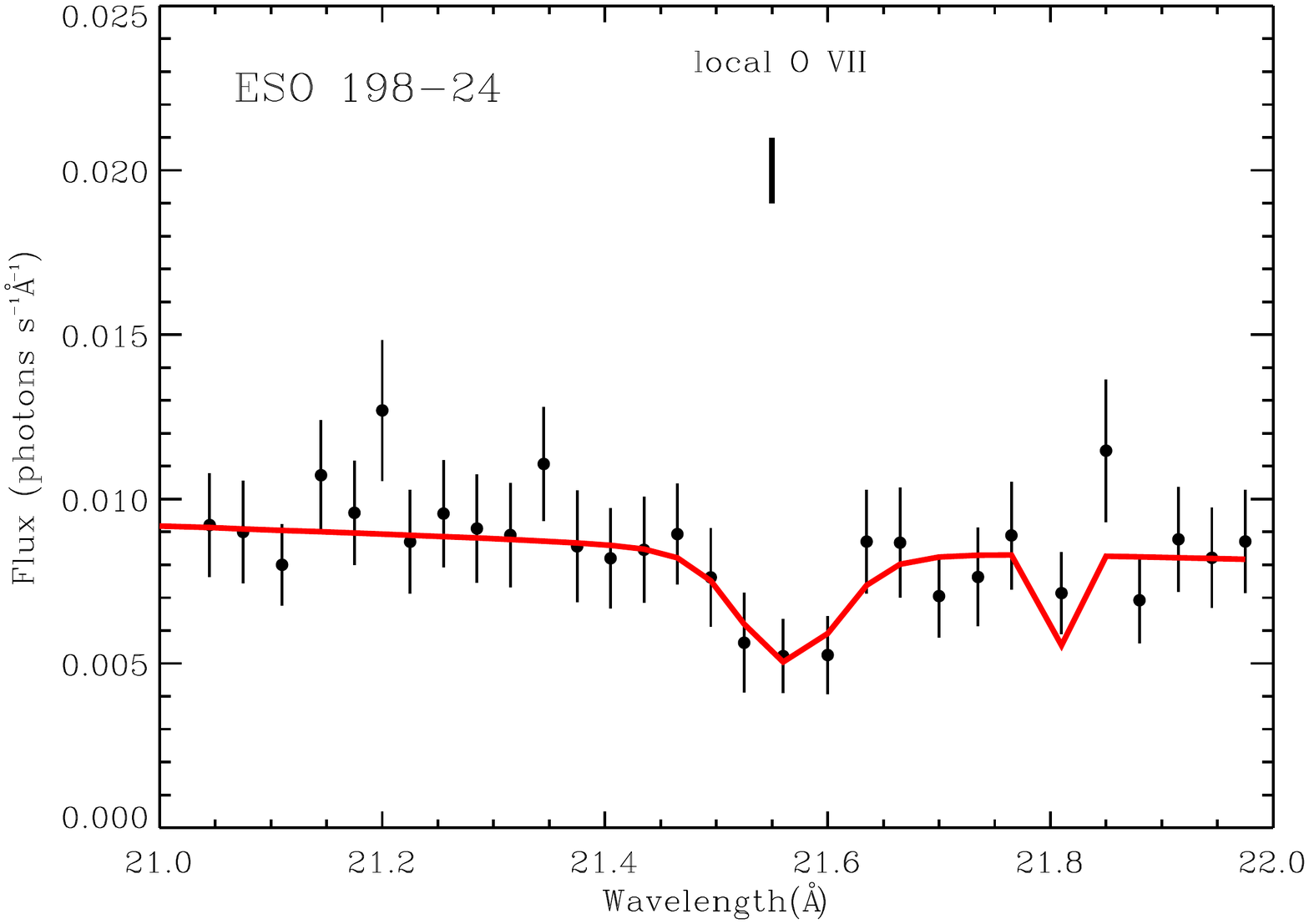}
\vskip-0.8cm
\caption{Same as the Figure~\ref{fig:spec1}, but for Fairall~9, NGC~7213, Mkn~841, 3C~120, Mkn~110, and ESO~198-24.}
\label{fig:spec6}
\end{figure*}

\begin{figure*}[t]
\center
\includegraphics[height=0.23\textheight,width=0.47\textwidth]{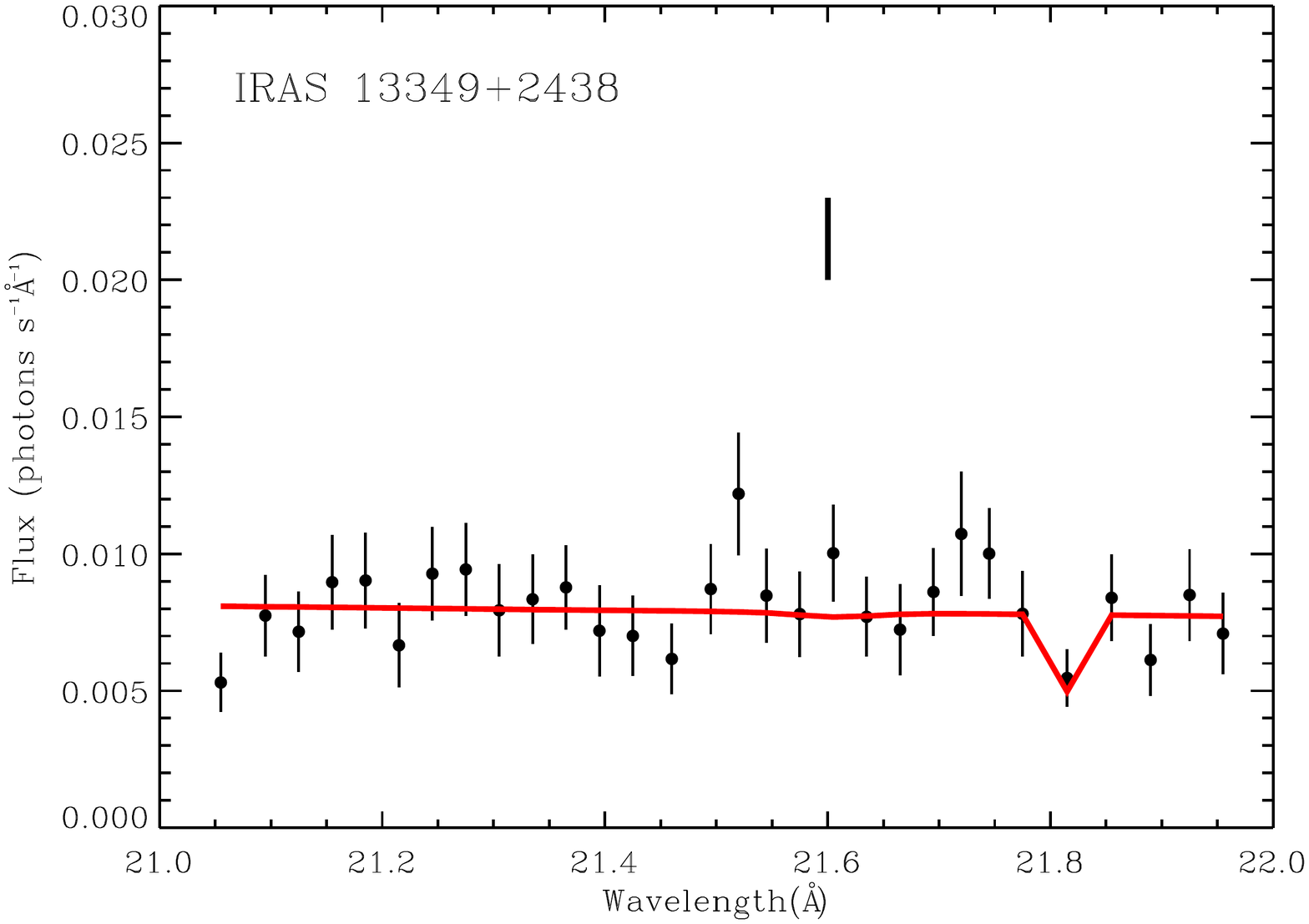}
\includegraphics[height=0.23\textheight,width=0.47\textwidth]{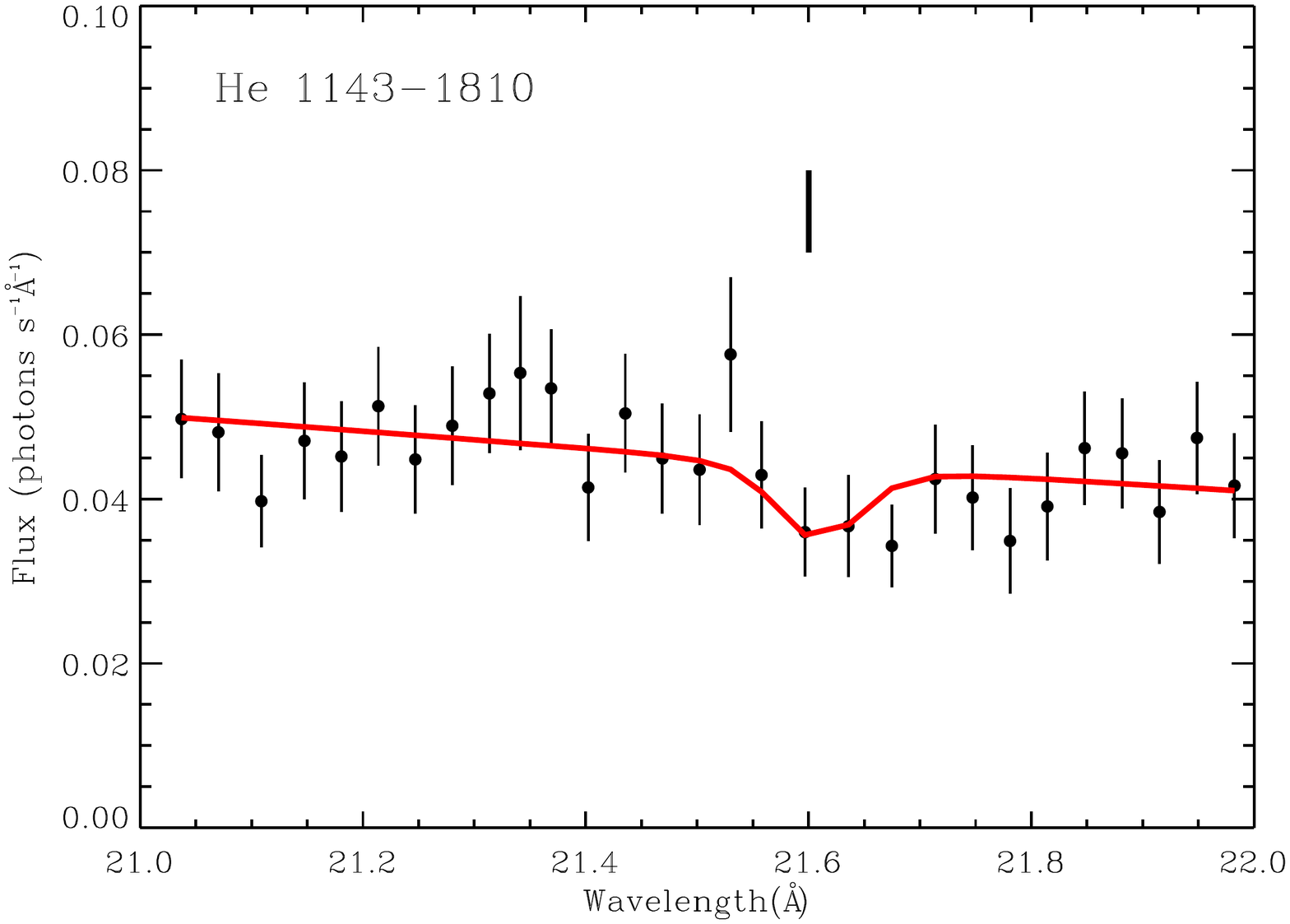}
\vskip-0.8cm
\includegraphics[height=0.23\textheight,width=0.47\textwidth]{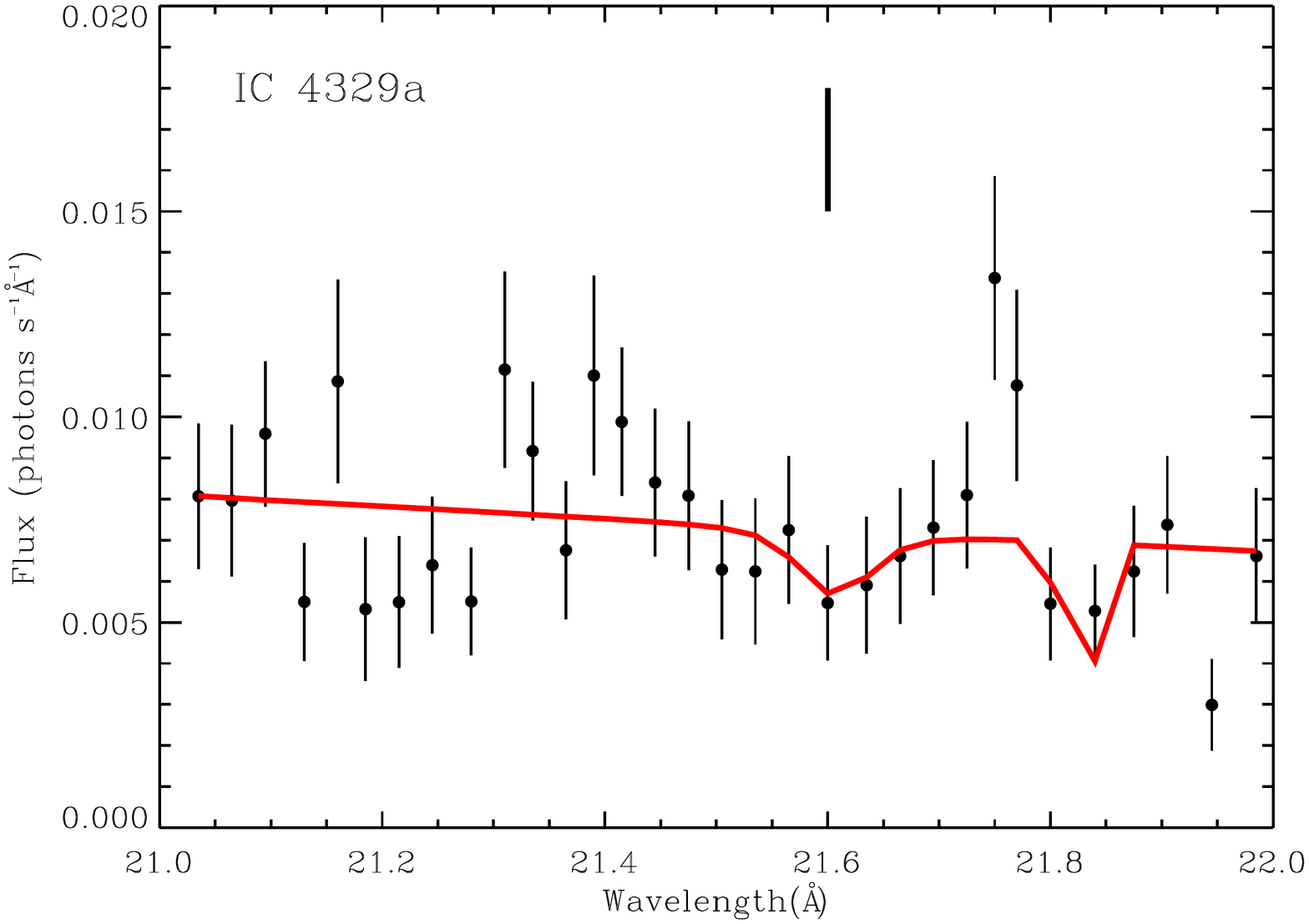}
\includegraphics[height=0.23\textheight,width=0.47\textwidth]{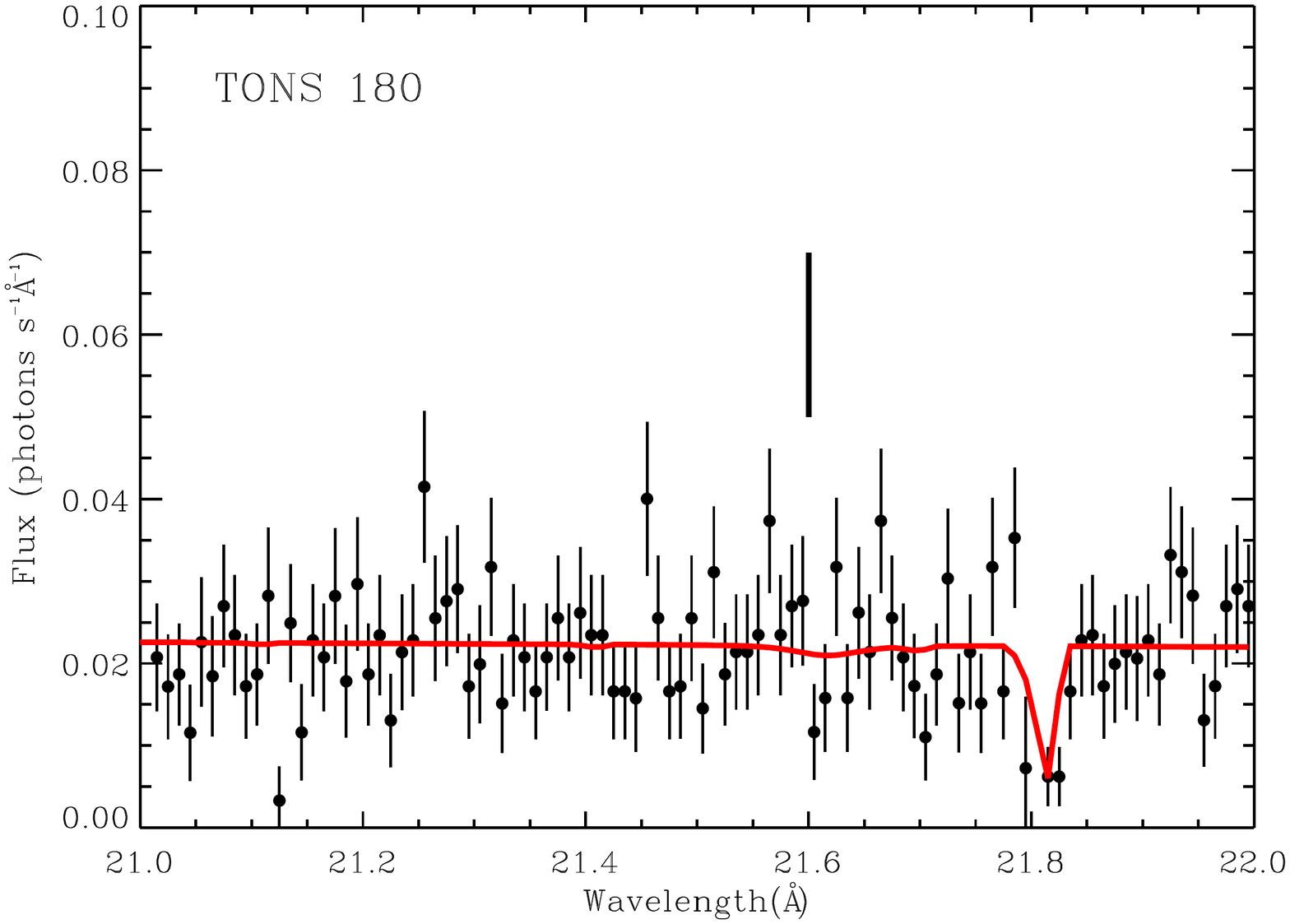}
\vskip-0.8cm
\includegraphics[height=0.23\textheight,width=0.47\textwidth]{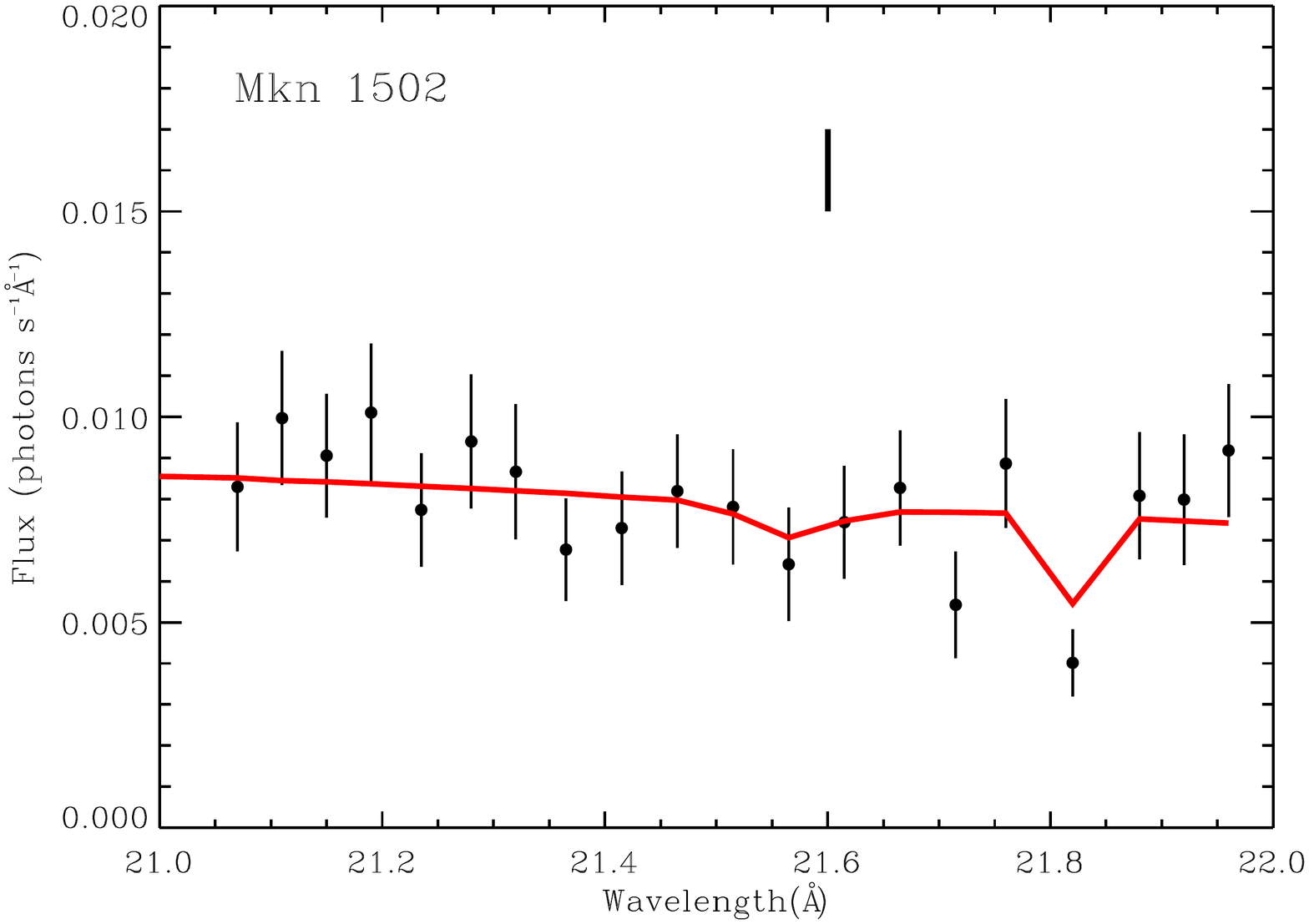}
\includegraphics[height=0.23\textheight,width=0.47\textwidth]{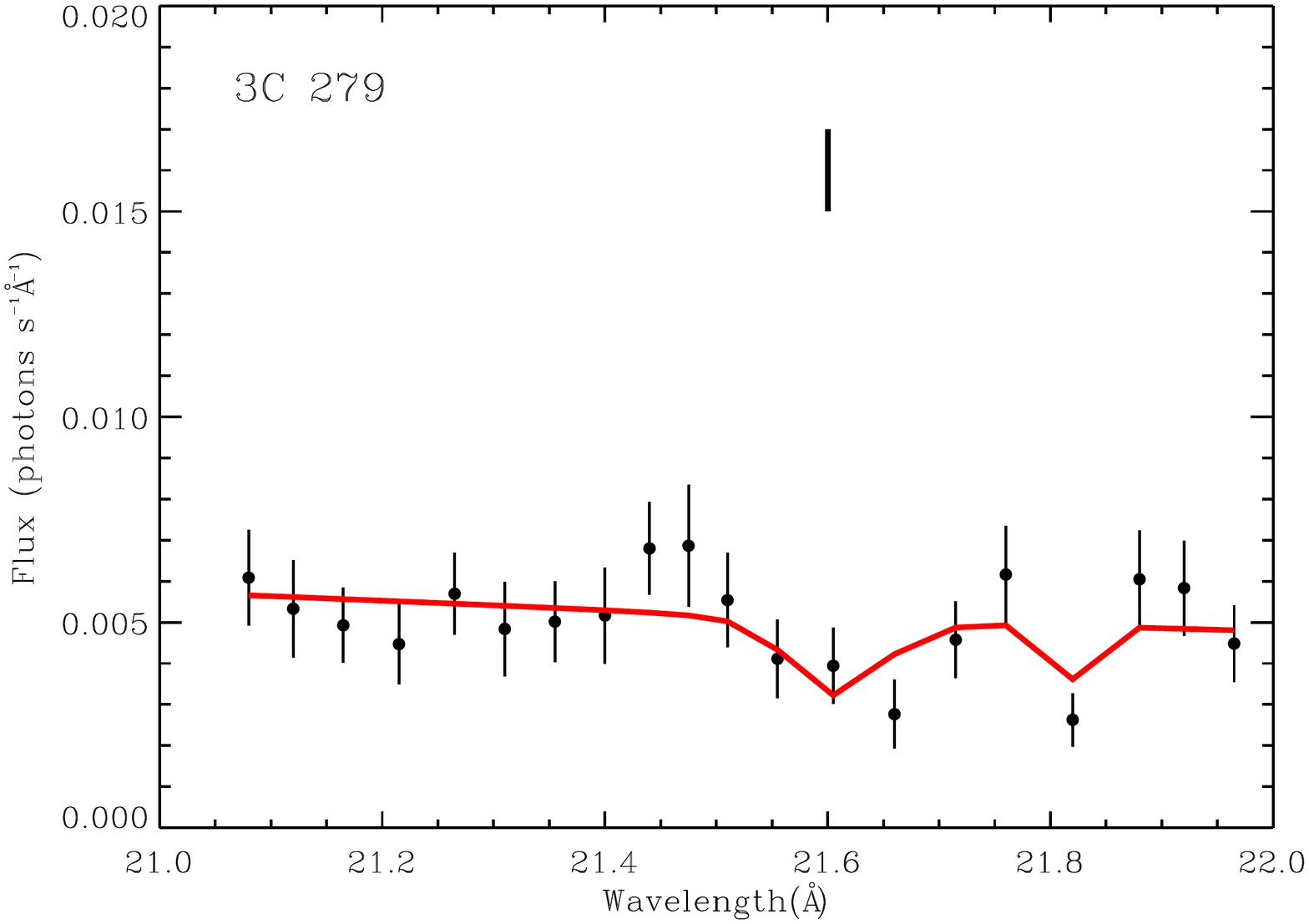}
\vskip-0.8cm
\caption{Same as the Figure~\ref{fig:spec1}, but for IRAS~13349+2438, He~1143-1810, IC~4329a, TONS~180, Mkn~1502, and 3C~279.}
\label{fig:spec7}
\end{figure*}

\begin{figure*}[t]
\center
\includegraphics[height=0.23\textheight,width=0.47\textwidth]{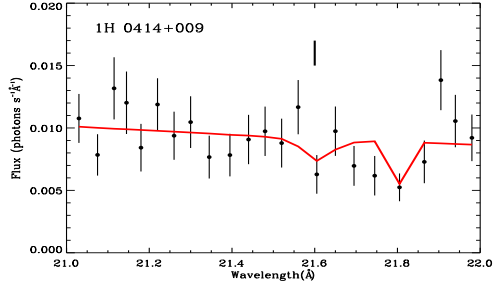}
\vskip-0.8cm
\caption{Same as the Figure~\ref{fig:spec1}, but for 1H~0414+009.}
\label{fig:spec8}
\end{figure*}

\begin{figure*}[t]
\center
\includegraphics[height=0.27\textheight,width=0.49\textwidth]{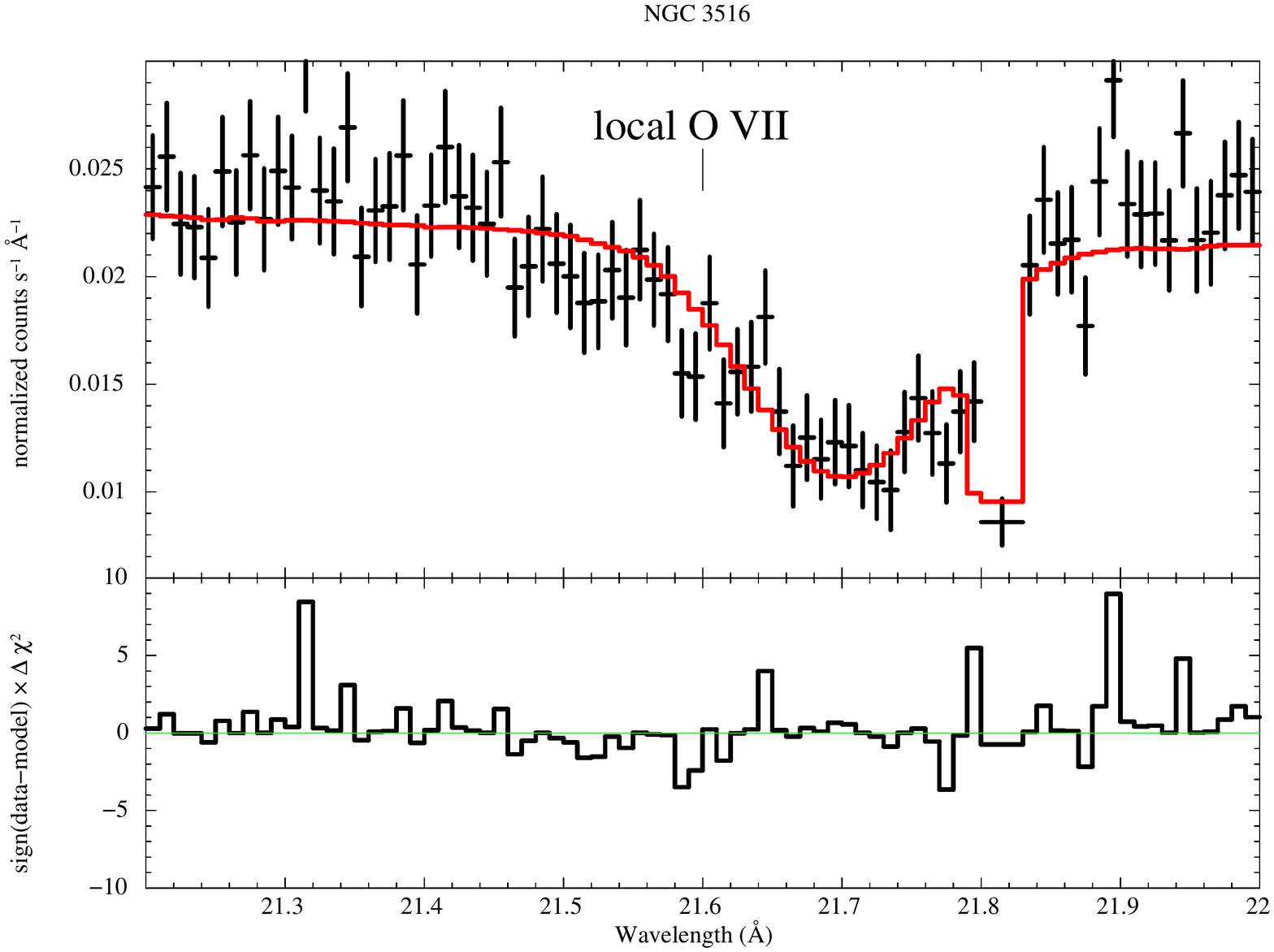}
\includegraphics[height=0.27\textheight,width=0.49\textwidth]{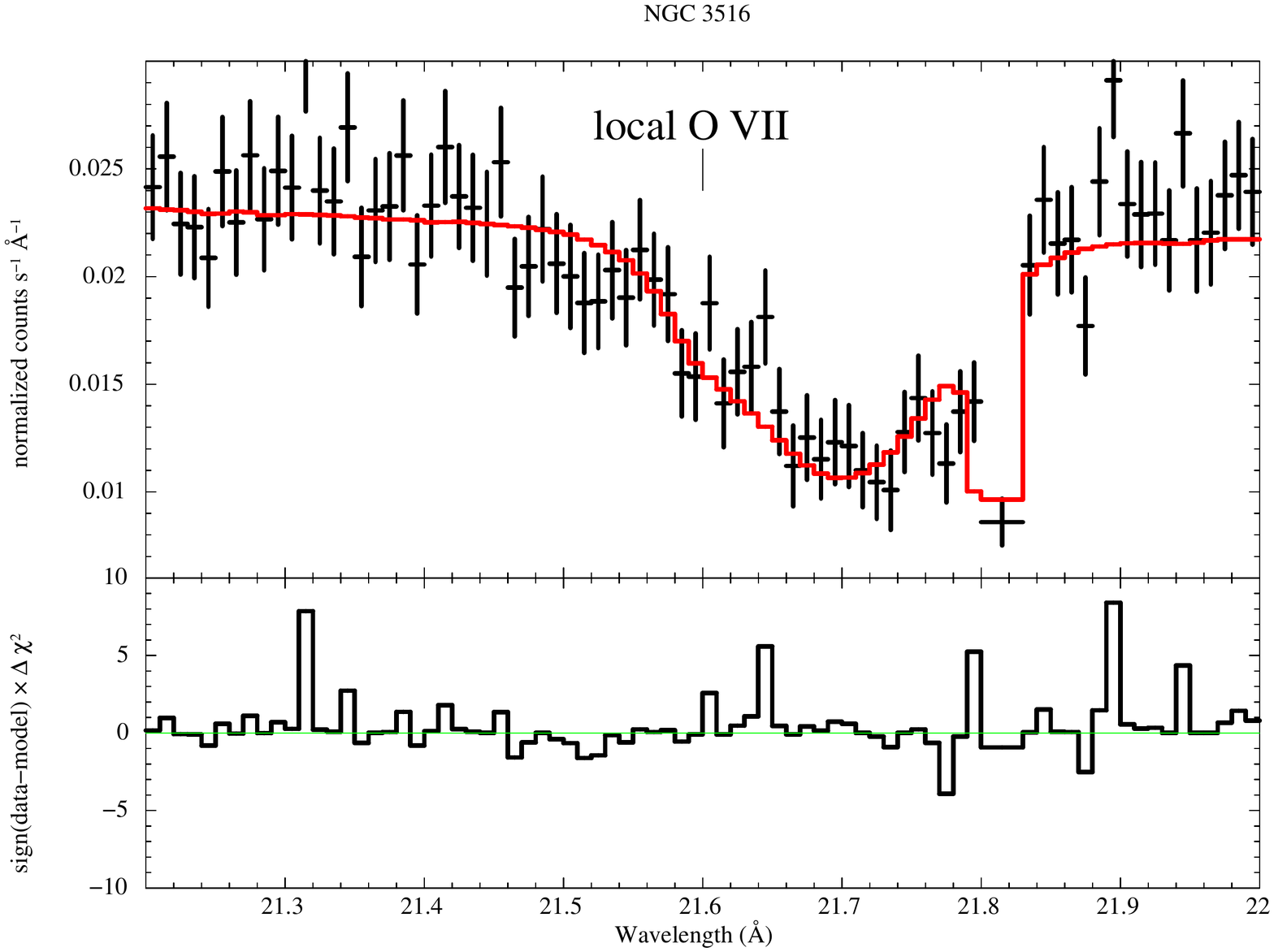}
\caption{Spectra of NGC~3516 between 21.2 and 22 \AA\, before including a $z\sim0$ \ion{O}{7} $K_{\alpha}$ absorption line (left figure) and after (right figure). The top panel in each figure is the spectrum, and the bottom is the $\Delta \chi^2$. The structure at $\sim$21.81 \AA\ is an instrumental feature. The broad absorption feature at $\sim$21.7 \AA\ is produced by the warm absorber intrinsic to NGC~3516.}
\label{fig:ngc3516}
\end{figure*}

\begin{figure*}[t]
\center
\includegraphics[height=0.3\textheight,width=0.49\textwidth]{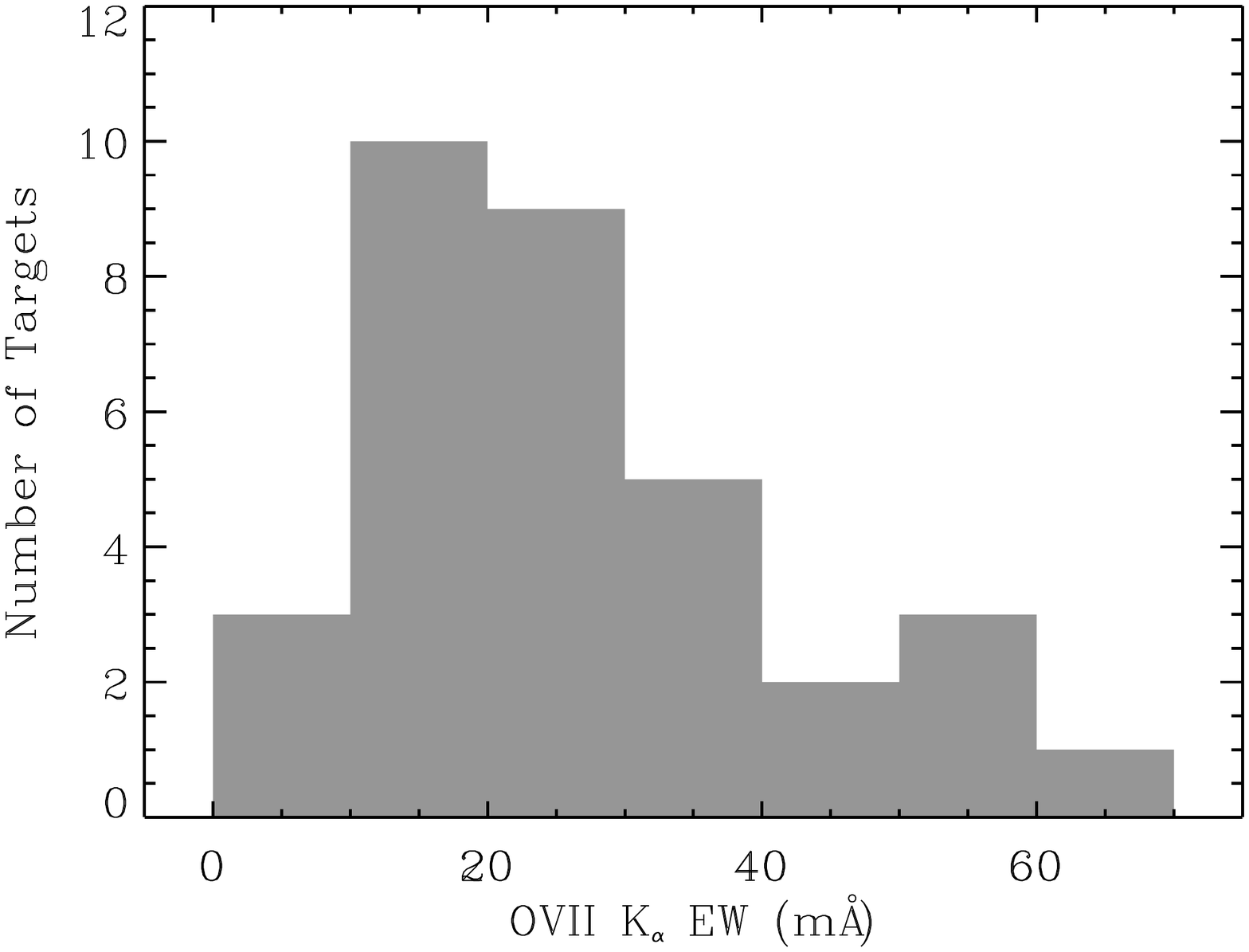}
\includegraphics[height=0.3\textheight,width=0.49\textwidth]{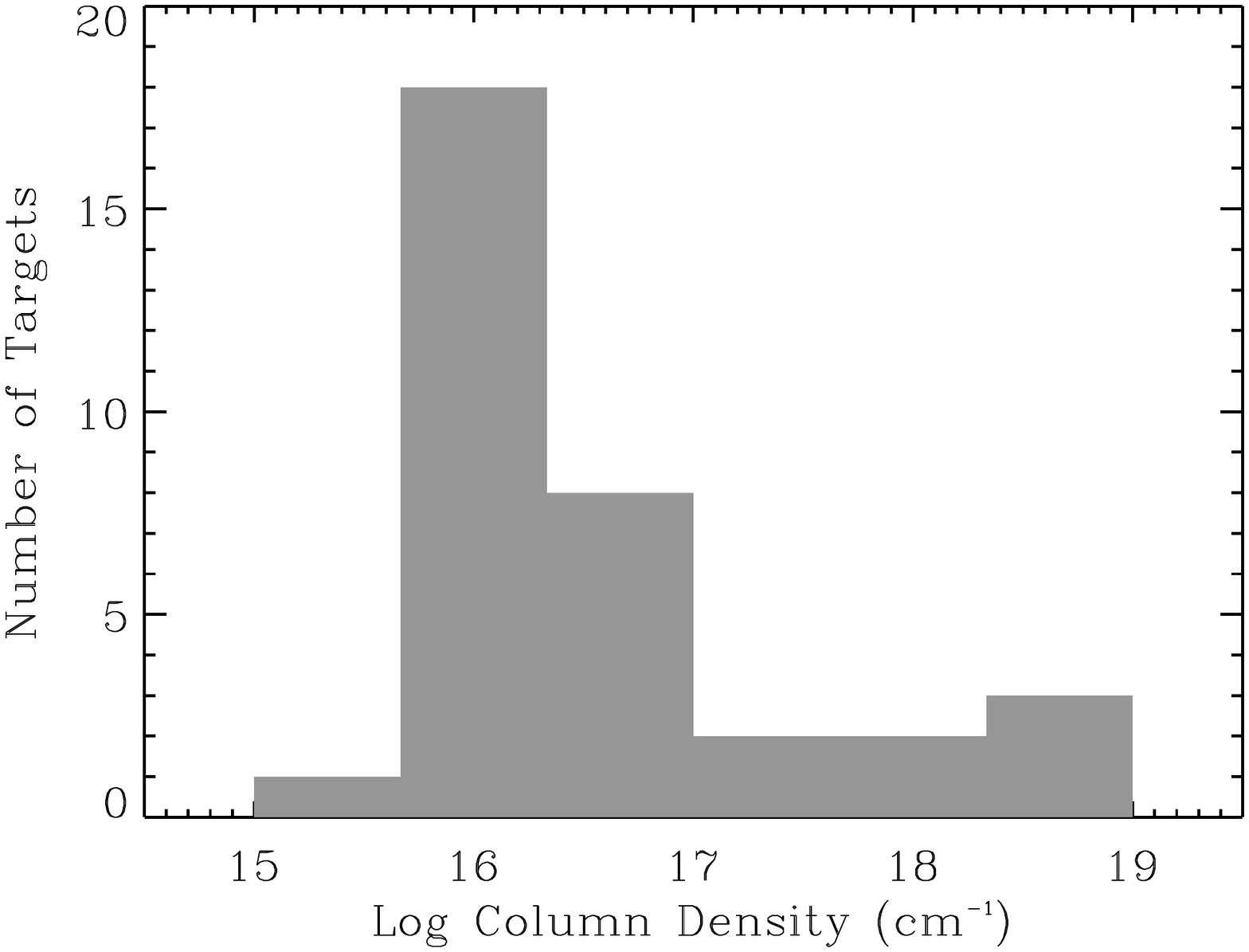}
\caption{Histogram distribution of the \ion{O}{7} $K_{\alpha}$ line EW (left panel) and column density (right panel).}
\label{fig:ewcol}
\end{figure*}

\begin{figure*}[h]
\center
\includegraphics[height=0.2\textheight,width=0.47\textwidth]{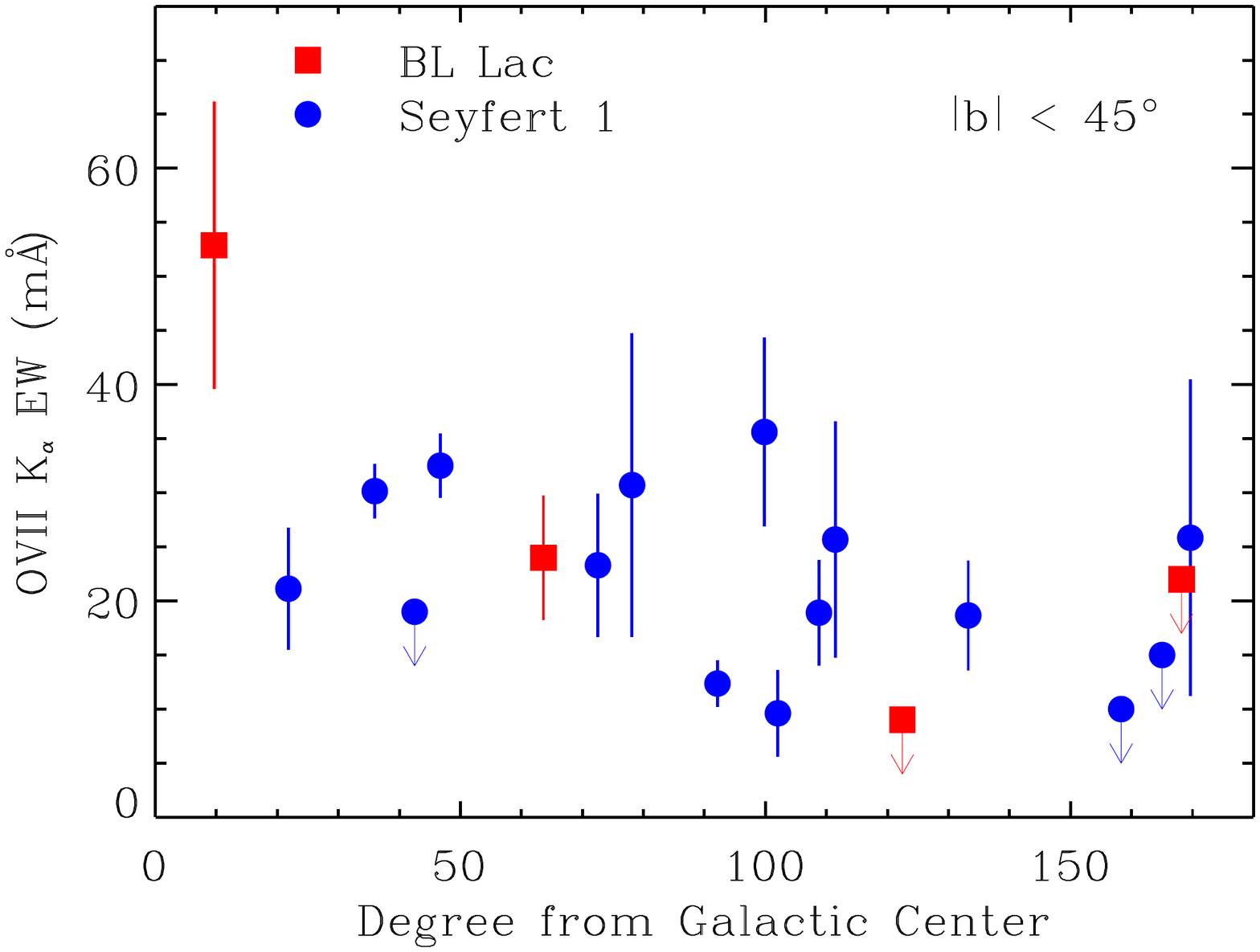}
\includegraphics[height=0.2\textheight,width=0.47\textwidth]{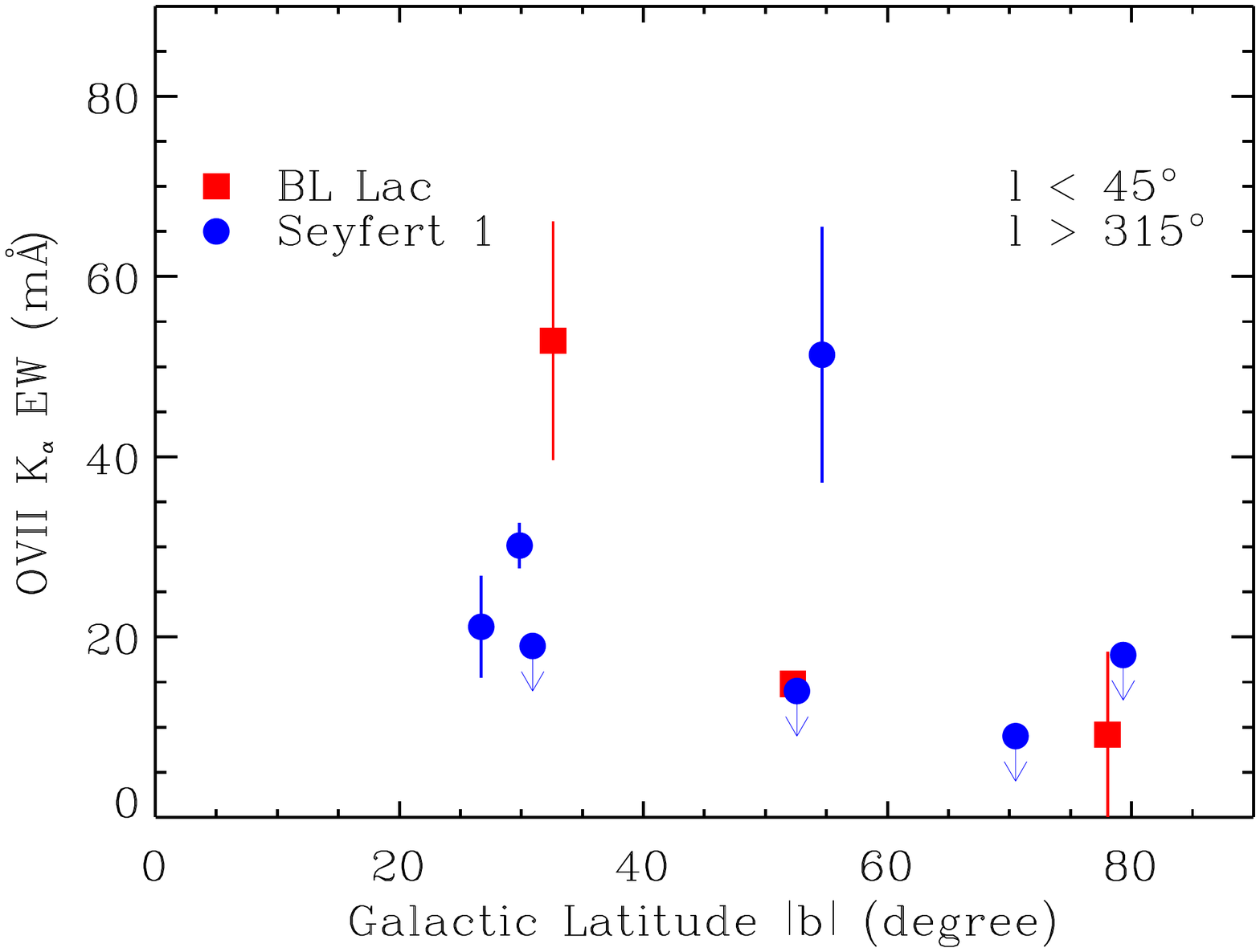}
\vskip-1cm
\includegraphics[height=0.2\textheight,width=0.47\textwidth]{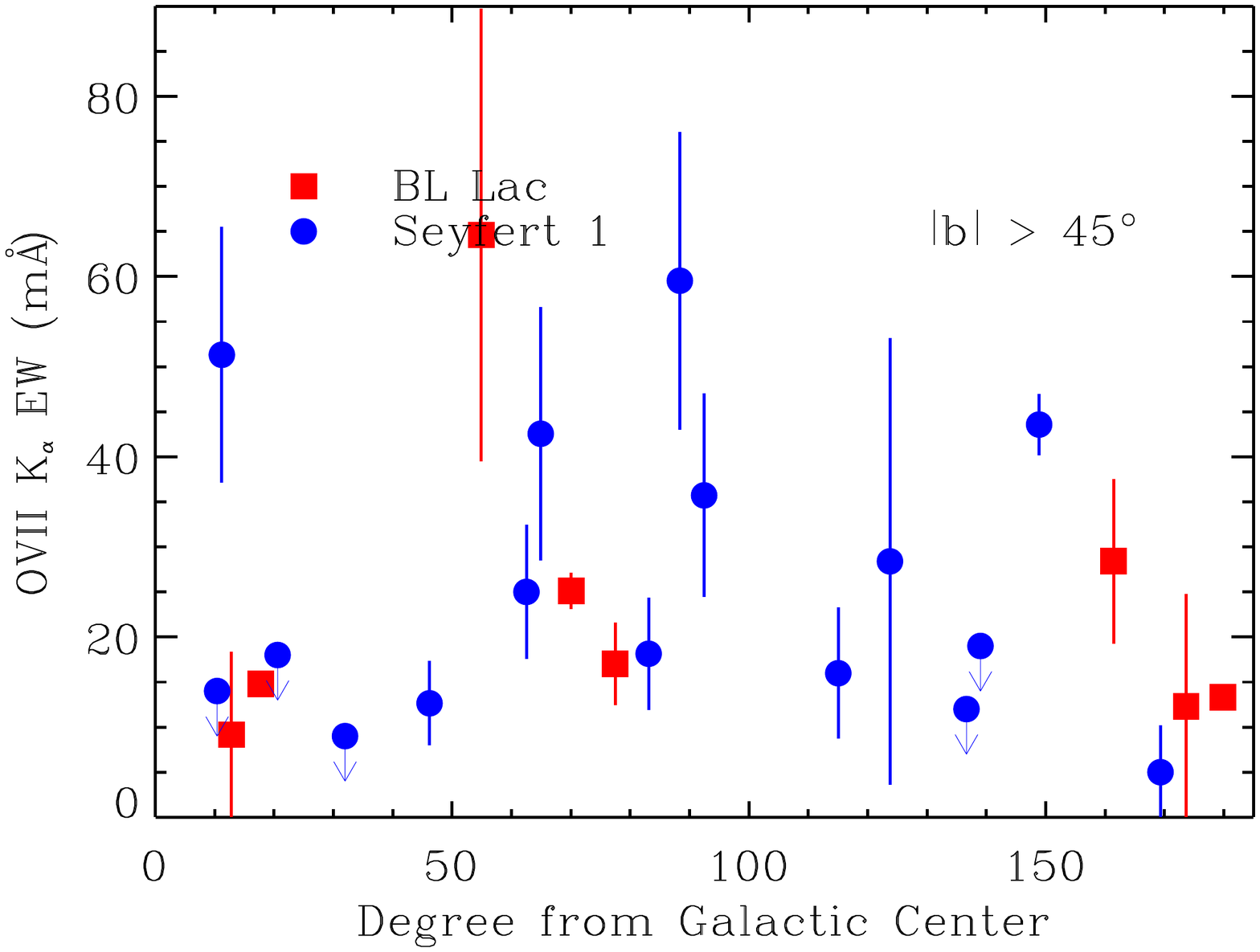}
\includegraphics[height=0.2\textheight,width=0.47\textwidth]{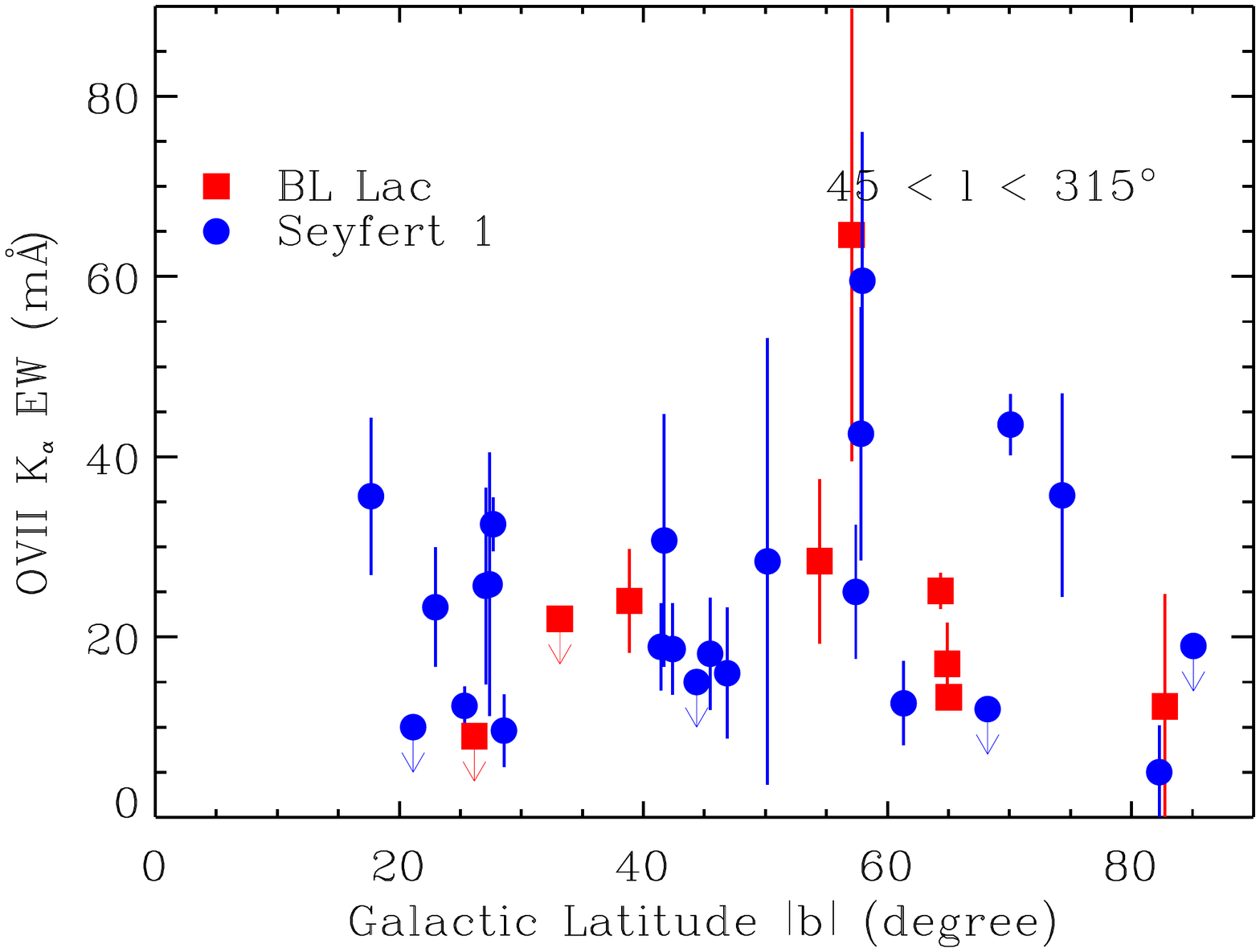}
\vskip-1cm
\includegraphics[height=0.2\textheight,width=0.47\textwidth]{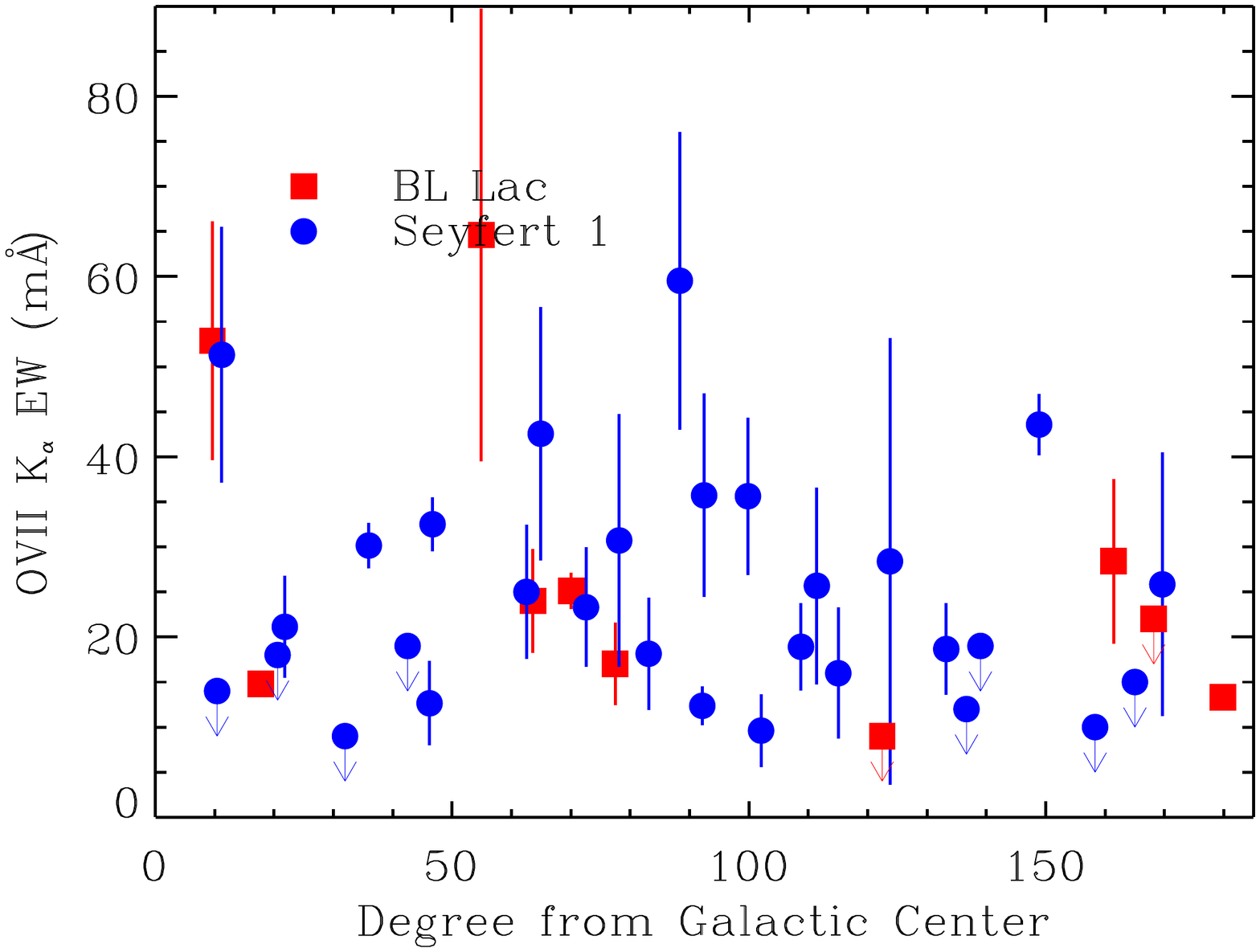}
\includegraphics[height=0.2\textheight,width=0.47\textwidth]{f54}
\vskip-0.5cm
\caption{\ion{O}{7} $K_{\alpha}$ line EW as a function of the Galactic longitude (left panels) and latitude (right panels). Red squares and blue circles represent BL Lac and Seyfert 1 galaxies, respectively. For the Galactic latitude plot, the $x$ axis is absolute latitude. For the Galactic longitude plot, the $x$ axis is the angle from the Galactic center. The left panels from top to bottoms are for targets with $|b| < 45^{\circ}$, $|b| > 45^{\circ}$, and all samples, respectively. The right panels from the top to bottoms are for targets within 45$^{\circ}$ from the Galactic center, more than 45$^{\circ}$ from the Galactic center, and all samples, respectively.}
\label{fig:lb}
\end{figure*}

\subsection{Absorption Line Analysis}

\begin{figure}[t]
\center
\includegraphics[height=0.35\textheight,width=0.5\textwidth]{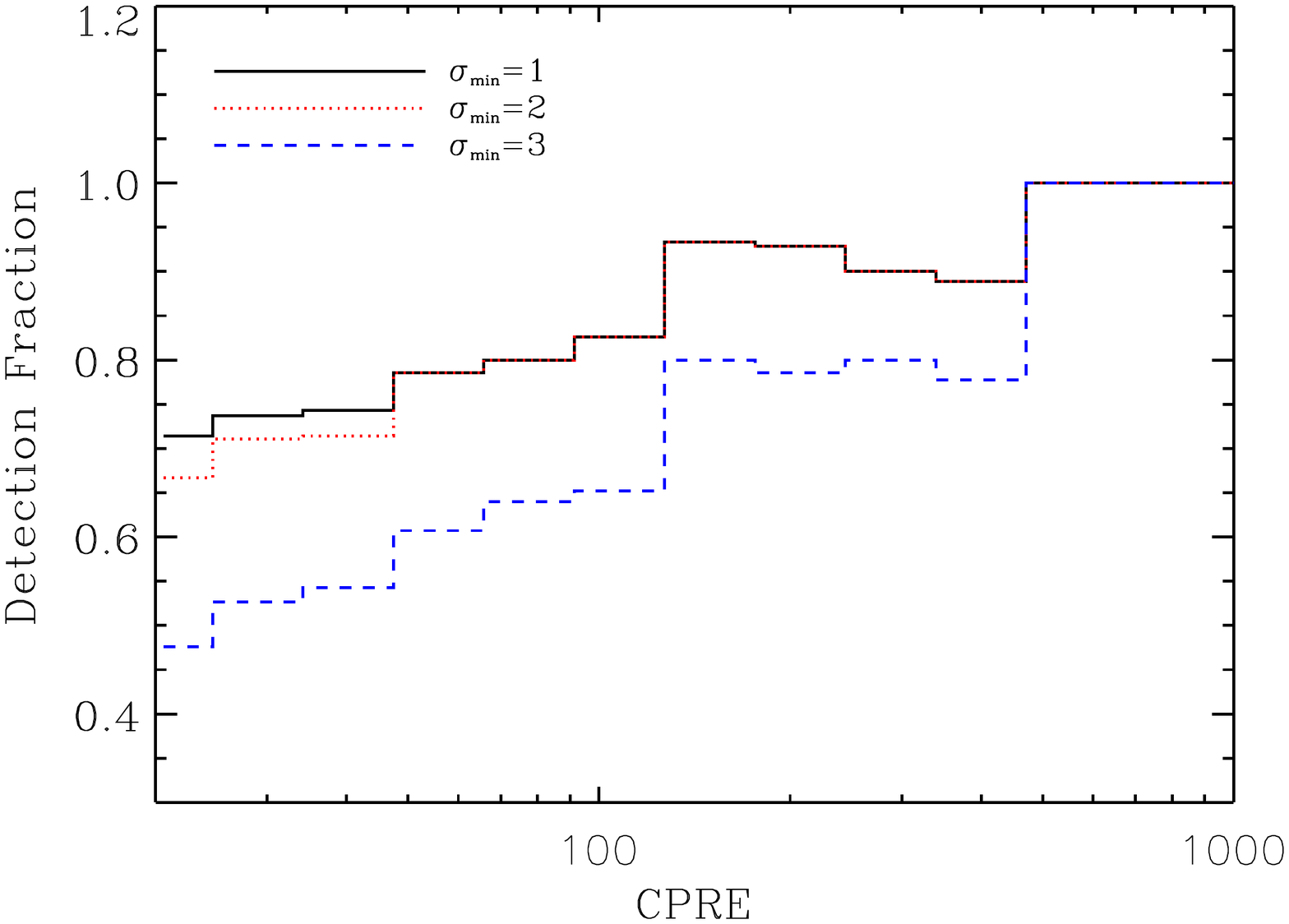}
\vskip-1cm		
\caption{Detection rate as a function of the CPRE. The dark solid, red dotted, and blue dashed lines represent the detection rates for lines with at least 1, 2, and 3$\sigma$ significance. The detection rates increase steadily for all the types to 100\% for the brightest targets.} 
\label{fig:det}
\end{figure}

We followed the data reduction process using the standard {\sl XMM}-Newton data analysis software SAS, version 12.0.1\footnote{See http://xmm.esa.int/sas/.}. Specifically, we made use the tool {\it rgsproc} to process the RGS data. We rejected events with flags of ``BAD\_SHAPE", ``ON\_BADPIX", ``ON\_WINDOW\_BORDER", and ``BELOW\_ACCEPTANCE". We also did not keep the cool pixels (``{\it keepcool = no}").

The above procedure outputs three files for each observation: A spectrum, a response file, and a background file. The reduced spectra were then analyzed with the X-ray software package XSPEC version 12.7.1\footnote{See http://heasarc.nasa.gov/xanadu/xspec/}. We fitted the continuum spectrum with a power law plus the absorption by the neutral hydrogen in our Galaxy. Since we are not interested in the broadband spectral properties, we selected only the spectral regions that contain the relevant spectral lines. The He-like oxygen has two transitions, the $K_{\alpha}$ line at 21.6019 \AA\ and the $K_{\beta}$ line at 18.6288 \AA\ (Verner et al.~1996). For most AGNs the $K_{\alpha}$ line is the only prominent feature, so we performed the fit of the \ion{O}{7} $K_{\alpha}$ line with a model that was developed in \citet{buote2009} and discussed also in \citet{fang2010}. We ignored the regions below 21 \AA, and above 22 \AA,\ as well as the region around an instrumental feature at $\sim21.8$ \AA.\ There are three free parameters in this Voigt-broadened line absorber model: the column density of the absorbing ion, the Doppler-$b$ parameter, and the redshift. This model reflects directly the physical properties of the absorbers and allows the line parameters to be tied consistently between the different spectra. The absolute wavelength uncertainty for RGS is about 100 $\rm km\ s^{-1}$.\footnote{See XMM-Newton Users Handbook at http://xmm.esac.esa.int/external/xmm\_user\_support/documentation/uhb/index.html}, so we limited the line center between with a redshift range of [-0.0004, 0.0004]. This range is also consistent with previous measurements in several high quality spectra. We also limited the Doppler-$b$ parameter in the range of 20 to 300 $\rm km\ s^{-1}$. The lowest temperature that can still produce a significant fraction of \ion{O}{7} is around $5\times10^5$ K under collisional ionization, this corresponds to a thermal velocity of $\sim 20\rm\ km\ s^{-1}$. The upper limit is adopted by assuming the absorbing gas shall not escape from our Galaxy. We performed the fit by minimizing the C-statistic (identical to maximizing the Poisson likelihood function) of \citet{cash1979}, which yields less biased best-fitting parameters \citep{humphrey2009}. 

Figure~\ref{fig:spec1}--\ref{fig:spec8} show the spectra of all  our targets. Here again we emphasize that the spectra for each exposure were combined only for plotting purposes. The red lines are the best-fit model. In fitting the spectra of a target with multiple exposures, we did not co-add the spectra to form a single spectrum and response. Rather, we simultaneously fitted all the exposures, each with its own response. We tied together the three line parameters (column density, Doppler-$b$ parameter, and line center shift) among all the observation for the same target while we let the continuum vary for each observation. This procedure takes longer time to find the best-fit parameters, however, it avoids introducing systematic errors that may arise from co-adding spectra with different responses (see \citealp{rasmussen2007} for details). 

Among all the 43 AGN spectra in our sample, the \ion{O}{7} $K_{\alpha}$ lines of 29 AGNs were discussed previously (see the last column of Table~\ref{tab:log}). We have included 14 additional AGNs that satisfy our selection criterion. Table~\ref{tab:log} columns 9--11 list the three fitting parameters and their 1$\sigma$ errors. Column (12) is the line equivalent width (EW), and (13) is the significance level. In total 21 targets show the \ion{O}{7} $K_{\alpha}$ line at more than the $3\sigma$ level, among which 7 were first detected in this work\footnote{We emphasize that the RGS spectra of some of these targets have been analyzed before, but this is the first time that detections of the local \ion{O}{7} $K_{\alpha}$ lines have been reported.}. We also list the $C$-statistic and the degree of freedom in column (14). In Table~\ref{tab:set} we listed the detection of \ion{O}{7} $K\alpha$ line at 3, 2, and 1$\sigma$ for different populations. 

For most AGNs we cannot constrain both the Doppler-$b$ parameter and velocity shift of the line center, so we also test another model in which we fix both values: We set the velocity shift of the line center at 0 $ \rm\ km\ s^{-1}$ and Doppler-b parameter at 100 $\rm\ km\ s^{-1}$. We listed the EW and the significance evaluated with this method in columns (15) and (16), respectively, in Table~\ref{tab:log}. While we see that both EW and the significance are slightly lower in this method, the results are essentiality consistent with those evaluated by allowing the line parameters to vary within a range. Unless all targets in actuality are properly described by these parameters, the derived line significances must be less in general, which is what we find. However, the differences are small and consistent with expectations that a range of $b$-parameters and velocity shifts are present in the sample.

Our default continuum model is a power law plus the Galactic neutral hydrogen absorption over approximately a 1 \AA\ range centered around the absorption lines of interest. The measured line properties crucially depend on the local continuum model, as well as the wavelength range over which the fitting is performed. To assess whether the power law model would be sufficient to characterize the continuum, we also investigated a more complex model, in particular a cubic-spline, to better match any fine details in the continuum. We found for most targets, such systematic errors are completely dominated by statistical errors. It only becomes important for the first several targets where the statistical errors are much smaller. In the case of Mkn~421, we found with a cubic-spline model the measured \ion{O}{7} K$\alpha$ line EW is $12.47\pm0.53$ m\AA.\  This is slightly lower than what we measured using the single power law model. Such variations are consistent with the large-scale uncertainties in the RGS effective area on 1\AA\ scale \citep{kaastra2006}. We also tested the systematic errors that may be caused by the wavelength range of our fitting. We found that as long as we restricted our fitting to within a wavelength size of $\sim$ 1\AA,\ the uncertainties caused by systematic errors are negligible.

A potential bias may occur when sometimes the fitting procedure tries to maximum the $C$-statistics. We tested our procedures by fitting simulated spectra. We define the bias as the difference between the input line EW and mean EW of the simulated spectra. We found in general for strong continuum with highly significant lines, the fitting procedure produced little or no statistically significant bias. For example, in the case of Mkn 421, the ratio between the bias and the standard deviation of the simulated EWs is less than 2\%. When the continuum is weaker and the line is less significant, this ratio becomes slightly larger, but in general less than $\sim$ 15\%. So our fitting procedure shall not produce statistically significant bias.

While our experience indicates that the such a model is often a reasonably good fit to the BL Lac-type targets, in the case of the Seyfert 1 galaxies, absorption features from the warm absorbers can sometimes  substantially complicate the analysis, and each case has to be treated differently. We use NGC~3516 as an example here. In Figure~\ref{fig:ngc3516} we show the combined spectrum of NGC~3516 between 21.2 and 22 \AA. This target was observed six times between 2001 and 2006, with a total exposure time of 285 $ksec$ \citep{mehdipour2010}. Here in the plot the data and fit are for illustration purposes only since the actual fit was performed on the individual observation simultaneously. In the left figure we see the broad \ion{O}{7} absorption line produced by the warm absorber intrinsic to the AGN, centered at $\sim$ 21.7 \AA. Using the power law plus the broad absorption line as our continuum model, In the bottom panel of the left figure, the $\Delta \chi^2$ plot clearly reveal the narrow, $z\sim0$ \ion{O}{7} absorption feature at around 21.6 \AA\footnote{While we use the $C$-statistics in this paper for less biased parameter fitting, here the conventional $\chi^2$ statistics is adopted to demonstrate the significance of the absorption feature.}. In the left figure we show the fit after including the local  \ion{O}{7} absorption line. The $\Delta \chi^2$ is improved substantially ($\Delta \chi^2 = 11$ for 77 degrees of freedom). 

Figure~\ref{fig:ewcol} shows the histogram distribution of the \ion{O}{7} $K_{\alpha}$ K$\alpha$ line EW (left panel) and column density (right panel). It appears that the most lines have an EW of $\sim$ 20 m\AA, with a range of 10 -- 30 m\AA. The column density distribution also centers around $\sim 10^{16}\rm cm^{-2}$, with a range of $10^{15.5}$ -- $10^{16.5} \rm cm^{-2}$.

In Figure~\ref{fig:lb} we show the dependence of the \ion{O}{7} absorption EW on the Galactic longitude (left figures) and latitude(right figures). Red squares and blue circles represent BL Lac and Seyfert 1 galaxies, respectively. For targets with a detection significance of at least $1\sigma$, we only plot the upper limits. To get a better understanding of the spatial distribution of the \ion{O}{7} absorbers, in the right panels of Figure~\ref{fig:lb} we divided the targets into two sectors based on their Galactic longitudes: Those close to the Galactic center within $45^{\circ}$ (the top right panel of Figure~\ref{fig:lb}) and those more than 45$^{\circ}$ away from the Galactic center (the middle right panel). The bottom left panel shows the dependence on the Galactic latitide of the entire sample. In the left panel of Figure~\ref{fig:lb} we also divided the targets into two sectors based on their Galactic latitudes: Those close to the Galactic plane with $|b| < 45^{\circ}$ (the top left panel of Figure~\ref{fig:lb}) and those that have $|b| > 45^{\circ}$ (the middle left panel). The bottom left panel shows the dependence on the Galactic longitude of the entire sample. 

We have calculated the correlation between the line EW and the Galactic latitude and longitude, using software package ASURV Rev 1.2 (\citealp{isobe1990, lavalley1992}), which implements the methods presented in \citet{isobe1986}. ASURV package is particular useful tool for correlation analysis of censored astronomical data, i.g., non-detections or detection limits (see, e.g., \citealp{keane2014}). We listed in Table~\ref{correlation} the probability of correlation by chance for three correlation tests: Cox hazard model, generalized Kendall's tau, and Spearman's rho. A correlation exists if the probability is less than 5\%. We have found in general the line EW has no correlation with either the Galactic latitude and the longitude. The only exception is the correlation between the line EW and the Galactic longitude for all the $|b|<45^{\circ}$ targets, in which all three tests give probabilities of less than 5\%; however, a close look reveals that the correlation is largely dominated by one target, PKS~2005-398. After removal of this target, the probabilities increase to above 5\%, and the correlation disappears.

\citet{gupta2002} suggested that most lines they detected are saturated, leading to larger ion column densities in their sample. The Voigt-fitting procedure we adopted allows us to extract directly the column density and Doppler-$b$ parameter information. Our study indicates that at least for the three brightest targets for which we can constrain the Doppler-$b$ parameter, the moderate variation of the Doppler-b parameters
demonstrated by Mkn~421 ($b \sim 71\rm\ km\ s^{-1}$), PKS~2155-304 ($\sim 74\rm\ km\ s^{-1}$) and
3C~273 ($\sim 129\rm\ km\ s^{-1}$) does suggest a likely complicate environment for the OVII production. While opacity correction may be important for Mkn~421 and PKS~2155-304 due to their small $b$ values, such correction shall become less important for targets such as 3C~273 with relatively larger $b$-values. The line saturation may need to be quantified on a case-by-case base when allowed by photon statistics.

In Figure~\ref{fig:det} we show the detection rate, or the sky covering fraction, as a function of the source flux. The sky covering fraction is defined as the ratio between the number of detections and the sample size. The solid black, dotted red, and dashed blue lines are for the detections with at least 3, 2, 1$\sigma$ significance, respectively. For the 3$\sigma$ case, the covering fraction increases steadily from $\sim$ 40\% for targets with a CPRE of more than 20 counts, to 100\% for targets with more than 600 CPRE. For 1$\sigma$ significance, the covering fraction becomes 100\% at $\sim$ 100 CPRE. Our study suggests that the \ion{O}{7} absorbers have a rather uniform distribution in the CGM/ISM.

\section{Discussion}

\subsection{Consistency Check with Previous Work}

The detection (or non-detection) of the \ion{O}{7} $K_{\alpha}$ absorbers in a large fraction of our targets have been reported previously. Here we compare our results with those obtained previously to assess the consistency of our work. 

We first compare our results of the three brightest targets (Mkn~421, PKS~2155-304, and 3C~273) with previous work since they have been studied multiple times by different groups, with both {\sl Chandra} and {\sl XMM}-Newton. We focus on the line EW since this is the observable that can be compared directly. We list the values of our work and several previous studies in Table~\ref{tab:comp}.  Columns 2, 3, and 4 are EWs measured in this work, with {\sl XMM}-Newton, and with {\sl Chandra}, respectively. Our results are largely consistent with previous work within the error ranges. However, a noticeable difference is that for Mkn~421 and PKS~2155-304, our results are systematically higher than \citet{williams2005} and \citet{williams2007} by $\sim$ 30\%. While they used {\sl Chandra}, for high quality data like these BL Lacs, such a discrepancy cannot be explained by the systematics between different instruments \citep{kaastra2006}. Our results are consistent with \citet{rasmussen2007}, in which they explained in detail the likely cause of this discrepancy with \citet{williams2005}. We refer the readers to this paper for further details.

For the rest of the targets, we mostly compare with results from \citet{bregman2007}, whose targets consist a subsample of ours. For most detections with more than $1\sigma$ significance, our results are consistent with theirs within $\sim$ 10\% or within the error ranges. A few noticeable difference are as follows: In NGC~4051, PKS~0558-504, MR~2251-178, our best-fit values are different from theirs by a factor of 2--3; in NGC~7469 we have a 2.9$\sigma$ detection while they had a non-detection; in PG~1116+215 we have a non-detection while they detected the \ion{O}{7} $K_{\alpha}$ line at 2.2$\sigma$. Except for NGC~7469, our exposure times are significantly larger than those of \citet{bregman2007} (3 -- 7 times longer), so our work should have a better statistical precision. For NGC~7469, we do not yet have a clear reason for the discrepancy. 

\subsection{Galactic Absorbers or AGN Outflows?}

Highly ionized, X-ray absorbing gas has been frequently detected in the nearby Seyfert galaxies (see, e.g., Reynolds et al.~1997; Tombesi et al.~2013). While these systems offer an excellent opportunity to probe the inner environment of the AGNs, their presence poses a serious challenge to the study of the CGM/ISM in our Galaxy. These absorbers, in the form of warm absorbers or ultra-fast outflows, often have outflow velocities ranging from a few hundred km $s^{-1}$ to mildly relativistic.  Sometimes such a velocity can exactly cancel the redshift effect of the AGN, making the identification of these absorbers particularly difficult. A careful case-by-case study is therefore necessary to determine whether these absorbers are intrinsic to the AGNs or local in the CGM medium in our Galaxy.

In our sample 31 background sources are Seyfert 1 galaxies, among which 20 show the detection of the local \ion{O}{7} $K_{\alpha}$ line at more than the 2$\sigma$ level. We made a dedicated study of the warm absorbers reported in these 20 targets. We found 6 Seyfert 1 galaxies have no reported warm absorbers; for the remaining 14 targets, we list them and information regarding their warm absorbers in Table~\ref{tab:seyfert} (also see the next subsection for comments on the targets with newly discovered $z\sim0$ \ion{O}{7} absorption line). Column 1 of Table~\ref{tab:seyfert} is the target name; column 2 is the recession velocity; column 3 is the reported velocities for various warm absorber outflows; and the last column is the reference in which the warm absorbers were discussed.  We found that in NGC~4051 we do have difficulty separating the local absorber from various components of the warm absorbers. In the case of PG~1211+143, while \citet{pounds2003} claimed that the \ion{O}{7} $K_{\alpha}$ line is actually produced by a relativistic  outflow with a velocity of $\sim 24,000 \rm\ km\ s^{-1}$, \citet{kaspi2006} found the absorbers can be explained as a much slower outflow but from other ion species. For all the other cases, it appears outflows cannot compensate the recession velocity, indicating the detected \ion{O}{7} $K_{\alpha}$ lines are most likely local at $z\sim0$.

There are further evidence that suggest most of the \ion{O}{7} absorbers should be local to our Galaxy. In a typical complex environment of the AGNs, the warm absorber/ultra-fast outflows are photo-ionized by the AGN, and exhibits absorption features in a variety of ion species in different ionization states. Sometimes emission lines are also detected. These absorbers/emitter typically coexist spatially, and form outflow components with different velocities (see NGC~7469 in our sample). However, for most reported \ion{O}{7} absorbers in our Seyfert 1 galaxy sample, we searched the entire spectrum for each data set and did not find other ion species (except in some cases \ion{Ne}{9} and \ion{O}{8}) share similar outflowing velocity if assuming they are intrinsic to the AGNs. Such a single-line absorber is unlikely to be produced in the rich environment of an AGN.

\subsection{Comments on Individual Target with Newly Discovered Absorption Line}

We briefly discuss the targets that have a newly detected, $z\sim0$ \ion{O}{7} absorber at more than the 2$\sigma$ level.

{\it Mkn~335 ---} This is a Seyfert 1 galaxy at $z=0.0258$. Previously neither X-ray or UV observations showed any intrinsic absorption (see, e.g., Reynolds et al.~1997; Zheng et al.~1995). However, In 2007 and later on, Mkn~335 went into a low state, providing an opportunity to study potential spectral features in its broad-line region (see, e.g, Grupe et al.~2007; Longinotti et al.~2008). Since then {\sl XMM}-Newton observations revealed various outflow components at a velocity of $\sim 5000\rm\ km\ s^{-1}$ (see, e.g., Longinotti et al.~2013), while {\sl Hubble} Space Telescope observations also indicated outflows in ultraviolet with veolcity of $\sim 5000\rm\ km\ s^{-1}$ (see, e.g., Longinotti et al.~2013). 

{\it NGC~7469 ---} This is also a Seyfert 1 galaxy at $z=0.0169$ \citep{vaucouleurs1991}. This target was extensively studied for its ionized outflow and received broadband coverage in optical, ultraviolet and X-ray. Outflows at two main velocity regimes, 600 and 2300 $\rm km\ s^{-1}$, were detected in both UV and X-ray observations (see, e.g., \citealp{scott2005, blustin2007}). Our detected feature at 21.6 \AA\ can also be interpreted as an outflow intrinsic to NGC~7469; however, that would imply a velocity of $\sim 5000\rm\ km\ s^{-1}$, which would be inconsistent with any of the known outflows. Furthermore, outflows often are associated with multiple features in emission and/or absorption. We examined the spectrum and cannot find any other lines from major ion species with similar velocity. Therefore, we conclude the feature at 21.6 \AA\ is most likely produced by \ion{O}{7} in our Galaxy.

{\it 1H~0707-495 ---} This is a narrow-line Seyfert 1 galaxy at $z=0.0411$ \citep{remillard1986}. it is a well-studied target in the X-ray because of its broad iron emission lines that have been interpreted as relativistic emission lines from an ionized accretion disk \citep{fabian2009}. Lately a mild relativistic highly ionized outflow was also suggested \citep{dauser2012}. The 21.6 \AA\  absorption feature was first noticed in \citet{blustin2009}; however, since they plotted the spectrum in the rest-frame of the AGN, this feature appeared at a wavelength of $\sim$ 20.7 \AA,\ leading them to conclude this is an unknown feature.\

{\it PG~1211+143 ---} This is a bright quasar at $z=0.0809$ \citep{marziani1996}. Based on the first $\sim 60\ ksec$ observation with {\sl XMM}-Newton, \citet{pounds2003} reported the detection of a high-velocity ionized outflow with velocity around $24,000\rm\ km\ s^{-1}$. Interestingly, as pointed out in \citet{mckernan2004}, this velocity can cancel the redshift of the quasar and put the absorber in our Galaxy. \citet{kaspi2006} reanalyzed this observation and concluded that most absorption can be explained by a mild outflow with velocity around $3,000\rm\ km\ s^{-1}$. This target was lately observed three more times with {\sl XMM}-Newton. Here our analysis is based on all the four observations with a total exposure time of $\sim 218\ ksec$. We confirmed the detection of the 21.6 \AA\ feature. We notice that in the Kaspi \& Behar model, while they can identify most high-velocity ion species in \citet{pounds2003} with a different ion species at a much lower velocity, they still cannot explain the features at 13.45 and 21.6 \AA. Coincidently, these are the rest-frame wavelength for the \ion{Ne}{9} and \ion{O}{7} $K_{\alpha}$ lines, respectively, suggesting these two features may indeed be produced by the hot gas local to the Milky Way.

{\it ESO~198-24 ---} This is one of the two Seyfert 1 galaxies ($z=0.0453$) in our sample that have no reported intrinsic absorption or warm absorber (Porquet et al.~2004). The only significant spectral feature that was detected was the Iron $K_{\alpha}$ line at $\sim 6.4\ kev$ (see, e.g, Guainazzi et al.~2003).

{\it Mkn~841 ---} This is also a Seyfert 1 galaxy at ($z=0.0364$) that has been extensively studied by many X-ray satellites. Its soft excess can be explained either by the reflection from the photoionized disk \citep{crummy2006}, or as ionized absorption in a relativistic wind (see, e.g., \citealp{gierlinski2004,petrucci2007}). \citet{longinotti2010} analyzed the high resolution RGS data, and found multiple component warm absorbers with velocities up to $\sim1600\rm\ km\ s^{-1}$.

{\it He1143-1810 ---} This is another Seyfert 1 galaxy ($z=0.0329$) that has no reported intrinsic absorption or warm absorber. \citet{cardaci2011} found a broad emission feature centering around 21.45 \AA, with a width of $\sim$ 2 \AA. They identified this feature as blended \ion{O}{7} emission triplets intrinsic to the source.

{\it 1ES\ 1028+511, PKS~2005-489, {\rm and} 3C~279 ---} All three targets are BL Lac-type objects in our sample. For 1ES~1028+511, previous {\sl Chandra} observations with HETGS did not reveal any absorption/emission features (Fang et al.~2005); however, Steenbrugge et al.~(2006) later reported tentative detections of narrow absorption features by 
intervening intergalactic medium, using {\sl Chandra} LETGS and {\sl XMM}-Newton RGS. No emission/absorption features were reported for PKS~2005-489 or 3C~279. 

\section{Summary}

The $z=0$ \ion{O}{7} absorption lines detected in the X-ray spectra of background AGN have a significant impact on our understanding of the Galactic gas content, both in the distant CGM environment or in the nearby Galactic disk. In this paper, we present an {\sl XMM}-Newton survey of these local \ion{O}{7} absorbers. We summarize our findings here.

\begin{itemize}

\item We studied 43 AGNs that satisfy our selection criteria, 12 targets are BL Lac-type, and 31 are Seyfert 1 Galaxies. A total of 21 targets show the detection of the $z=0$ \ion{O}{7} absorption line with more than the $3\sigma$ significance, among which 7 were newly discovered in this work.

\item We have fitted the $z=0$ \ion{O}{7} $K_{\alpha}$ transition with a Voigt-profile based line model. We find that most $K_{\alpha}$ lines have an EW of $\sim$ 20 m\AA, with a range of 10 -- 30 m\AA. The column density distribution also centers around $\sim 10^{16}\rm cm^{-2}$, with a range of $10^{15.5}$ -- $10^{16.5} \rm cm^{-2}$.

\item We correlate the line equivalent width with the Galactic coordinates and do not find any strong correlations between the line $EW$s with either the Galactic longitude or latitude.

\item The sky covering fraction, defined as the ratio between the number of detections and the sample size, increase from at about 40\% for all targets to 100\% for the brightest targets, suggesting a uniform distribution of the \ion{O}{7} absorbers.

\item Some AGNs have warm absorbers that may complicate the analysis of the local X-ray absorbers. A case-by-case study indicates that while in very few targets we do have difficulty separating local and intrinsic absorbers, for the majority the 21.6 \AA\ absorbers appear to be located in our Galaxy.

\end{itemize}

The study of the $z=0$ \ion{O}{7} absorbers plays an important role in understanding the hot gas in and around our Galaxy, since \ion{O}{7} samples a broad range of temperature around $10^6$ K. When combined with a study of the x-ray emission, an X-ray absorption study of the Galactic X-ray binaries, as well as a number of other investigations, we expect our study can help provide important clues to the properties of the CGM medium, the multi-phase model of the ISM, as well as questions such as the ``missing Galactic baryons". We plan to present our theoretical modeling in a later paper.

\acknowledgments
We thank the referee for very helpful comments and suggestions. TF was partially supported by the National Natural Science Foundation of China under grant No.~11273021, by ``the Fundamental Research Funds for the Central Universities" No.~2013121008, and by the Strategic Priority Research Program ``The Emergence of Cosmological Structures" of the Chinese Academy of Sciences, Grant No. XDB09000000. RM was partially supported by the National Natural Science Foundation of China under the grant No.~11333004.


\clearpage

\clearpage
\begin{deluxetable}{lcccccccccccrcccc}
\setlength{\tabcolsep}{0.03in}
\tablewidth{0pt}
\tabletypesize{\scriptsize}
\rotate
\tiny
\tablecaption{The AGN Sample}
\tablehead{Name & Type$^a$ & $l$ & $b$ & $z$ & $\rm N_H$$^{b}$ & Exposure   & CPRE & Log N(O VII)    & $b$ & Velocity                  & EW & S/N & $C$/dof & EW & S/N & Note$^c$ \\ 
                  &       &    &       &          &       ($10^{20}\rm cm^{-2}$)                  & ($ksec$) &      & ($\rm cm^{-2}$)    & ($\rm km\ s^{-1}$) & ($\rm km\ s^{-1}$)  & (m\AA)              &  &  & (m\AA) & \\ 
(1) & (2) & (3) & (4) & (5) & (6) & (7) & (8) & (9) & (10) & (11) & (12) & (13) & (14) & (15) & (16) & (17)} 
\startdata
Mkn~421  & 1 & 179.832 & 65.032 & 0.0300 & 1.31 & 929 & 14554 & $16.15^{+0.91}_{-0.34}$  & $63^{+81}_{-24}$  & $46^{+48}_{-43}$  & $13.13\pm0.67$ & 19.6 & 2544/2386 & $12.95\pm0.59$ & 21.9 & 1$^d$ \\ 
PKS~2155-304  & 1 & 17.730 & -52.246 & 0.1160 & 1.71 & 1086 & 3591 & $16.15^{+1.48}_{-0.39}$  & $74^{+225}_{-54}$  & $31^{+88}_{-90}$  & $14.82\pm1.20$ & 12.4 & 2567/2535 & $14.83\pm1.12$ & 13.2 & 1,2,3 \\ 
3C~273  & 1 & 289.951 & 64.360 & 0.1583 & 1.79 & 769 & 1333 & $16.52^{+1.48}_{-0.47}$  & $114^{+185}_{-66}$  & $142^{+90}_{-90}$  & $25.12\pm2.11$ & 11.9 & 1728/1692 & $24.53\pm1.82$ & 13.5 & 1,4 \\ 
Ark~564  & 2 & 92.138 & -25.337 & 0.0247 & 6.38 & 614 & 1281 & $16.38^{+1.29}_{-0.85}$  & $45^{+254}_{-25}$  & $24^{+96}_{-144}$  & $12.36\pm2.18$ & 5.7 & 691/768 & $11.92\pm2.14$ & 5.6 & 5 \\ 
Mkn~509  & 2 & 35.971 & -29.855 & 0.0344 & 4.09 & 876 & 1254 & $17.83^{+0.60}_{-1.73}$  & $72^{+227}_{-52}$  & $50^{+70}_{-90}$  & $30.15\pm2.55$ & 11.8 & 703/738 & $29.09\pm2.15$ & 13.5 & 6 \\ 
NGC~4051  & 2 & 148.883 & 70.085 & 0.0023 & 1.38 & 677 & 551 & $17.11^{+1.15}_{-0.70}$  & $159^{+127}_{-56}$  & $120^{+0}_{-49}$  & $43.57\pm3.43$ & 12.7 & 562/406 & $41.28\pm3.79$ & 10.9 & 6 \\ 
PKS~0558-504  & 2 & 257.962 & -28.569 & 0.1372 & 4.55 & 759 & 544 & $16.95^{+0.70}_{-1.75}$  & $24^{+275}_{-4}$  & $-118^{+238}_{-2}$  & $9.61\pm4.03$ & 2.4 & 458/408 & $9.46\pm3.79$ & 2.5 & 5 \\ 
MCG~-6-30-15  & 2 & 313.292 & 27.680 & 0.0077 & 4.09 & 447 & 505 & $16.57^{+1.96}_{-0.42}$  & $155^{+144}_{-135}$  & $120^{+0}_{-240}$  & $32.51\pm3.00$ & 10.8 & 377/346 & $32.04\pm3.35$ & 9.6 & 7 \\ 
Mkn~766  & 2 & 190.681 & 82.270 & 0.0129 & 1.35 & 669 & 401 & $15.82^{+1.53}_{-0.82}$  & $27^{+272}_{-7}$  & $120^{+0}_{-240}$  & $5.00\pm5.21$ & 1.0 & 337/299 & $5.61\pm4.06$ & 1.4 & $\star$ \\ 
NGC~3516  & 2 & 133.236 & 42.403 & 0.0088 & 3.23 & 481 & 292 & $17.83^{+0.40}_{-2.28}$  & $20^{+280}_{-0}$  & $-120^{+240}_{-0}$  & $18.66\pm5.10$ & 3.7 & 177/136 & $14.69\pm4.85$ & 3.0 & 5 \\ 
H~1426+428  & 1 & 77.487 & 64.899 & 0.1291 & 1.36 & 312 & 277 & $16.30^{+1.84}_{-0.83}$  & $75^{+224}_{-55}$  & $120^{+0}_{-227}$  & $17.02\pm4.56$ & 3.7 & 166/211 & $15.33\pm4.90$ & 3.1 & 5 \\ 
MR~2251-178  & 2 & 46.197 & -61.325 & 0.0640 & 2.71 & 393 & 250 & $15.69^{+2.23}_{-0.69}$  & $202^{+97}_{-182}$  & $-120^{+240}_{-0}$  & $12.64\pm4.69$ & 2.7 & 148/138 & $11.04\pm5.47$ & 2.0 & 5 \\ 
Mkn~335  & 2 & 108.763 & -41.424 & 0.0258 & 3.99 & 412 & 227 & $15.90^{+2.30}_{-0.31}$  & $300^{+0}_{-280}$  & $-81^{+201}_{-39}$  & $18.91\pm4.87$ & 3.9 & 274/274 & $17.80\pm4.96$ & 3.6 & $\star$ \\ 
Mkn~501  & 1 & 63.600 & 38.859 & 0.0337 & 1.74 & 171 & 212 & $16.75^{+1.62}_{-0.98}$  & $86^{+213}_{-66}$  & $68^{+52}_{-188}$  & $24.00\pm5.77$ & 4.2 & 156/141 & $22.91\pm5.18$ & 4.4 & 8 \\ 
ESO~141-55  & 2 & 338.183 & -26.711 & 0.0371 & 5.14 & 232 & 190 & $15.92^{+2.43}_{-0.33}$  & $300^{+0}_{-280}$  & $-120^{+223}_{-0}$  & $21.13\pm5.65$ & 3.7 & 116/136 & $17.62\pm5.72$ & 3.1 & 9 \\ 
NGC~7469  & 2 & 83.099 & -45.467 & 0.0163 & 4.96 & 163 & 140 & $15.88^{+2.42}_{-0.49}$  & $300^{+0}_{-280}$  & $-120^{+240}_{-0}$  & $18.14\pm6.23$ & 2.9 & 88/87 & $16.02\pm6.62$ & 2.4 & 5 \\ 
NGC~3783  & 2 & 287.456 & 22.947 & 0.0097 & 9.10 & 304 & 137 & $16.00^{+2.46}_{-0.37}$  & $300^{+0}_{-280}$  & $120^{+0}_{-240}$  & $23.30\pm6.65$ & 3.5 & 147/103 & $19.82\pm6.61$ & 3.0 & 10 \\ 
Mkn~279  & 2 & 115.042 & 46.865 & 0.0305 & 1.82 & 173 & 134 & $17.65^{+0.62}_{-2.65}$  & $20^{+280}_{-0}$  & $120^{+0}_{-240}$  & $15.98\pm7.28$ & 2.2 & 101/99 & $13.62\pm8.02$ & 1.7 & 11 \\ 
NGC~5548  & 2 & 31.960 & 70.496 & 0.0172 & 1.83 & 146 & 119 & ... & ... & ... & $ < 9$ & ... & ... & ... & ... & ...$^e$ \\ 
H~2356-309  & 1 & 12.839 & -78.035 & 0.1654 & 1.33 & 242 & 119 & $15.97^{+2.11}_{-0.97}$  & $44^{+255}_{-24}$  & $29^{+91}_{-149}$  & $9.17\pm9.18$ & 1.0 & 77/79 & $9.20\pm7.40$ & 1.2 & 12 \\ 
PKS~0548-322  & 1 & 237.566 & -26.145 & 0.0690 & 2.19 & 178 & 117 & ... & ... & ... & $ < 9$ & ... & ... & ... & ... & ... \\ 
NGC~4593  & 2 & 297.483 & 57.403 & 0.0090 & 2.37 & 103 & 109 & $16.91^{+1.65}_{-1.15}$  & $89^{+210}_{-69}$  & $-42^{+162}_{-78}$  & $25.00\pm7.46$ & 3.4 & 118/78 & $24.93\pm7.43$ & 3.4 & 13 \\ 
1H~0707-495  & 2 & 260.169 & -17.672 & 0.0406 & 6.06 & 446 & 109 & $18.43^{+0.38}_{-1.88}$  & $20^{+280}_{-0}$  & $-120^{+240}_{-0}$  & $35.62\pm8.73$ & 4.1 & 67/87 & $29.46\pm8.36$ & 3.5 & $\star$ \\ 
Ark~120  & 2 & 201.695 & -21.131 & 0.0327 & 11.86 & 111 & 99 & ... & ... & ... & $ < 10$ & ... & ... & ... & ... & ... \\ 
1ES~1028+511  & 1 & 161.439 & 54.439 & 0.3604 & 1.25 & 251 & 86 & $16.13^{+2.58}_{-0.43}$  & $300^{+0}_{-280}$  & $-120^{+218}_{-0}$  & $28.41\pm9.13$ & 3.1 & 59/69 & $24.35\pm9.36$ & 2.6 & $\star$ \\ 
PG~1116+215  & 2 & 223.360 & 68.209 & 0.1765 & 1.29 & 361 & 68 & ... & ... & ... & $ < 12$ & ... & ... & ... & ... & ... \\ 
3C~390.3  & 2 & 111.438 & 27.074 & 0.0561 & 4.16 & 117 & 60 & $16.06^{+2.62}_{-0.75}$  & $300^{+0}_{-280}$  & $120^{+0}_{-240}$  & $25.68\pm10.93$ & 2.3 & 31/48 & $20.13\pm11.07$ & 1.8 & 5 \\ 
PG~1211+143  & 2 & 267.552 & 74.315 & 0.0809 & 2.73 & 218 & 57 & $16.61^{+2.31}_{-0.71}$  & $173^{+126}_{-153}$  & $120^{+0}_{-240}$  & $35.71\pm11.30$ & 3.2 & 62/51 & $35.89\pm13.48$ & 2.7 & $\star$ \\ 
1H~1219+301  & 1 & 186.359 & 82.735 & 0.1836 & 1.74 & 29 & 54 & $15.71^{+2.71}_{-0.71}$  & $300^{+0}_{-280}$  & $-120^{+240}_{-0}$  & $12.28\pm12.48$ & 1.0 & 25/40 & $9.60\pm12.13$ & 0.8 & $\star$ \\ 
PKS~2005-489  & 1 & 350.373 & -32.601 & 0.0710 & 5.03 & 62 & 48 & $18.79^{+0.20}_{-1.47}$  & $20^{+280}_{-0}$  & $120^{+0}_{-240}$  & $52.88\pm13.28$ & 4.0 & 39/34 & $51.60\pm16.69$ & 3.1 & $\star$ \\ 
Fairall~9  & 2 & 295.073 & -57.826 & 0.0470 & 3.05 & 162 & 47 & $17.13^{+1.86}_{-1.18}$  & $156^{+143}_{-136}$  & $-116^{+236}_{-4}$  & $42.55\pm14.09$ & 3.0 & 39/32 & $41.50\pm16.79$ & 2.5 & 6 \\ 
NGC~7213  & 2 & 349.588 & -52.580 & 0.0058 & 2.09 & 172 & 47 & ... & ... & ... & $ < 14$ & ... & ... & ... & ... & ... \\ 
Mkn~841  & 2 & 11.209 & 54.632 & 0.0364 & 2.36 & 111 & 45 & $18.75^{+0.24}_{-1.33}$  & $20^{+280}_{-0}$  & $-120^{+240}_{-0}$  & $51.31\pm14.22$ & 3.6 & 39/35 & $51.71\pm16.83$ & 3.1 & $\star$ \\ 
3C~120  & 2 & 190.373 & -27.397 & 0.0330 & 10.89 & 126 & 44 & $16.95^{+1.79}_{-1.69}$  & $84^{+215}_{-64}$  & $120^{+0}_{-240}$  & $25.84\pm14.65$ & 1.8 & 26/34 & $21.39\pm12.96$ & 1.7 & 6 \\ 
Mkn~110  & 2 & 165.011 & 44.364 & 0.0353 & 1.52 & 47 & 42 & ... & ... & ... & $ < 15$ & ... & ... & ... & ... & ... \\ 
ESO~198-24  & 2 & 271.639 & -57.948 & 0.0455 & 3.09 & 154 & 31 & $18.75^{+0.24}_{-2.69}$  & $98^{+201}_{-78}$  & $-120^{+218}_{-0}$  & $59.53\pm16.54$ & 3.6 & 17/24 & $61.05\pm19.23$ & 3.2 & $\star$ \\ 
IRAS~13349+2438  & 2 & 20.603 & 79.317 & 0.1076 & 1.12 & 153 & 31 & ... & ... & ... & $ < 18$ & ... & ... & ... & ... & ... \\ 
He~1143-1810  & 2 & 281.854 & 41.710 & 0.0329 & 3.50 & 31 & 31 & $16.17^{+2.71}_{-0.75}$  & $300^{+0}_{-280}$  & $120^{+0}_{-240}$  & $30.71\pm14.03$ & 2.2 & 14/24 & $24.10\pm15.97$ & 1.5 & $\star$ \\ 
IC~4329a  & 2 & 317.496 & 30.920 & 0.0161 & 4.38 & 148 & 26 & ... & ... & ... & $ < 19$ & ... & ... & ... & ... & 5 \\ 
TONS~180  & 2 & 138.995 & -85.070 & 0.0620 & 1.49 & 48 & 26 & ... & ... & ... & $ < 19$ & ... & ... & ... & ... & ... \\ 
Mkn~1502  & 2 & 123.749 & -50.175 & 0.0589 & 4.94 & 106 & 21 & $16.75^{+2.23}_{-1.75}$  & $106^{+193}_{-86}$  & $-120^{+240}_{-0}$  & $28.39\pm24.78$ & 1.1 & 16/12 & $17.00\pm27.69$ & 0.6 & ... \\ 
3C~279  & 1 & 305.104 & 57.062 & 0.5362 & 2.26 & 152 & 21 & $16.89^{+2.10}_{-0.87}$  & $299^{+0}_{-279}$  & $120^{+0}_{-240}$  & $64.62\pm25.15$ & 2.6 & 12/12 & $60.93\pm30.24$ & 2.0 & $\star$ \\ 
1H~0414+009  & 1 & 191.815 & -33.159 & 0.2870 & 10.47 & 90 & 21 & ... & ... & ... & $ < 22$ & ... & ... & ... & ... & $\star$ \\ 
\enddata
\label{tab:log}
\tablecomments{(a) 1 -- BL Lac, 2 -- Seyfert 1; (b) Galactic neutral hydrogen column density \citep{dickey1990}; (c) The references where the $z\sim0$ O VII absorption line was first reported. A ``$\star$" sign means the target is first reported in this paper. 
1.\ \citet{rasmussen2003},
2.\ \citet{fang2002b},
3.\ \citet{nicastro2002},
4.\ \citet{fang2003},
5.\ \citet{bregman2007a},
6.\ \citet{mckernan2004},
7.\ \citet{fang2006},
8.\ \citet{ren2014},
9.\ \citet{miller2013},
10.\ \citet{williams2006a},
11.\ \citet{kaspi2002},
12.\ \citet{buote2009},
13.\ \citet{steenbrugge2003a}; (d) For the three brightest targes, Mkn~421, PKS~2155-304, and 3C~273, we did not evalute the line parameters by fixing both the Doppler-$b$ parameter and the line center shift; (e) For targets with a significance of less than 1$\sigma$ we list the 3$\sigma$ upper limits of EW.}
\end{deluxetable}

\clearpage

\begin{deluxetable}{lcccc}
\tablewidth{0pt}
\tablecaption{Target Detection}
\tablehead{ & $>3\sigma$ &  $>2\sigma$ & $>1\sigma$ & Total }
\startdata
This work & 7 & 9 & 11 & 11 \\
\hline
BL Lac & 2 & 3 & 4 & 4 \\ 
\hline
Seyfert 1 & 5 & 6 & 7 & 7 \\
\hline\hline
All & 21 & 28 & 33 & 43 \\
\hline
BL Lac & 7 & 8 & 10 & 12 \\
\hline
Seyfert 1 & 14 & 20 & 23 & 31
\enddata
\label{tab:det}
\tablecomments{Statistics of the detected targets. Top three rows show the detected targets in this work; and bottom three rows show all the targets in our sample. Columns 2, 3, 4 list the targets detected with a S/N larger than 3, 2, and 1$\sigma$, respectively. The last column is the total number of targets in each group.}
\end{deluxetable}

\clearpage

\begin{deluxetable}{llccc}
\tablewidth{0pt}
\tablecaption{Correlation Tests$^a$}
\tablehead{Correlation & Subsample & Cox Hazard & Kendall Tau & Spearman Rho}
\startdata
EW vs. Degree from GC & $|b| < 45^{\circ}$, BL Lac & 6\% & 7\%& 10\% \\
                                        & $|b| < 45^{\circ}$, Seyfert 1 & 14\% & 29\% & 22\% \\
                                        & $|b| < 45^{\circ}$, Total & 2\% & 4\% & 4\% \\
                                        & $|b| < 45^{\circ}$, Total$^b$ & 5\% & 12\% & 10\% \\
\hline
EW vs. Degree from GC & $|b| > 45^{\circ}$, BL Lac & 86\% & 90\%& 88\% \\
                                        & $|b| > 45^{\circ}$, Seyfert 1 & 93\% & 89\% & 94\% \\
                                        & $|b| > 45^{\circ}$, Total & 89\% & 98\% & 90\% \\
\hline
EW vs. Degree from GC & BL Lac & 33\% & 22\%& 22\% \\
                                        & Seyfert 1 & 44\% & 47\% & 46\% \\
                                        & Total & 25\% & 18\% & 22\% \\
\hline
EW vs. $|b|$ & $l < 45^{\circ}$ or $l > 315^{\circ}$, BL Lac & 11\% & 12\%& 16\% \\
                     & $l < 45^{\circ}$ or $l > 315^{\circ}$, Seyfert 1 & 28\% & 40\% & 27\%\\
                     & $l < 45^{\circ}$ or $l > 315^{\circ}$, Total & 7\% & 17\% & 13\% \\
\hline
EW vs. $|b|$ & $315 > l > 45^{\circ}$, BL Lac & 46\% & 83\%& 87\% \\
                     & $315 > l > 45^{\circ}$, Seyfert 1 & 49\% & 96\% & 94\%\\
                     & $315 > l > 45^{\circ}$, Total & 86\% & 97\% & 89\% \\
\hline
EW vs. $|b|$ & BL Lac & 82\% & 39\%& 61\% \\
                     & Seyfert 1 & 16\% & 56\% & 56\%\\
                     & Total & 24\% & 33\% & 42\% \\                                  
\enddata
\label{correlation}
\tablecomments{a. Correlation tests using (1) Cox Hazard, (2) Kendall Tau, and (3) Spearman Rho methods. Columns 3---5 gives the probability that the correlation between eh two parameters is obtained by chance, and a less than 5\% probability suggests the two parameters are correlated. b. Total sample excluding the bright BL Lac object PKS~2005-398.}
\end{deluxetable}

\clearpage

\begin{deluxetable}{lcll}
\tablewidth{0pt}
\tablecaption{Comparison with Previous Observations}
\tablehead{Name & This Work & {\sl XMM}-Newton & {\sl Chandra} \\
 & (m\AA) & (m\AA) & (m\AA)}
\startdata
Mkn~421 & $13.13\pm0.67$ & $13.1\pm1.0$ [1] & 11-13 [2]  \\
                &                            & $11.8\pm0.8$ [3] & $9.1\pm1.1$ [4] \\ \\
\hline \\
PKS~2155-304 & $14.82\pm1.20$ & $13.7\pm1.9$ [3] & $13.3\pm2.8$ [5] \\
                          &                            & $16.3\pm3.3$ [6] & $11.6\pm1.6$ [7]  \\  \\
\hline \\
3C~273 & $25.12\pm2.11$ &  $24.6\pm3.3$ [3] & $28.4^{+12.5}_{-6.2}$ [8] \\
              &                            & $26.3\pm4.5$ [6] &        \\                        
\enddata
\label{tab:comp}
\tablecomments{[1]: \citet{rasmussen2007}; [2] \citet{kaastra2006}; [3] \citet{bregman2007}; [4]  \citet{williams2005}; [5] \citet{hagihara2010}; [6] \citet{rasmussen2003}; [7] \citet{williams2007}; [8] \citet{fang2003}   }
\end{deluxetable}

\clearpage
\begin{deluxetable}{lccl}
\tablewidth{0pt}
\tablecaption{Seyfert 1 Galaxies with Warm Absrobers}
\tablehead{Name & $cz^{a}$ & Warm Absorber & Reference\\
 & ($\rm km\ s^{-1}$) & ($\rm km\ s^{-1}$) & }
\startdata
Ark~564   & $7472$  & 200--2000 & \citet{smith2008} \\
Mkn~509 & $10358$      & $\sim 1000$ & \citet{kaastra2012} \\
NGC~4051 & $800$ & 200--600, 2340, 4600 & \citet{pounds2013} \\
                   &                                &  9000, 0.12$c^b$            & \\
MCG-6-30-15 & $2610$ & 0, 1900 & \citet{holczer2010}                   \\
NGC~3516 & $2201$& $\sim$ 100, 1000 & \citet{mehdipour2010} \\
MR~2251-178 & 19194 &  $<130, \sim 480, 15600$ & \citet{reeves2013} \\
Mkn~335 & $7740$ & 5000 & \citet{longinotti2013} \\
NGC~7469 & $4619$ & 600, 2300 & \citet{blustin2007}  \\
Mkn~279 & 9150 & 0--1500 & \citet{ebrero2010} \\
NGC~3783 & $2409$ & 450--750 & \citet{kaspi2002} \\
                   &                                        & 1360-2080 & \\
NGC~4593 & $2651$ & 200--600 & \citet{steenbrugge2003a}                    \\
1H~0707-495 & $11880$ & 0.11--0.18$c$ & \citet{dauser2012} \\
PG~1211+143 & $24450$ & $3000^c$ & \citet{kaspi2006} \\
                        &                                        & $24000^c$ & \citet{pounds2003} \\
Mkn~841         &  10920 & $<1600$ & \citet{longinotti2010}                        
\enddata
\label{tab:seyfert}
\tablecomments{(a): Recession velocity based on redshift. (b): $c$ is the speed of light. (c): This is one warm absorber, but \citet{kaspi2006} and \citet{pounds2003} have different interpretations.}
\end{deluxetable}

\end{document}